\documentclass[11pt]{article}
\pdfoutput=1
\usepackage{epsfig}
\usepackage{graphicx}
\usepackage{multicol}
\usepackage{amsmath,amssymb}
\numberwithin{equation}{section}
\usepackage{cite}
\usepackage{graphicx,wrapfig,color}
\usepackage{multicol}
\usepackage{tabularx}

\numberwithin{equation}{section}

\usepackage[utf8x]{inputenc}

\usepackage{hyperref}         % hypertext links %%ARXIV
\usepackage{graphicx}
\usepackage{times,psfrag,rotating}
\usepackage{lscape}
\usepackage{aas_macros}

\usepackage{amsmath}
\usepackage{amssymb}
\usepackage{enumerate}
\usepackage[caption=false]{subfig}
\usepackage{epstopdf}
\usepackage{color}
\usepackage{array}
\usepackage{multirow, multicol}
\newcolumntype{A}{ >{\centering\arraybackslash} m{3.5cm} }
\newcolumntype{B}{ >{\centering\arraybackslash} m{2.5cm} }
\newcolumntype{C}{ >{\centering\arraybackslash} m{2cm} }
\newcolumntype{D}{ >{\centering\arraybackslash} m{1.5cm} }
\newcolumntype{E}{ >{\centering\arraybackslash} m{2.5cm} }

\newcommand{\pri}[1]{{\color{black}#1}}     
\newcommand{\suk}[1]{{\color{black}#1}}   
\def\bel#1{\begin{equation} \label{#1}}

\def\be{\begin{equation}}
\def\ee{\end{equation}}
\def\bea{\begin{eqnarray}}
\def\eea{\end{eqnarray}}

\def\ltap{\ \raise.3ex\hbox{$<$\kern-.75em\lower1ex\hbox{$\sim$}}\ }
\def\gtap{\ \raise.3ex\hbox{$>$\kern-.75em\lower1ex\hbox{$\sim$}}\ }
\def\gl{\ \raise.5ex\hbox{$>$}\kern-.8em\lower.5ex\hbox{$<$}\ }
\def\roughly#1{\raise.3ex\hbox{$#1$\kern-.75em\lower1ex\hbox{$\sim$}}}

\newcommand{\comments}[1]{}

\newcommand{\ben}{\begin{enumerate}}
\newcommand{\een}{\end{enumerate}}
\newcommand{\bi}{\begin{itemize}}
\newcommand{\ei}{\end{itemize}}
\newcommand{\ba}{\begin{align}}
\newcommand{\ea}{\end{align}}

\def\NG{{\rm NANOGrav}~}
\def\OMG1{{$\Omega_{\rm GW}^{(1)}$}~}
\def\OMG2{{$\Omega_{\rm GW}^{(2)}$}~}

\def\beq{\begin{equation}}
\def\eeq{\end{equation}}
\def\bea{\begin{eqnarray}}
\def\eea{\end{eqnarray}}

\def\ln#1{\mathrm{ln}\left(#1\right)}

\begin{document}
\thispagestyle{empty}

\begin{center}

%\vspace*{0.5cm}
{\LARGE\bf Implications of the NANOGrav result on primordial gravitational waves in nonstandard cosmologies\\}
\bigskip

{\large Sukannya Bhattacharya}\,$^{a}$, \, \, 
{\large Subhendra Mohanty}\,$^{a}$, \, \, 
{\large Priyank Parashari }\,$^{a,b}$, \, \, \\

\bigskip 
\bigskip
{\small
$^a$Theoretical Physics Division, Physical Research Laboratory, Navrangpura, Ahmedabad - 380009, India. \\[2mm]
$^b$ Indian Institute of Technology, Gandhinagar, 382355, India \\[2mm]

}
\end{center}
\begin{center}
\it{
sukannya@prl.res.in, mohanty@prl.res.in, parashari@prl.res.in}
\end{center}
\bigskip 
 %
%
%\vspace*{1cm}
\begin{center} 
{\bf Abstract} 
\end{center}
%\vspace*{-0.35in}
\begin{quotation}
\noindent 
{\small
Recently, the \NG collaboration has reported the evidence for a common-spectrum stochastic process, which might be interpreted as the first ever detection of stochastic gravitational wave (GW) background. We discuss the possibility of the signal arising from the first and second order GWs in nonstandard cosmological history. We show that the \NG observation can be explained by the first order GWs in the nonstandard thermal history with an early matter dominated era, whereas the parameter space required to explain the \NG observation in the standard cosmology or in the nonstandard epoch of kination domination is ruled out by the BBN and CMB observations. For the second order GWs arising from the large primordial scalar fluctuations, we study the standard radiation domination and two specific nonstandard cases with a few forms of the primordial power spectrum $P_{\zeta}(k)$ to achieve abundant primordial black hole (PBH) production. We find that the \NG observation can be explained with standard radiation domination for all of these $P_{\zeta}(k)$. Furthermore, a dustlike epoch leads to abundant PBH formation for a lower amplitude of $P_{\zeta}(k)$ than the radiation dominated case, and complies with the \NG observation only for a few of the all $P_{\zeta}(k)$ forms considered here, where the peak wavenumber is larger than the wavenumber range probed by the \NG . In this nonstandard epoch, for a broad power spectrum, PBHs are produced in a wide mass range in the planetary mass regime. A nonstandard epoch of kination domination cannot produce enough PBH for any of the $P_{\zeta}(k)$ if the \NG result is to be satisfied.
}
\end{quotation}

\newpage
\tableofcontents
\section{Introduction}
\setcounter{footnote}{0}
\label{intro}
The North American Nanohertz Observatory for Gravitational Waves (\NG) searches for an isotropic stochastic gravitational wave background (SGWB) by analyzing the cross-power spectrum of pulsar timing residuals~\cite{Arzoumanian:2018saf}. The recent $12.5$ year pulsar timing array (PTA) data released by the \NG has reported the discovery of a stochastic common-spectrum process~\cite{Arzoumanian:2020vkk}, which can be fitted into a power law in a narrow range of frequencies. At this moment, they report that the angular correlations are inconclusive and therefore, the collaboration does not claim a positive signal of gravitational waves (GWs). However, if the signal is interpreted as an upper bound on GW background, we can explore the primordial universe with the \NG constraints.
%The recent analysis of the $12.5$ year pulsar timing array data collected by the North American Nanohertz Observatory for Gravitational Waves (NANOGrav) has found evidence for a common spectrum process, which can be a stochastic gravitational wave (GW) signal[ref NANOGrav]. This signal, observed in a narrow range of frequencies around $f_0=5.5$ nHz, can be fitted into a power law $\Omega_{\rm GW}= \Omega_{{\rm GW},0}(f/f_0)^{\zeta}$ with 1$\sigma$ limits of the observed amplitude and exponent given by $\Omega_{{\rm GW},0}\equiv \Omega_{{\rm GW}}(f_0)\in (3\times 10^{-10},2\times 10^{-9})$ and $\zeta \in (-1.5,0.5)$. At this point, even though NANOGrav do not claim this as a GW detection due to the absence of the quadrupole correlations in the signal, if it is interpreted to be a gravitational wave background then it is interesting to analyse the primordial production of GW with this signal. 
Assuming the \NG signal to be a GW background, the source of such a signal can be supermassive black hole merger events~\cite{Sesana:2004sp,Arzoumanian:2020vkk}, cosmic strings~\cite{Ellis:2020ena,Blasi:2020mfx,Buchmuller:2020lbh,Samanta:2020cdk}, phase transition of a hidden sector\cite{Nakai:2020oit,Addazi:2020zcj,Ratzinger:2020koh,Li:2020cjj}, inflationary perturbations in the first~\cite{Vagnozzi:2020gtf} or second orders~\cite{Vaskonen:2020lbd,DeLuca:2020agl,Kohri:2020qqd,Domenech:2020ers} etc.~\cite{Pandey:2020gjy}.
%. [PICK REFS FROM KOHRI TERADA].

The LIGO/VIRGO detection of GWs from the binary black hole mergers has opened up a new window towards the early universe cosmology~\cite{Abbott:2016blz,Abbott:2016nmj,TheLIGOScientific:2016pea,Abbott:2016bqf,Abbott:2017vtc,Abbott:2017oio,Abbott:2017gyy}. The probability that the observed black holes of supersolar masses are of primordial origin has also rekindled the interest in the primordial black holes (PBHs)~\cite{Fernandez:2019kyb}. Other than such merger events, GWs of a stochastic nature can also originate from the primordial tensor perturbations generated during inflation~\cite{Caprini:2018mtu,Christensen:2018iqi}. The amplitude of the primordial tensor power spectrum at the cosmic microwave background (CMB) pivot scale ($k_{\rm pivot}=0.05$ Mpc$^{-1}$) is constrained by the Planck 2018 and BICEP2/Keck Array~\cite{Ade:2018gkx,Aghanim:2019ame} observations in terms of the tensor-to-scalar ratio $r<0.056$ with $95\%$ confidence limit. \pri{However, if some feature in the inflationary dynamics induces a blue-tilt in the tensor power spectrum at smaller scales, GWs of larger amplitude are possible~\cite{Cook:2011hg,Barnaby:2011qe,Wang:2014kqa,Kuroyanagi:2014nba}. Since the \NG observation are at much higher frequencies, $f\in (2.5\times10^{-9},1.2\times10^{-8})$, as compared to the CMB frequencies, a blue tilted tensor power spectrum may be consistent with both the CMB and \NG observations.}
%~\cite{Cook:2011hg,Barnaby:2011qe,Wang:2014kqa,Kuroyanagi:2014nba}. 

However, in the second and higher orders of perturbation theory, GWs are sourced by the first order curvature perturbations~\cite{Ananda:2006af,Baumann:2007zm,Espinosa:2018eve,Kohri:2018awv}. 
If the curvature perturbations have a red-tilted power spectrum throughout inflation, such second order GWs are suppressed with respect to the first order contribution. \pri{Observation of the temperature and polarization anisotropies in the CMB sky has constrained the primordial scalar power spectrum to have a red tilt ($n_s=0.9649$) and tiny amplitude $A_s \sim \mathcal{O}(10^{-9})$ at the CMB pivot scale~\cite{Ade:2018gkx,Aghanim:2019ame}. However, if there is a large running of $n_s$, the primordial scalar power spectrum can have a blue-tilt at smaller scales, and then the primordial curvature perturbations can be large at these scales, which can induce second order GW of large amplitude (see~\cite{Clesse:2015wea,Garcia-Bellido:2017mdw,Germani:2017bcs,Ballesteros:2017fsr,Bhaumik:2019tvl,Mishra:2019pzq,Ballesteros:2020qam,Palma:2020ejf,Fumagalli:2020adf,Braglia:2020eai,Ozsoy:2020kat,Aldabergenov:2020yok,Braglia:2020taf} for relevant physical models). Such large curvature perturbations, leading to density contrast larger than the critical value for gravitational collapse, will lead to a copious amount of PBH production.}
%
%However, if the primordial scalar power spectrum has a blue-tilt at smaller scales then the primordial curvature perturbations can be large at these scales (see~\cite{Clesse:2015wea,Garcia-Bellido:2017mdw,Germani:2017bcs,Ballesteros:2017fsr,Bhaumik:2019tvl,Mishra:2019pzq,Ballesteros:2020qam,Palma:2020ejf,Fumagalli:2020adf,Braglia:2020eai,Ozsoy:2020kat,Aldabergenov:2020yok,Braglia:2020taf} for relevant physical models), which can induce second order GW of large amplitudes.
%} On the other hand, if the curvature perturbations are \suk{such that the density contrast is} larger than its critical value for gravitational collapse, \blue { then it leads a copious amount of PBH can be produced.}
 Hence, the phenomenologies of PBH and GW in the second order are always related~\cite{Hawking:1971ei,Carr:1974nx,Carr:1975qj}. PBH abundance with the \NG result has been studied in~\cite{Vaskonen:2020lbd,DeLuca:2020agl,Kohri:2020qqd,Domenech:2020ers}.
 %\footnote{For comparison of our results to these papers, see Sec.~\ref{ressec}}. 

Big Bang Nucleosynthesis (BBN) constraints demand that the universe was radiation dominated at least at the temperature $T_{\rm BBN}\simeq 5$ MeV. CMB constraint on $r$ reveals that the upper bound on the energy scale of inflaton potential is $V\simeq 10^{16}$ GeV~\cite{Akrami:2018odb,Akrami:2019izv}. 
For a standard slow roll-inflation, the energy scale of inflation remains nearly constant until the end of inflation. The epoch between the end of inflation and BBN, which can span a large number of e-folds, is not well constrained by the observations. In the standard theoretical description, the universe quickly becomes radiation dominated after the end of inflation, and the standard radiation epoch sustains until the matter radiation equality ($T_{\rm eq}$). However, it is possible that the equation of state $w$ of the universe deviates from radiation after the end of inflation and before BBN\footnote{For a review, see~\cite{Allahverdi:2020bys}.}.
An epoch of nonstandard post inflationary evolution can be motivated by slow reheating ($w$ slowly changes from $0$ to $1/3$)~\cite{Carr:2018nkm}; early matter dominated epoch ($w=0$) with energy dominated by the moduli field~\cite{Coughlan:1983ci,Banks:1993en,deCarlos:1993wie,Conlon:2005jm,Cicoli:2016olq,Bhattacharya:2017ysa,Maharana:2017fui}; kinetic energy domination ($w=1$) in case of the quintessential inflation models~\cite{Peebles:1998qn}; general stiff domination~\cite{DiMarco:2018bnw} etc. The dynamics of formation and evolution of PBH in such epochs is also different and their abundance is related to the amplitude of the second order GW. Therefore, any deviation from the standard radiation epoch before BBN affects the PBH and GW dynamics. 

In this paper, we inspect the first and second order primordial GWs in a non-standard post-inflationary epoch (i.e., equation of state $w\neq 1/3$) that is active at the scales probed by the \NG . \suk{The evolution of the first order tensor perturbations upon horizon entry depends on the background as $\propto a^{3(1+w)-4}$. Thus, for a standard radiation dominated (RD) epoch, the scale dependence of the first order GW spectrum is only due to the primordial tensor tilt $n_t$. However, for $w\neq 1/3$, the slope of the resulting GW spectra depends on $w$. Here, we study the first order GW spectra for $w=1/3$, $0$, and $1$ and find that only a fraction of the parameter space for the $w=0$ scenario is consistent with the \NG data, when the constraints from CMB, BBN, and LIGO are included. We compare our analysis for the first order GW spectrum in RD with ref.~\cite{Vagnozzi:2020gtf}.

The critical value $\delta _c$ of the density contrast strongly depends on the background in which the scalar perturbations enter the horizon and collapse to form PBHs. This has motivated many studies of PBH formation in nonstandard epochs~\cite{1981SvA....25..406P,1982AZh....59..639P,Hwang:2012bi,Alabidi:2013lya,Harada:2016mhb,Carr:2017edp,Harada:2017fjm,Inomata:2019zqy,Inomata:2019ivs,Figueroa:2019paj,Matsubara:2019qzv,Bhattacharya:2019bvk}. In this paper, to analyse second order GW spectra for abundant PBH formation, we have considered two nonstandard epochs: $w=1$, i.e. a kination epoch and $w=1/9$, which is a dustlike epoch. Since $\delta _c$ for both of these nonstandard epochs are smaller than $\delta _c^{\rm RD}$ (see Fig.3 in ref.~\cite{Harada:2013epa}), we expect to reach a given PBH abundance with much lower amplitudes of the primordial scalar perturbations than that required in RD. This work focuses on determining whether those values of the primordial scalar amplitude can explain the \NG observation. We follow the formation of PBH and evolution of the scalar induced GW spectra in second order in $w=1$ and $w=1/9$ epochs for a few functional forms of the primordial curvature power spectrum. We find that although abundant PBH formation in a kination epoch is not consistent with the \NG detection, for the dustlike epoch, large PBH abundance can explain the \NG data as a second order GW spectrum. The standard case of PBH formation and second order GW evolution in RD has also been studied here and compared with the contemporary works with a similar approach. We find our results for the RD epoch to be consistent with ref.~\cite{Kohri:2020qqd}, however, disagree with those of ref.~\cite{Vaskonen:2020lbd} due to reasons discussed later.}

The rest of the paper is organised as follows: in Sec.~\ref{GWsec}, we represent the result from the \NG in terms of the theoretical quantity: the GW spectrum $\Omega_{\rm GW}$. In Sec.~\ref{GW1sec}, we discuss the effect of a general equation of state $0\leq w\leq 1$ on the first order GW and constrain the relevant parameters using the power law nature of the observed GW signal, fitted by \NG using the frequency in the first five  bins. Sec.~\ref{GW2sec} relates the second order GW to the observation and translates the constraints in terms of PBH abundance. In Sec.~\ref{ressec}, we discuss our results and conclude.

\section{NANOGrav Result and Gravitational waves}
\label{GWsec}
PTA experiments typically describe the obtained result for GW background in terms of the characteristic strain spectrum $h_c(f)$. The quantity $h_c(f)$ is typically fitted with a power law dependence on frequency $f$~\cite{Zhao:2013bba,Liu:2015psa,Arzoumanian:2018saf}
\begin{eqnarray}
h_c(f) = A_{\rm GWB} \left ( \frac{f}{f_{\rm yr}} \right )^{\alpha_{_{\rm GWB}}}\,,
\label{eq:strain}
\end{eqnarray}
with $\alpha_{_{\rm GWB}}=(3-\gamma)/2$, where $\gamma$ is the timing-residual cross-power spectral density. The reference frequency is $f_{\rm yr} = {\rm yr}^{-1}=3.1\times 10^{-8}$ Hz. 
The recent 12.5 yr observation from the \NG~\cite{Arzoumanian:2020vkk} measured
 the characteristic strain in the frequency range $f\in (2.5\times10^{-9},1.2\times10^{-8})$ Hz and found the evidence of a stochastic common-spectrum process (CP). The data is fitted as a power law and the corresponding GW energy density $\Omega_{\rm GW}(f)$ is given as~\cite{Zhao:2013bba,Liu:2015psa,Arzoumanian:2018saf}:
\begin{eqnarray}
\Omega_{\rm GW}(f) = \frac{2\pi^2}{3H_0^2}f^2h_c^2(f)=\frac{2\pi^2f_{\rm yr}^2}{3H_0^2}A_{\rm CP}^2\bigg(\frac{f}{f_{\rm yr}}\bigg)^{5-\gamma _{\rm CP}}\,.
\label{eq:hcGW}
\end{eqnarray}
PTA data from \NG give a joint $A_{_{\rm CP}}$-$\gamma_{_{\rm CP}}$ posterior distribution.
%Results for GW searches by PTA experiments generally given in the joint $A_{_{\rm CP}}$-$\gamma_{_{\rm CP}}$ posterior distributions.
Therefore, early universe theories with a prediction for a stochastic GW signal can be constrained using this result in terms of the amplitude $A_{_{\rm CP}}$ and slope $\gamma_{_{\rm CP}}$.

In a perturbed spatially flat Friedmann-Lema\^{i}tre-Robertson-Walker (FLRW) universe, the tensor perturbations $h_{ij}$ are traceless and transverse. The tensor modes are given as:
%%
%\begin{eqnarray}
%ds^2 = a^2(\eta) \left [ (1-2\phi)d\eta^2 - (\delta_{ij}+2 \psi+ h_{ij})dx^idx^j \right ]\,,
%\label{eq:flrw}
%\end{eqnarray}
%where $a$ is the scale factor, $\eta$ is the conformal time, $\phi$ and $\psi$ are the scalar perturbations and $h_{ij}$ are the tensor perturbations in the metric. Tensor perturbations $h_{ij}$ are traceless and transverse. Moreover, we write 
\begin{equation}
h_{ij} (\eta,{\bf x}) = \sum_{\lambda} \int  \frac{d^3 {\bf k}}{(2\pi)^{3}} e^{i {\bf k.x}} \epsilon_{ij}^\lambda({\bf k})h_{{\bf k}}^\lambda({\eta}) ,
\end{equation}
where $\eta$ is conformal time, ${\bf k}$ is the comoving wavenumber and $ \epsilon_{ij}^\lambda({\bf k})$ are the polarization tensors for the two polarization states $\lambda = + ,\times $. The tensor power spectrum $\mathcal{P}_h$ is defined as:
\begin{equation}
\frac{k^3}{2\pi^{2}} \langle h_{{\bf k}}^\lambda({\eta})h_{{\bf k'}}^{\lambda'}({\eta}) \rangle = \delta_{\lambda \lambda'} \delta^3({\bf k}+{\bf k'})\mathcal{P}_h(k,\eta)\,.
\end{equation}
The GW energy density spectrum $\Omega_{{\rm GW}}$ is defined as the GW energy density $\rho_{{\rm GW}}$ in a comoving wavenumber interval $(k, k + dk)$, normalised with the critical density $\rho_{c}(\eta)$, 
\begin{equation}
\label{originalGWPh}
\Omega_{{\rm GW}} (\eta ,k) \equiv \frac{\rho_{\rm GW}(\eta,k)}{\rho_c (\eta)} = \frac{1}{24}\bigg(\frac{k}{ \mathcal{H}}\bigg)^2 \mathcal{P}_h(k,\eta)\, .
\end{equation}
Observationally relevant quantity is the GW spectrum at present $\Omega_{{\rm GW}}(\eta_0 ,k)$. 
%Furthermore, tensor power spectrum can be derived using perturbation theory at different order in perturbations.  
%At first order of perturbation theory, which is the dominant contribution for the simplest early universe scenarios, the scalar, vector and tensor fluctuations evolve independent of each other. However, their evolution are coupled for second and higher order perturbations. In the second order, tensor perturbations are sourced by scalar perturbations.
As discussed in Sec.~\ref{intro}, for a given theory, GW spectrum can be evaluated at first and higher orders of perturbation theory. Thus, theoretical predictions for GW resulting in different orders can be constrained with the \NG observation. When the scalar fluctuations give rise to PBH, then the PBH abundance can also be probed in terms of the observed GW spectra, \suk{where the second order contribution is dominant over the first order.}
%assuming subdominant contribution from the first order.

In this work, we focus on analysing the first and second order GW in a nonstandard cosmological history, where the background evolution deviates from the standard radiation domination (RD) before BBN. We separately derive the predictions for different orders and compare them with the \NG observation to discuss bounds on the relevant primordial quantities.

\section{Primordial gravitational waves in first order}
\label{GW1sec}
\begin{figure}[!b]
\begin{center}
\includegraphics[width=0.6\textwidth]{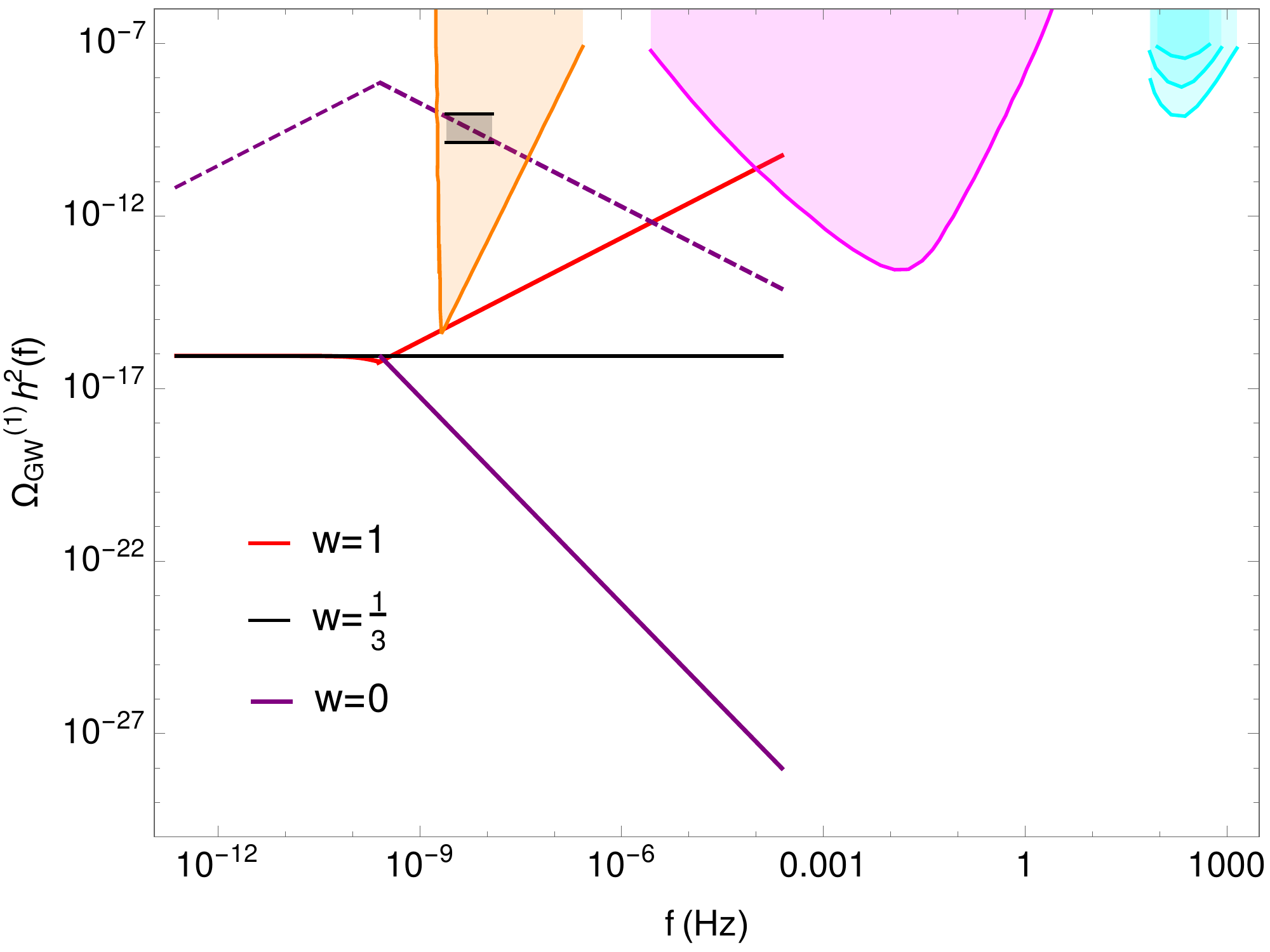}
\caption{ \small{First order GW spectrum for the three different cosmological histories are shown here. We have fixed $n_t = 0$, $H_{\rm inf} = 6.2 \times 10^{13}$ GeV and $f(T_1)=10^{-10}$ Hz. LIGO runs for O1,O2 and O5 (future) are given as cyan curves (top to bottom); magenta shaded region represnents the proposed sensitivity for the LISA survey; black shaded region represents the $1\sigma$ bound from the \NG 12.5 year observation; orange shaded part represents the proposed sensitivity from the future SKA survey. Purple dashed line shows the GW spectrum in cosmologcal history with an early MD epoch and $n_t=1$. }}\label{fig:first_GW_kd_rd_md}
\end{center}
\end{figure}

Inflation predicts a stochastic GW background arising from the tensor fluctuations~\cite{Caprini:2018mtu,Christensen:2018iqi}. \pri{In the absence of anisotropic stresses, the evolution of these tensor fluctuations is source free at the first order of perturbations.} GW background depends on the primordial tensor power spectrum, which can be written as:
\begin{equation}
\mathcal{P}_h(k,\eta)= T(k,\eta) \Delta ^2_{h,{\rm inf}}(k),\label{omtot}
\end{equation}
where $\Delta ^2_{h,{\rm inf}}(k) =\frac{2}{\pi^{2}}\bigg(\frac{H_{\rm inf}}{M_{\rm P}}\bigg)^2 \bigg(\frac{k}{k_{p}}\bigg)^{n_t}$ is the primordial tensor power spectrum at horizon re-entry and $T(k,\eta)$ is the transfer function. Here $n_t$ is the spectral index, $k_{p}$ is a pivot scale and $H_{\rm inf}$ is the Hubble rate when mode $k_{p}$ exited the horizon during inflation. For a mode $k$ re-entering the horizon at time $\eta(k)$, the transfer function depends on the background evolution from the time of its horizon re-entry until the time of its observation. 
In the case of standard cosmological evolution history, radiation domination (RD) starts just after inflation ends (assuming instant reheating) followed by the matter domination (MD). However, there can be a pre-BBN epoch with equation of state (EoS) other than the radiation ($w\neq1/3$) in alternative cosmological histories. In such alternative cosmological histories, the transfer function will be different for the modes re-entering the horizon during the nonstandard epoch. In this work, we have assumed a period with EoS $w\neq1/3$ from the end of inflation till the onset of RD at temperature $T_1$. GW energy density in such a nonstandard cosmological thermal history at the first order of perturbations is \cite{Bernal:2019lpc,Figueroa:2019paj,Bernal:2020ywq}:
\begin{eqnarray}
\Omega_{\rm GW,0}^{(1)}(k)&=&\frac{\Omega_{\rm rad, 0}}{12\pi^2}\bigg(\frac{g_{*,k}}{g_{s,k}}\bigg)\bigg(\frac{g_{s,0}}{g_{s,k}}\bigg)^{4/3}\bigg(\frac{H_{\rm inf}}{M_{\rm P}}\bigg)^2\frac{\Gamma ^2 (\alpha +1/2)}{2^{2(1-\alpha}\alpha ^{2\alpha}\Gamma ^2(3/2)}\mathcal{W}(\kappa)\kappa^{2(1-\alpha)} \bigg(\frac{k}{k_{p}}\bigg)^{n_t},\label{GW_1_gen}
\end{eqnarray}
where $\alpha=\frac{2}{1+3w}$, $\kappa=\frac{k}{k(T_1)}=\frac{f}{f(T_1)}$, and
\begin{equation}
\mathcal{W}(\kappa)= \frac{\pi \alpha}{2\kappa}\bigg[\bigg(\kappa J_{\alpha+1/2}(\kappa)-J_{\alpha-1/2}(\kappa)\bigg)^2+\kappa^2J^2 _{\alpha-1/2}(\kappa)\bigg],\label{Bessel_GW1}
\end{equation}
\pri{where $J_{i}$ and $\Gamma(\alpha)$ are the Bessel function of order $i$ and the gamma function respectively. $k(T_1)$ and $f(T_1)$ are the wavenumber and frequency corresponding to the temperature ($T_1$) of the universe at the time of the onset of RD epoch respectively. } Also, $\Omega_{\rm rad, 0}=9\times10^{-5}$ is the radiation energy fraction in the present universe, $g_{*}$ and $g_{s}$ are the number of relativistic degrees of freedom for the energy and entropy respectively. We also find that $\mathcal{W}(\kappa) \approx \alpha$ for $f> f(T_1)$. It is clear from Eq.\eqref{GW_1_gen} that the GW spectrum today depends on $H_{\rm inf}$, $w$, $n_t$, and $f(T_1)$ (for detailed discussion see~ref~\cite{Bernal:2019lpc,Figueroa:2019paj,Bhattacharya:2019bvk}). First order GW spectrum for the three different values of EoS,  $w=0,1/3$, and $1$, are shown in Fig.~\ref{fig:first_GW_kd_rd_md} where we have fixed $n_t = 0$, $H_{\rm inf} = 6.2 \times 10^{13}$ GeV and $f(T_1)=10^{-10}$ Hz. In order to do the analysis with the \NG result, it is useful to relate the \NG observational parameters $A_{_{\rm CP}}$ and $\gamma_{_{\rm CP}}$ to the set of theoretical parameters: ($H_{\rm inf}$, $w$, $n_t$, $f(T_1)$). Using the Eq.s~\eqref{eq:strain},  \eqref{eq:hcGW} and \eqref{GW_1_gen}, we get
\begin{eqnarray}
 A_{\rm CP} = \bigg( \frac{3 H_0^2}{2 \pi^2}\frac{\Omega _{\rm rad, 0}}{12\pi ^2}\bigg(\frac{g_{*,k}}{g_{s,k}}\bigg)\bigg(\frac{g_{s,0}}{g_{s,k}}\bigg)^{4/3} \frac{\Gamma ^2 (\alpha +1/2)}{2^{2(1-\alpha}\alpha ^{2\alpha}\Gamma ^2(3/2)} \alpha  \bigg)^{1/2}\frac{H_{\rm inf}}{M_{\rm P}} \bigg(
 \frac{f_{\rm yr}}{f_{p}}\bigg)^{n_t/2} \frac{{f_{\rm yr}}^{\alpha}}{{f_{T1}}^{1-\alpha}} \,, \label{eq:acp}
\end{eqnarray}
and
\begin{eqnarray}
\gamma_{_{\rm CP}} = 3+2\alpha-n_t\,.
\label{eq:gammacp}
\end{eqnarray}
Evidently, Eq.s~\eqref{eq:acp} and \eqref{eq:gammacp} show that $A_{\rm CP}$ depends on $H_{\rm inf}$, $w$, $n_t$, and $f(T_1)$, whereas $\gamma _{\rm CP}$ depends only on $w$ and $n_t$.
\begin{figure}[!t]
  \subfloat[\label{fig:acp_w}]{
	\begin{minipage}[c][1\width]{
	   0.5\textwidth}
	   \centering
	   \includegraphics[width=1.05\textwidth]{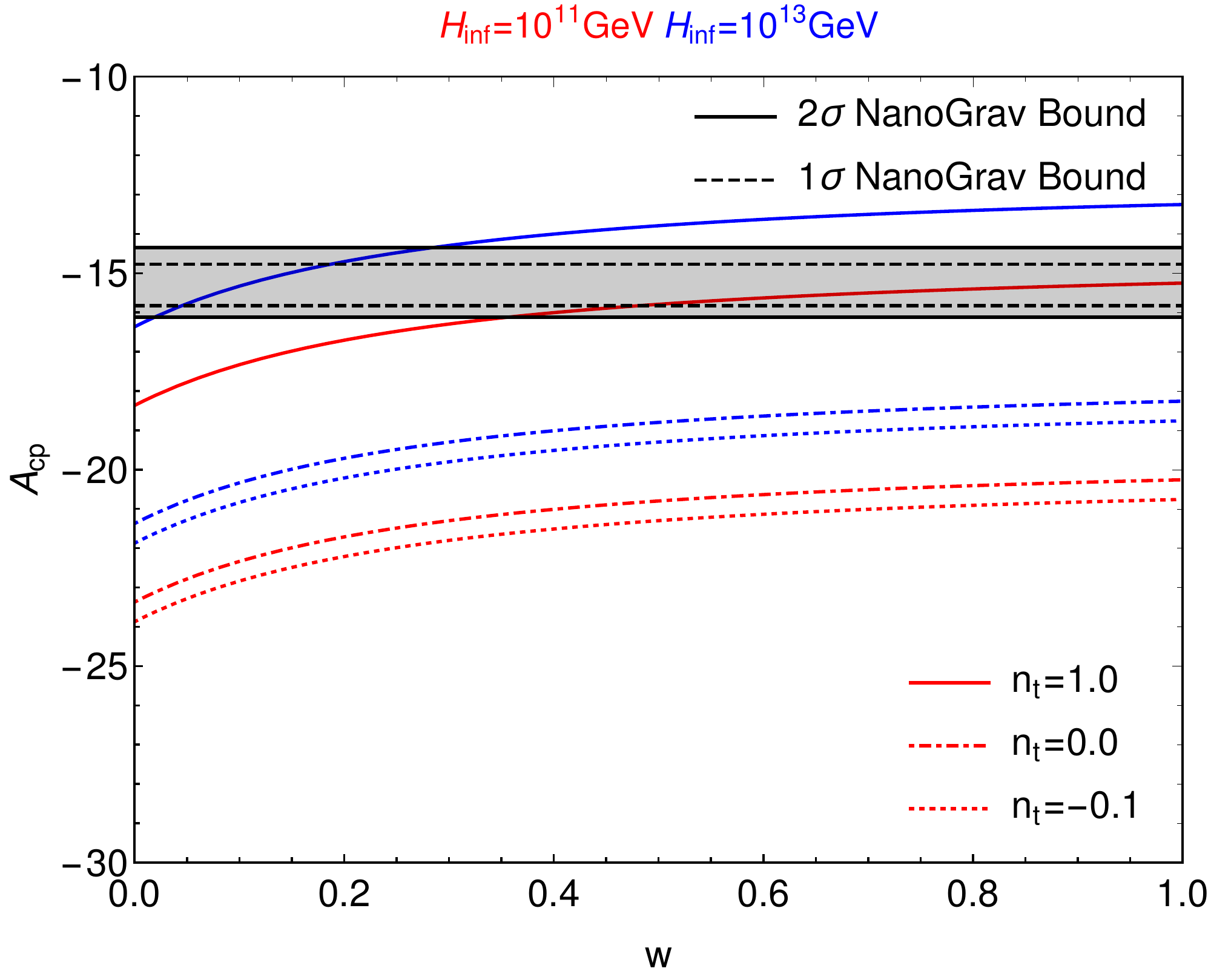}
	\end{minipage}}
 \hfill 	
  \subfloat[\label{fig:gammacp_w}]{
	\begin{minipage}[c][1\width]{
	   0.5\textwidth}
	   \vspace{0.3cm}
	   \centering
	   \includegraphics[width=0.98\textwidth]{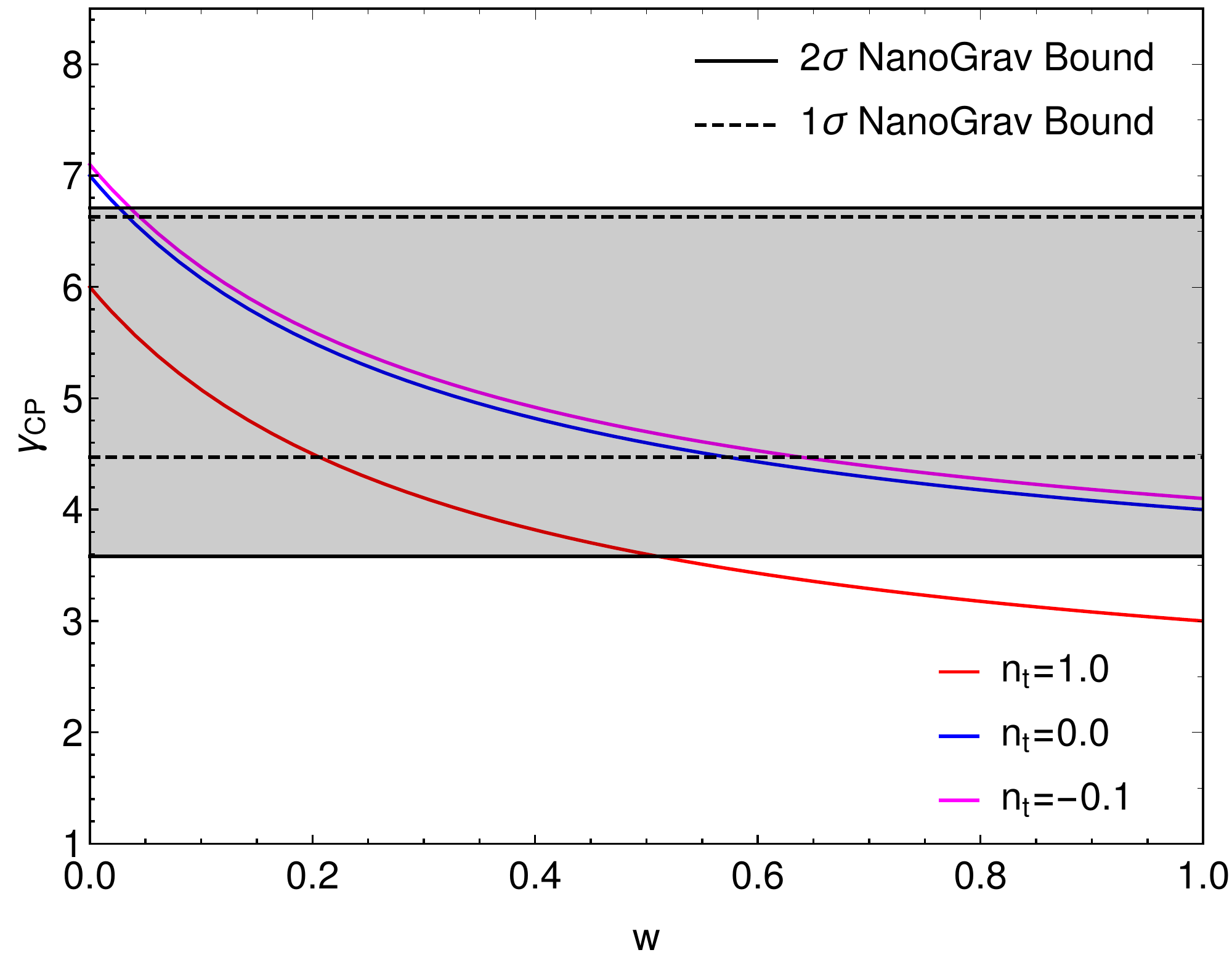}
	\end{minipage}}
\caption{\small{(a) $A_{{\rm CP}}$ is plotted against EoS $w$ for $H_{\rm inf} =10^{11}$ and $10^{13}$ GeV, $n_t=1,0,-0.1$. (b) $\gamma _{{\rm CP}}$ is plotted against EoS $w$ for $n_t=1,0,-0.1$. Gray shaded band in both the plots is the allowed region from the NANOGrav observation. }}
\end{figure}
In Fig.~\ref{fig:acp_w}, $A_{\rm CP}$ is plotted against EoS $w$ for $H_{\rm inf} =10^{11}$ and $10^{13}$ GeV, $n_t=1,0,-0.1$, and $f(T_1)=10^{-10}$ Hz, which corresponds to temperature $T_1=10$ MeV. We have also plotted $\gamma _{\rm CP}$ against EoS $w$ for $n_t=1,0,-0.1$ in Fig.~\ref{fig:gammacp_w}. Gray shaded bands in Fig.s~\ref{fig:acp_w} and~\ref{fig:gammacp_w} represent the $1\sigma$ and $2\sigma$ region allowed from the NANOGrav 12.5-yr result. 
%\suk{This sentence is unnecessary: It is clear from the Fig~\ref{fig:acp_w} that a positive $n_t$ or a large $H_{\rm inf}$ value is required to get $A_{\rm CP}$ in the \NG allowed range.} 
GWs generated during inflation leave their imprint on the CMB temperature  and polarisation anisotropy, which put a constraint on tensor-to-scalar ratio $r<0.056$ at $95\%$ confidence limit at the CMB pivot scale $k_{\rm pivot}=0.05$ Mpc$^{-1}$ from the joint analysis of BICEP2/Keck Array and Planck CMB data\cite{Akrami:2018odb}. This bound, in turn, puts a constraint on the value of $H_{{\rm inf}} < 6.2 \times 10^{13}$ GeV at $95\%$ confidence limit~~\cite{Akrami:2018odb,Akrami:2018vks,Aghanim:2018eyx,Aghanim:2019ame,Ade:2018gkx}.
% The Planck CMB and BICEP2 observations put a constraint on $H_{\rm inf} $ $\leq 6.2 \times 10^13$ GeV. 
Therefore, it is evident from  Fig.~\ref{fig:acp_w} that a positive value of $n_t$ is required to explain the allowed $A_{\rm CP}$ values. On the other hand, it can be seen from Fig.~\ref{fig:gammacp_w} that $\gamma_{\rm CP}$  value decreases with increasing $n_t$ and $w$ values. Therefore, for large values of $w$, $\gamma_{\rm CP}$ goes outside of the allowed region from the \NG observation.
\begin{figure}[]
  \subfloat[\label{fig:Hinf_nt_allowed_rd_mdw1}]{
	\begin{minipage}[c][1\width]{
	   0.5\textwidth}
	   \centering
	   \includegraphics[width=0.95\linewidth,height=0.8\linewidth]{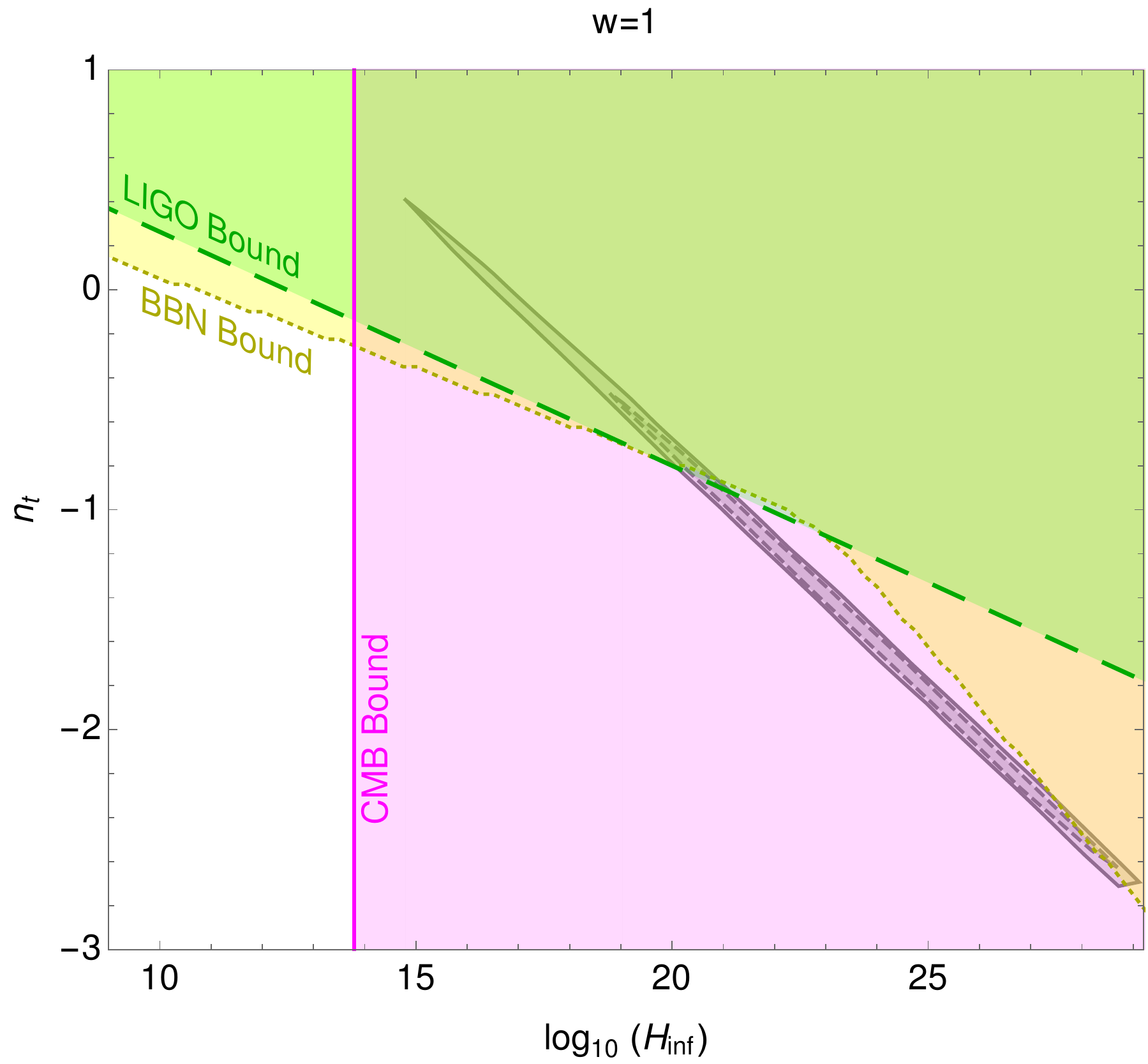}
	\end{minipage}}
 \hfill 	
  \subfloat[\label{fig:Hinf_nt_allowed_rd_mdw13}]{
	\begin{minipage}[c][1\width]{
	   0.5\textwidth}
	   \centering
	   \includegraphics[width=0.95\linewidth,height=0.8\linewidth]{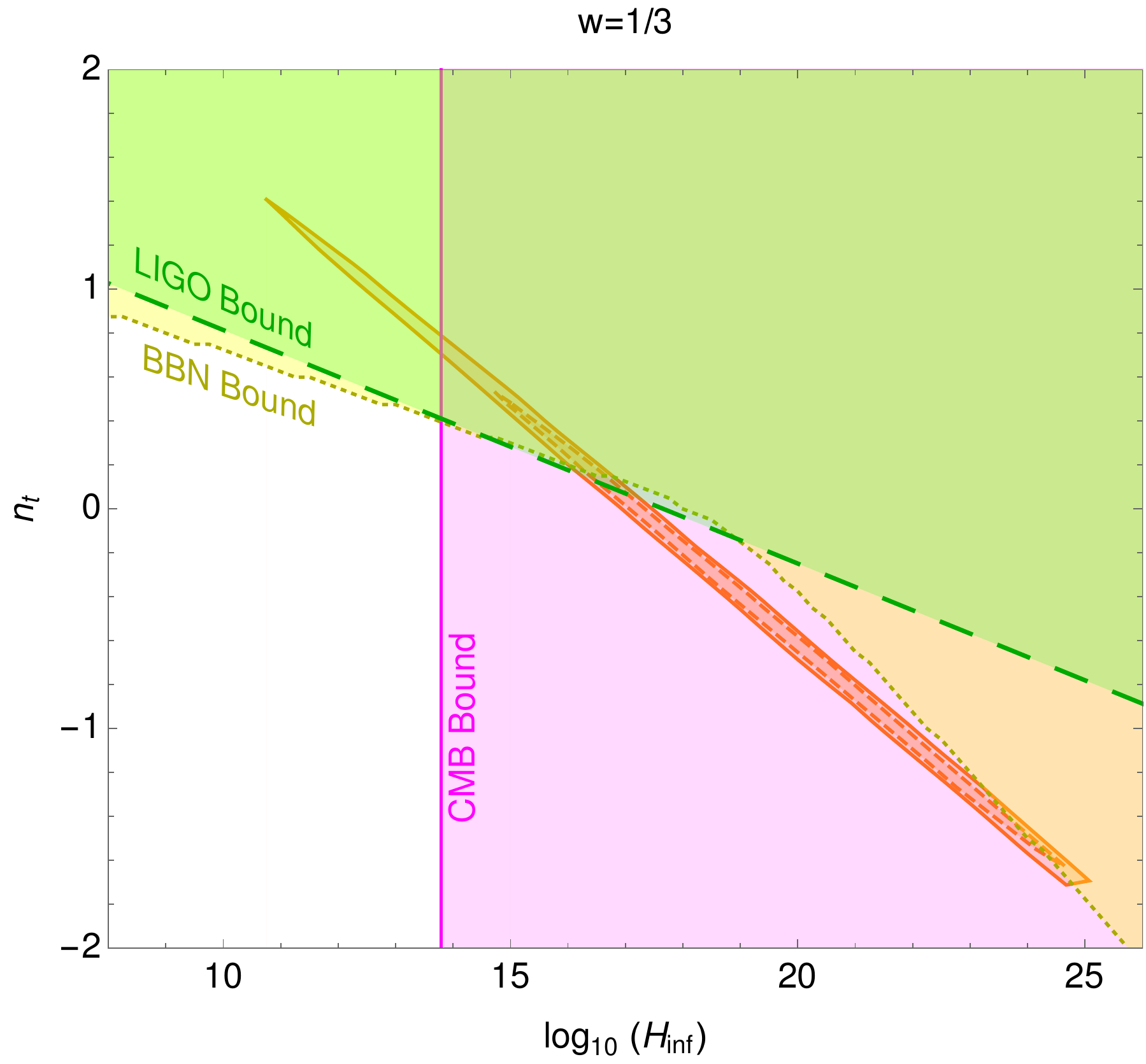}
	\end{minipage}}\\
	\subfloat[\label{fig:Hinf_nt_allowed_rd_mdw0}]{
	\begin{minipage}[c][1\width]{
	   0.5\textwidth}
	   \centering
	   \includegraphics[width=0.95\linewidth,height=0.8\linewidth]{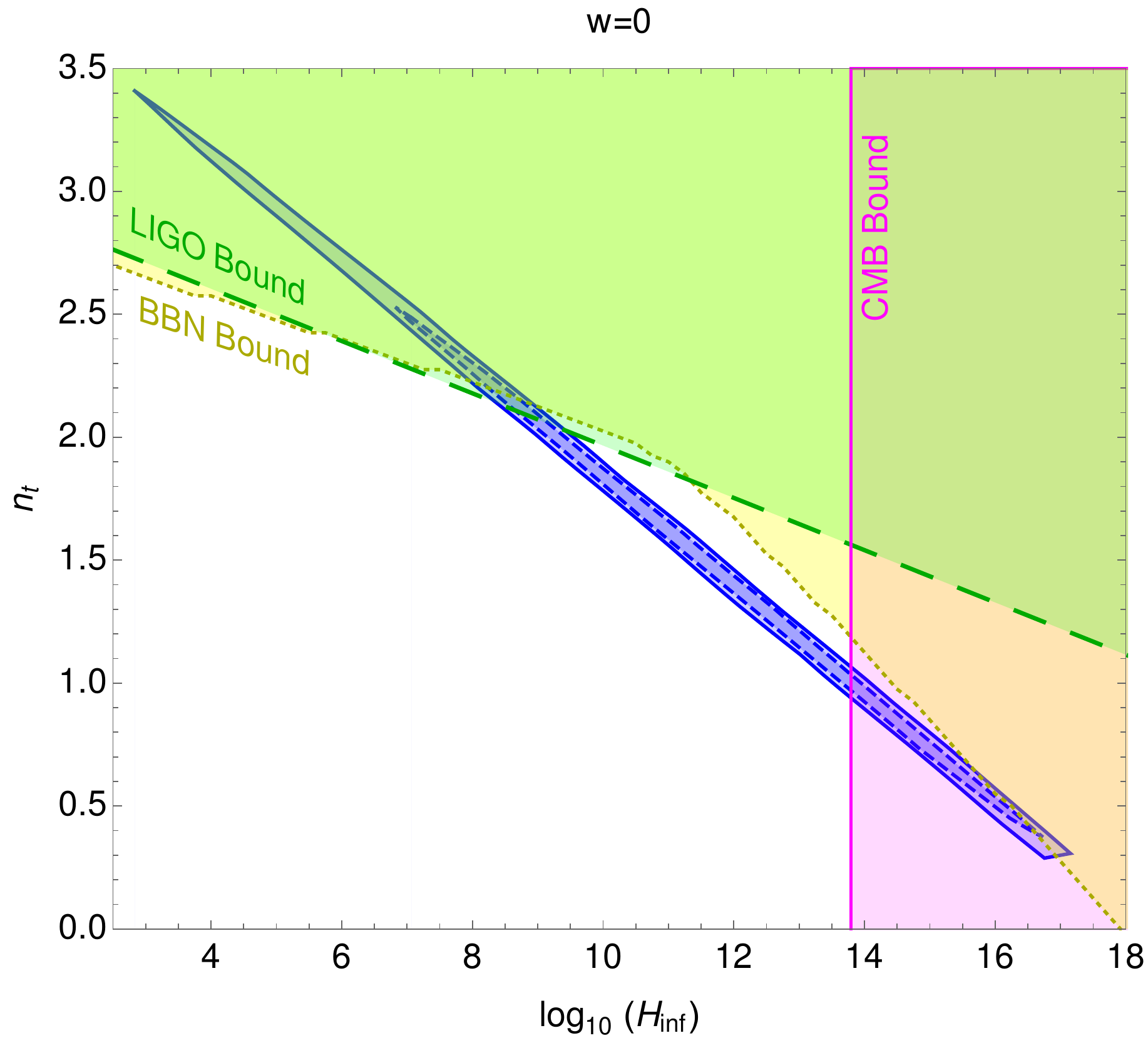}
	\end{minipage}}
 \hfill 	
  \subfloat[\label{fig:Hinf_nt_allowed_rd_mdw0acp}]{
	\begin{minipage}[c][1\width]{
	   0.5\textwidth}
	   \centering
	   \includegraphics[width=0.95\linewidth,height=0.8\linewidth]{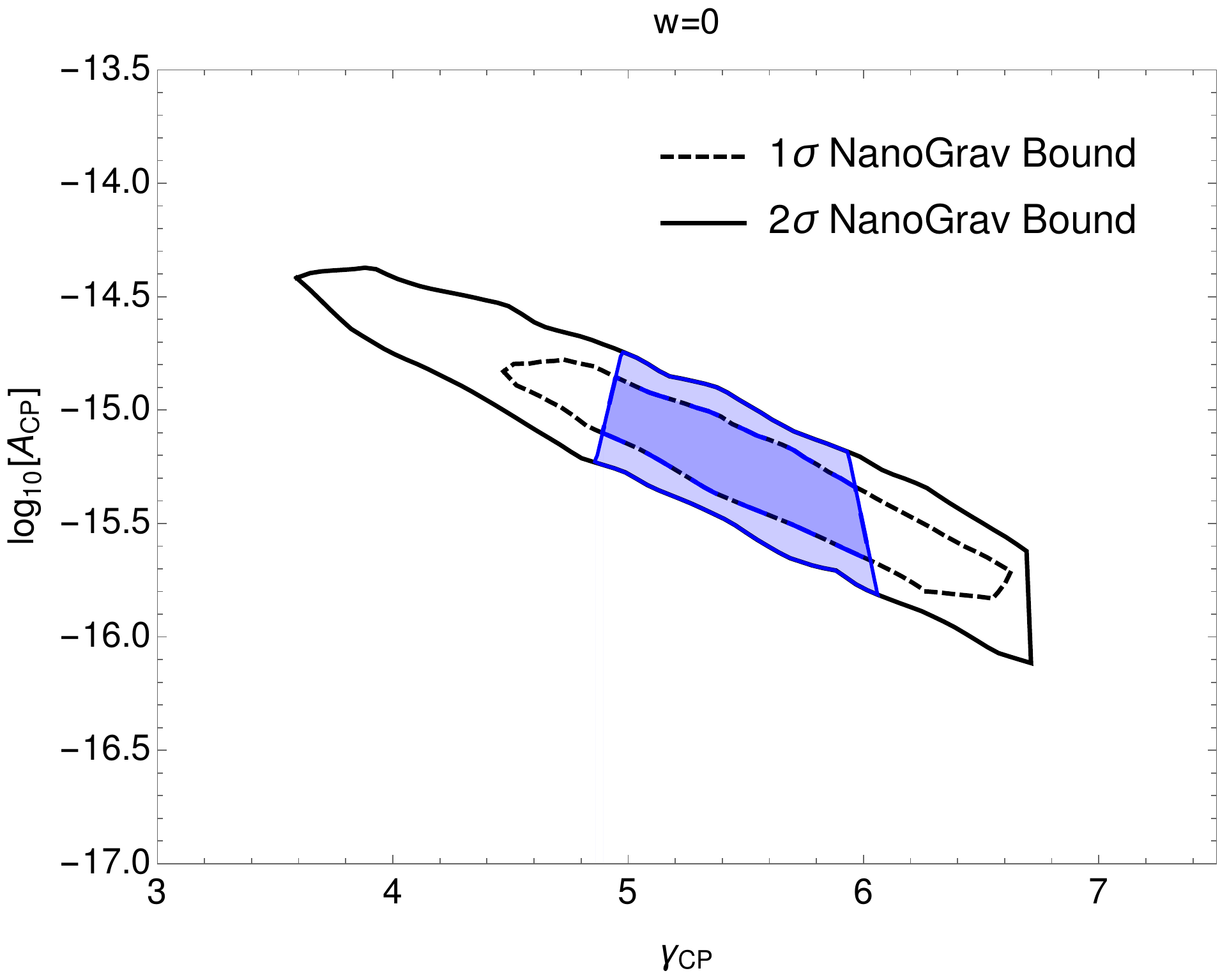}
	\end{minipage}}
\caption{ \small{ \pri{Allowed $H_{{\rm inf}}-n_t$ parameter space required to explain the NANOGrav result by the first order GW for different cases of cosmological thermal histories along with the bounds from different observations are shown here. Magenta shaded region represent the $H_{\rm inf}$ values excluded by the Planck CMB and BICEP2 observations. Green and yellow shaded regions represent the parameter space excluded by the LIGO constraints and by the contribution of GWs to $\Delta N_{\rm eff}$ respectively. (a) Gray shaded contours show the 1-$\sigma $ and 2-$\sigma$ region of $H_{{\rm inf}}-n_t$ parameter space required to explain the NANOGrav result by the first order GW in a pre-BBN kination dominated era. (b) Orange shaded contours represent the 1-$\sigma $ and 2-$\sigma$ region of $H_{{\rm inf}}-n_t$ parameter space required to explain the NANOGrav result by the first order GW in the standard cosmological thermal history. (c)  Blue shaded contours show the 1-$\sigma $ and 2-$\sigma$ region of $H_{{\rm inf}}-n_t$ parameter space required to explain the NANOGrav result by the first order GW in a pre-BBN matter dominated era. 
(d) 1-$\sigma $ and 2-$\sigma$ contours of $\gamma_{{\rm CP}}-A_{{\rm CP}}$ values as reported by the \NG obvservation are shown here in black dashed and solid lines respectively. Blue shaded region shows the $\gamma_{{\rm CP}}-A_{{\rm CP}}$ parameter space which corresponds to the \NG allowed $H_{{\rm inf}}-n_t$ parameter space and consistent with the CMB, BBN and LIGO bounds for the nonstandard cosmological history with a pre-BBN MD era.}}}\label{fig:Hinf_nt_allowed_rd_md}
\end{figure}

\pri{In this work, our aim is to find the allowed region of the $H_{{\rm inf}}-n_t$ parameter space required to explain the NANOGrav observation for different values of the EoS. It is clear from Eq.s~\eqref{eq:acp} and \eqref{eq:gammacp} that there are four parameters $H_{\rm inf}$, $w$, $n_t$, and $f(T_1)$ which affect $A_{\rm CP}$ and $\gamma _{\rm CP}$. We do the analysis for three different values of the EoS, $w=0,1/3$, and $1$. For each value of $w$, we fix the end of the nonstandard epoch at $T_1=10$ MeV, which corresponds to the frequency $f(T_1)=10^{-10}$ Hz. Therefore, we are now left with only two free parameters: $H_{\rm inf}$ and $n_t$, and we do a parameter scan for them. We vary $n_t \in (-3,5)$ and $ \log_{10}\big(\frac{H_{{\rm inf}}}{\rm GeV}\big) \in (2,30)$ and compute the corresponding $A_{{\rm CP}}$ and $\gamma_{{\rm CP}}$ values. Out of the resulting points in the $A_{{\rm CP}}$-$\gamma_{{\rm CP}}$ plane, only those which lie within the 2-$\sigma$ region of the NANOGrav results have been retained. 
These allowed values of $A_{{\rm CP}}$ and $\gamma_{{\rm CP}}$ are shown in Fig.~\ref{fig:Hinf_nt_allowed_rd_md} for the three different values of $w$ in there different shaded contours (gray for $w=1$, orange for $w=1/3$, and blue for $w=0$).
 
In Fig.~\ref{fig:Hinf_nt_allowed_rd_md}, $H_{{\rm inf}}$ values excluded from the CMB observations are shown in the magenta shaded region. It is evident from Fig.s~ \ref{fig:Hinf_nt_allowed_rd_mdw13} and~\ref{fig:Hinf_nt_allowed_rd_mdw0} that there is some part of $H_{{\rm inf}}-n_t$ parameter space which is allowed by the CMB bound for both  $w=1/3$ and $w=0$ cases. However, the parameter space required to explain the \NG observation for $w=1$ is fully excluded by the bound on $H_{{\rm inf}}$ from CMB (see~Fig.~\ref{fig:Hinf_nt_allowed_rd_mdw1}). Therefore, looking at the allowed values of $n_t$ for $w=1/3$ and $w=0$ in this figure, a blue tilted primordial tensor power spectrum is required to explain the NANOGrav observation in case of the standard cosmological history as well as in the presence of an early MD era. 
The simplest inflationary models, like single field slow roll inflation, predict a slghtly red tilted tensor spectrum spectrum. However, a blue tilted power spectrum can arise in inflation models like G-inflation~\cite{Kobayashi:2010cm}, super-inflation models~\cite{Baldi:2005gk}, partcle produntion during inflation~\cite{Cook:2011hg,Mukohyama:2014gba}, inflation with non-Bunch Davies initial state~\cite{Ashoorioon:2014nta} and inflation with beyond slow roll~\cite{Gong:2014qga}.}
% \begin{figure}[!t]
%  \subfloat[\label{fig:acp_gammacp_allowed_rd}]{
%	\begin{minipage}[c][1\width]{
%	   0.5\textwidth}
%	   \centering
%	   \includegraphics[width=1\textwidth]{Acp_vs_gammacp_var_nt_hinf_fixed_RD.pdf}
%	\end{minipage}}
% \hfill 	
%  \subfloat[\label{fig:acp_gammacp_allowed_md}]{
%	\begin{minipage}[c][1\width]{
%	   0.5\textwidth}
%	   \centering
%	   \includegraphics[width=1\textwidth]{Acp_vs_gammacp_var_nt_hinf_fixed_MD.pdf}
%	\end{minipage}}
%\caption{\small{(a) Orange shaded region shows the $\gamma_{{\rm CP}}-A_{{\rm CP}}$ parameter space which corresponds to the allowed $H_{{\rm inf}}-n_t$ parameter space in the standard cosmological thermal history. (b) Blue shaded region shows the $\gamma_{{\rm CP}}-A_{{\rm CP}}$ parameter space which corresponds to the allowed $H_{{\rm inf}}-n_t$ parameter space in the nonstandard cosmological thermal history with a pre-BBN MD era.} }
%\end{figure}

As GW contributes to the radiation energy density in the universe, it will contribute to the effective number of relativistic degrees of freedom ($N_{\rm eff}$) in the universe. An overly abundant GW can change the expansion rate too much in the early universe and therefore, can affect the abundance of light elements produced during the BBN~\cite{Allen:1997ad,Smith:2006nka,Boyle:2007zx,Kuroyanagi:2014nba,Ben-Dayan:2019gll}. BBN and Planck CMB observations constrain the contribution of the additional relativistic degrees of freedom, $\Delta N_{\rm eff}$, which requires
\begin{eqnarray}
\int df\,\frac{\Omega_{\rm GW}(f)h^2}{f} \leq 5.6 \times 10^{-6} \Delta N_{\rm eff}\, ,
\label{eq:bbncons}
\end{eqnarray}
where lower limit of the integration is equal to the frequency corresponding to the mode entering the horizon at the time of BBN and the upper limit is set to $10^7\,{\rm Hz}$, approximately corresponding to the comoving horizon at the time of the end of the inflation at temperature $T \approx 10^{15}\,{\rm GeV}$~\cite{Maggiore:1999vm,Kuroyanagi:2014nba,Liu:2015psa}. Planck CMB measurements and BBN observations severely constrain the $\Delta N_{\rm eff} \lesssim 0.4$~\cite{Aver:2015iza,Cooke:2017cwo,Aghanim:2018eyx,Hsyu:2020uqb,Aiola:2020azj}. We then compute the allowed region in the $H_{{\rm inf}}-n_t$ parameter space for which Eq.~(\ref{eq:bbncons}) is satisfied with the observed value of $\Delta N_{\rm eff}$. \pri{ The resulting constraint on $H_{{\rm inf}}-n_t$ parameter space for all the three $w$ under consideration are shown by the yellow dotted line in Fig.~\ref{fig:Hinf_nt_allowed_rd_md}, where the region above that line (shaded in yellow colour) is ruled out. 
%Purple dotted line gives the constraints on first order GWs in the standard cosmological thermal history, whereas red dotted line represents the bound on first order GWs in nonstandard thermal history with an early MD era. 
In addition to the bound on GWs from their contribution to the $N_{\rm eff}$, non-observation of the stochastic GW background in LIGO O1 and O2 runs can also put the constraints on the $H_{{\rm inf}}-n_t$ parameter space~\cite{LIGOScientific:2019vic}. The constraints for all three $w$ are shown by the green dashed lines in Fig.~\ref{fig:Hinf_nt_allowed_rd_md}, where the region above the lines (green shaded region) are excluded.  To calculate the bound from LIGO, the primordial power spectrum is assumed to retain the power law form until the LIGO frequencies. It is evident from Fig.~\ref{fig:Hinf_nt_allowed_rd_mdw13} that the entire region of $H_{{\rm inf}}-n_t$ parameter space allowed by the NANOGrav and CMB bounds in the standard RD history is ruled out when the bound on $\Delta N_{\rm eff}$ from BBN and Planck CMB observations and the LIGO bound are taken into account.  However, there is a finite region left in the $H_{{\rm inf}}-n_t$  parameter space which can explain the NANOGrav result and simultaneously evades the constraints from LIGO, BBN, and CMB in the case of nonstandard thermal history with an early MD epoch (see Fig.~ \ref{fig:Hinf_nt_allowed_rd_mdw0}).  The $\gamma_{{\rm CP}}-A_{{\rm CP}}$ parameter space corresponding to the $H_{{\rm inf}}-n_t$ parameter space allowed by the \NG, CMB, BBN, and LIGO observation in a nonstandard cosmological history with a pre-BBN MD era is shown by the blue shaded region in Fig.~\ref{fig:Hinf_nt_allowed_rd_mdw0acp}). }

\section{Induced primordial gravitational waves from curvature perturbations}
\label{GW2sec}
Scalar and tensor perturbations do not evolve independently at the second order of perturbation theory. \suk{In certain single field and multi-field models of inflation, the scalar perturbations can become large at small length scales owing to various mechanisms in the inflationary dynamics~\cite{Garcia-Bellido:2017mdw,Ballesteros:2017fsr,Ballesteros:2020qam,Bhaumik:2019tvl,Germani:2017bcs,Mishra:2019pzq,Clesse:2015wea,Braglia:2020eai,Braglia:2020taf,Fumagalli:2020adf,Palma:2020ejf,Ozsoy:2020kat,Aldabergenov:2020yok}. The resulting GWs induced by the large scalar perturbations can be large enough to be relevant in view of the (proposed) sensitivities of the current and future GW detectors. Such large scalar fluctuations can also lead to abundant formation of PBHs. Therefore, it is interesting to check the implications for PBH abundance assuming the observed \NG signal to be second order GW induced by the first order scalar perturbations. In this section, we analyse such implications for the standard and nonstandard post-inflationary evolutions for three different profiles of the small scale growth of the scalar power spectra. The difference between the analysis in this section and the previous one is that here, the primordial scalar perturbations have large amplitude at small scales, whereas, in the previous section the small scale primordial tensor perturbations (first order) were large. This difference directly relates to different features in the inflationary dynamics that lead to growth in either the scalar or the tensor power spectra.} 

Evolution of the second order tensor perturbation is given as
\begin{equation}
\label{equ:hhh}
h_k'' + 2 {\cal H} h_k' + k^2 h_k=  {\cal S_{\mathbf{k}}}(\eta)\, , 
\end{equation}
where ${\cal S}(\mathbf{k}, \eta)$ is the source term which depends on the first order scalar perturbations. The scalar induced second order tensor power spectrum is~\cite{Kohri:2018awv},
\begin{align}\label{eq:pgamma}
\overline{\mathcal{P}_h(k,\eta)}=2\int_0^\infty dv \int_{|1-v|}^{1+v} du\left[\frac{4v^2-\left(1+v^2-u^2\right)^2}{4uv}\right]^2 P_{ \zeta}(kv)P_{\zeta}(ku)\overline{I^2}(v,u,x)\,,
\end{align}
where the bar denotes the oscillation average.
Here $v\equiv q/k$, $u\equiv|\mathbf{k}-\mathbf{q}|/k$, $P_{ \zeta}$ is the primordial curvature power spectrum, and $x=k\eta$. The integration kernel $I(u,v,x)$ contains the source information (complete expression can be found in ref~\cite{Espinosa:2018eve,Kohri:2018awv}) which can be written with a simple analytical expression for very late time, i.e. $x \gg 1$. For the modes entering the horizon during RD era,
\begin{align}
\overline{I_{RD}^2}(x\gg 1,u,v)=\frac{9(u^2+v^2-3)^2}{32u^6v^6x^2}&\Bigg\{ {\pi}^2(u^2+v^2-3)^2\Theta(u+v-\sqrt{3}) \nonumber\\&+  \bigg(-4uv + (u^2+v^2-3) \ln {\frac{3-(u+v)^2}{3-(u-v)^2} } \bigg)^2\Bigg\}\,.
\label{IRD}
\end{align}
%In nonRD epochs, there can be an epoch of EoS $w\neq 1/3$ staring from the end of inflation until the onset of RD era at temperature $T_1$. 
For a general EoS ($0 < w \leq 1$), the expression for $I(u,v,x)$ is given as~\cite{Domenech:2019quo}
\begin{align}\label{eq:kernel3}
I(x\gg1,u,v)&=2^{\beta}\frac{3\sqrt{2}}{w\pi\alpha^3}\frac{1+w}{1+3w}\Gamma^2[\beta+2]{\left(uvx\right)^{-\beta-1/2}}\nonumber\\&\times\left\{\frac{\pi}{2}\sin\left(x-\frac{\beta\pi}{2}-\frac{\pi}{4}\right){I}_J(u,v,w)+\cos\left(x-\frac{\beta\pi}{2}-\frac{\pi}{4}\right){I}_Y(u,v,w)\right\}\, ,
\end{align}
where $I_J(u,v,w)$ and $I_Y(u,v,w)$ are expressed in terms of the associated Legendre polynomials and the Legendre polynomials on the cut (complete expression can be found in ref~\cite{Domenech:2019quo}). In general, it is numerically challenging to calculate the second order tensor power spectrum for an arbitrary value of EoS ($0 < w \leq 1$) due to the complicated dependence of $I_J$ and $I_Y$ on $w$. Hence, we chose two different values to interpret the dynamics in nonRD epochs: (i) $w=1$, where the energy density falls faster than radiation and (ii) $w=\frac{1}{9}$, where it falls slower. Rest of the analysis is carried out for these two nonstandard epochs in addition to the standard RD epoch. EoS $w=1$ can arise in models of quintessential inflation~\cite{Peebles:1998qn}, where, after the end of inflation, the inflaton fast rolls (kinetic energy dominated (KD) universe) to reach the potential for quintessence dark energy. An epoch with $w=1/9$ is a nearly dust-like EoS, which can arise during very slow reheating process of the inflaton or the moduli such that the effective EoS can lie between MD and RD for some considerable time. The particular choice of $w=1/9$ is because the expressions for $I_J$ and $I_Y$ are much simplified in this case and can be solved analytically for $x\gg 1$. The nonstandard epochs terminate at $T_1\geq T_{\rm BBN}$ and we choose $T_1=10$ MeV.

The integral $I_{{\rm KD}}$ for $w=1$ is estimated as:
%\begin{align}\label{eq:IJJ}
%I_J(u,v,w)\equiv Z^{\beta-1/2}\left[\mathsf{P}^{-\beta+1/2}_{\beta-1/2}(y)+\frac{3(1+w)}{2}\mathsf{P}^{-\beta+1/2}_{\beta+3/2}(y)\right]\Theta(u+v-w^{-1/2})
%\end{align}
%and
%\begin{align}\label{eq:IYY}
%I_Y(u,v,w)\equiv&Z^{\beta-1/2}\left[
%	\mathsf{Q}^{-\beta+1/2}_{\beta-1/2}(y)+\frac{3(1+w)}{2}\mathsf{Q}^{-\beta+1/2}_{\beta+3/2}(y)\right]\Theta(u+v-w^{-1/2})\nonumber\\&
%	-\tilde Z^{\beta-1/2}\left[{\cal Q}^{-\beta+1/2}_{\beta-1/2}(\tilde y)+3(1+w){\cal Q}^{-\beta+1/2}_{\beta+3/2}(\tilde y)\right]\Theta(w^{-1/2}-u-v)\,.
%\end{align}
\begin{align}\label{IKD}
\overline{I_{KD}^2}(x\gg1,u,v)&=\frac{9}{16\pi u^4v^4x}\left\{\frac{\left(3(u^2+v^2-1)^2-4u^2v^2\right)^2}{{4u^2v^2-(u^2+v^2-1)^2}}+9\left(u^2+v^2-1\right)^2\right\}\,.
\end{align}
Similarly, the expression for $I_{1/9}$ can be derived for $w=\frac{1}{9}$ and given as
\begin{align}
\label{I19D}
\overline{I^2_{1/9}}(x\gg1,u,v)&=\frac{25}{16\pi u^8v^8{x}^{3}}\Bigg\{\left(u^2+v^2-9\right)^2\left[5\left(u^2+v^2-9\right)^2-6u^2v^2\right]^2\nonumber\\&
+{\left|4u^2v^2-(u^2+v^2-9)^2\right|}\left(4u^2v^2+5(u^2+v^2-9)^2\right)^2\nonumber\\&+
2\left(u^2+v^2-9\right)\left[5\left(u^2+v^2-9\right)^2-6u^2v^2\right]\sqrt{\left|4u^2v^2-(u^2+v^2-9)^2\right|}\nonumber\\&
\times\left(4u^2v^2+5(u^2+v^2-9)^2\right)\Theta(\sqrt{9}-u-v)\Bigg\}\,.
\end{align}
Now, using the expressions for the integration kernels in Eq.s~\eqref{IRD},~\eqref{IKD} and~\eqref{I19D} with the Eq.s~\eqref{originalGWPh} and~\eqref{eq:pgamma}, we can determine the \OMG2 for standard RD and the two nonstandard EoS. For nonstandard EoS, the respective $I_{w}$ contribute to \OMG2 until the temperature $T_1$ (corresponding to the comoving time $\eta_1$) is reached, after which, standard evolution with RD follows\footnote{Here, we do not explicitly solve for the continuity of \OMG2 at $\eta _1$, since the scales corresponding to the transition are outside the \NG relevant scales. Moreover, we consider the transition from the nonstandard to RD epoch to be instantaneous. In a more generic analysis with a gradual transition between these two epochs, it is important to account for the continuous transition of \OMG2 at $\eta _1$.}. Thus, for a general $w$ dominated epoch~\cite{Domenech:2019quo}, 
\begin{equation}
 \Omega_{{\rm GW}}^{(2)}(k,\eta _0)= 0.387 ~\Omega _{{\rm rad},0} \bigg(\frac{g_s^4 g_*^{-3}}{106.75}\bigg)^{-1/3} \Omega_{{\rm GW}}^{(2)} (k,\eta _1),
\label{omg2fullnow}
\end{equation}
where 
\begin{equation}
 \Omega_{{\rm GW}}^{(2)} (k,\eta _1)=\frac{(1+3w)^2}{48}(k\eta _1)^2 \overline{\mathcal{P}_h(k,\eta _1)}. \label{omg2trans}
\end{equation}
Therefore, a general $I_{w}$ contributes to Eq.~\eqref{omg2trans} and the full second order GW can be calculated then using Eq.~\eqref{omg2fullnow} once the primordial power spectrum $P_{\zeta}(k)$ is specified.

\subsection{PBH formation}
\label{secPBH}
If the amplitude of the primordial fluctuatiions is such that when the modes re-enter the horizon at the post-inflationary epoch, the density fluctuations $\delta$ are larger than the critical density for collapse ($\delta_c$), then PBH can be produced with mass $M=\epsilon M_H$, where $M_H$ is the horizon mass at collapse. $\epsilon$ signifies the efficiency of collapse and is typically of order $1$; here we consider $\epsilon = 0.33$. The PBH formation process depends on the background EoS through the dependence of $\delta_c$ on $w$ and the evolution of PBH abundance in this epoch depends on the modified expansion rate. Taking the numerically evaluated expression in~\cite{Harada:2013epa}:
\begin{equation}
\delta _c(w)=\frac{3(1+w)}{(5+3w)}\sin ^2 \bigg(\frac{\pi \sqrt{w}}{(1+3w)}\bigg). \label{deltac_w}
\end{equation}
Considering the Press-Schechter formalism for gravitational collapse, present abundance of PBH in an interval of mass $M$ to $M+dM$ for general $w$ is given as~\cite{Bhattacharya:2019bvk}
\begin{equation}
\psi_w(M)=\frac{\epsilon}{T_{\rm eq}}\bigg(\frac{g_s(T_1)}{g_s(T_{\rm eq})}\bigg)^{\frac{1}{3}}\bigg(\frac{\Omega_m h^2}{\Omega_c h^2}\bigg)\bigg(\frac{90M_P^2}{\pi ^2g_*(T_1)}\bigg)^{\frac{w}{1+w}}(4\pi \epsilon M_P^2)^{\frac{2w}{1+w}}T_1^{\frac{1-3w}{1+w}}\frac{\beta (M)}{M^{\frac{3w+1}{1+w}}},\label{psiM}
\end{equation}
where $\epsilon$ is the fraction of the total horizon mass that contributes to the PBH mass~\cite{Carr:1975qj} and $M_P$ is the reduced Planck mass. The PBH mass fraction is given as: 
\begin{equation}
\beta(M)={\rm erfc} \bigg[\frac{\delta_c(w)}{\sqrt{2\sigma_{\delta}^2}}\bigg], \label{beta_w}
\end{equation}
where $\sigma_{\delta}^2$ is the variance of the density power spectrum and calculated as: 
\begin{equation}
\sigma_{\delta}^2=\frac{4(1+w)^2}{(5+3w)^2}\int \frac{dk}{k}(kR)^4 W^2(k,R)P_{\zeta}(k).
\label{vardel}
\end{equation}
We choose the window function $W^2(k,R)=\exp(\frac{-k^2R^2}{4})$ to smooth the perturbations on the comoving scale $R$ at formation. The mass $M$ of the PBH produced is related to the comoving wavenumber $k$ at formation for general EoS as: 
\begin{equation}
M(k)=4\pi\epsilon {M_P}^2\bigg(\frac{\pi ^2 g_*^{\rm eq}}{45M_P^2}\bigg)^{\frac{1}{3w+1}}\bigg(\frac{g_s^{\rm eq}}{g_s(T_1)}\bigg)^{\frac{3w-1}{3(3w+1)}}(a_{\rm eq}T_{\rm eq})^{\frac{3(1+w)}{3w+1}}T_1^{-\frac{3w-1}{3w+1}}k^{-\frac{3(1+w)}{3w+1}}. \label{Mk_w} 
\end{equation}
$\psi_w(M)$ can be calculated from Eq.~\eqref{psiM} using Eq.s~\eqref{deltac_w},~\eqref{beta_w},~\eqref{vardel} and~\eqref{Mk_w}. The ratio of the PBH abundance for a general $w$ dominated epoch and that for a RD epoch is: 
\begin{equation}
g_{\psi}(M,w)\equiv \frac{\psi_w(M)}{\psi_{\rm RD}(M)}=\frac{\beta (M)}{\beta ^{\rm RD}(M)}\frac{g_s(T)^{w-1/3}}{g_s(T_1)^{w-1}g_{s}(T_{\rm eq})^{2/3}}\bigg(\frac{T}{T_1}\bigg)^{3w-1}.\label{gain01}
\end{equation}
The dependence of $\delta_c$ on $w$ is such that $\delta_c^{(w=1)},\delta_c^{(w=1/9)} < \delta_c^{(w=1/3)}$ and therefore, the gain $g_{\psi}(M,w)>1$ for both the nonstandard epochs. This means that to form same total abundance $f_{\rm PBH}=\int dM \psi_w(M)$, the two nonstandard epochs will require a lower value of $A_s$ than that required in RD. In this work, we check if the low value of $A_s$ required for abundant PBH formation in each nonstandard epoch can satisfy the \NG bounds. 
%\suk{There are some repeatations in motivating PBH and large GW from enhanced $P_{\zeta}(k)$ in the last paragraph and end of section 2. Sort it out.}
\subsection{Primordial scalar power spectrum}
\suk{An enhanced $P_{\zeta}(k)$ at scales smaller than that observed in CMB surveys, can be motivated in a variety of inflation models. In single-field inflation models, presence of an inflection point or a bump can lead to ultra slow-roll of the inflaton, which results in enhancement of $P_{\zeta}(k)$~\cite{Garcia-Bellido:2017mdw,Ballesteros:2017fsr,Ballesteros:2020qam,Bhaumik:2019tvl,Germani:2017bcs,Mishra:2019pzq}. For multi-field models of inflation, the inflaton coupled to a secondary field (e.g. mild waterfall phase in hybrid inflation~\cite{Clesse:2015wea} or noncanonical coupling~\cite{Braglia:2020eai,Braglia:2020taf}) or rapid turns in field space~\cite{Fumagalli:2020adf,Palma:2020ejf} can induce large $P_{\zeta}(k)$.
% A large amplitude of the scalar power spectrum can lead to large density perturbations in the postinflationary epoch, which can collapse and form PBHs. As it is evident from the discussion and expressions above, large $P_{\zeta}(k)$ leads to large secondary GW spectra. In the rest of this section, we explore the implications of the \NG result for PBH abundance for different nonstandard postinflationary histories. 
However, in the rest of this paper, we proceed with an agnostic approach towards the exact inflationary model that gives rise to an enhanced $P_{\zeta}(k)$; rather, we consider three well-motivated functional forms for $P_{\zeta}(k)$ and analyse the PBH abundance and relevant second order GW spectrum for each case. Throughout the analysis of this paper, we consider the scalar fluctuations to be Gaussian, and any nontrivial non-Gaussianity can affect the results for PBH abundance and SGWB spectra and therefore alter the implications derived here.}
\subsubsection{Gaussian power spectrum}
\label{GPP}
\begin{figure}[!htbp]
\centering{
\includegraphics[width=0.6\textwidth]{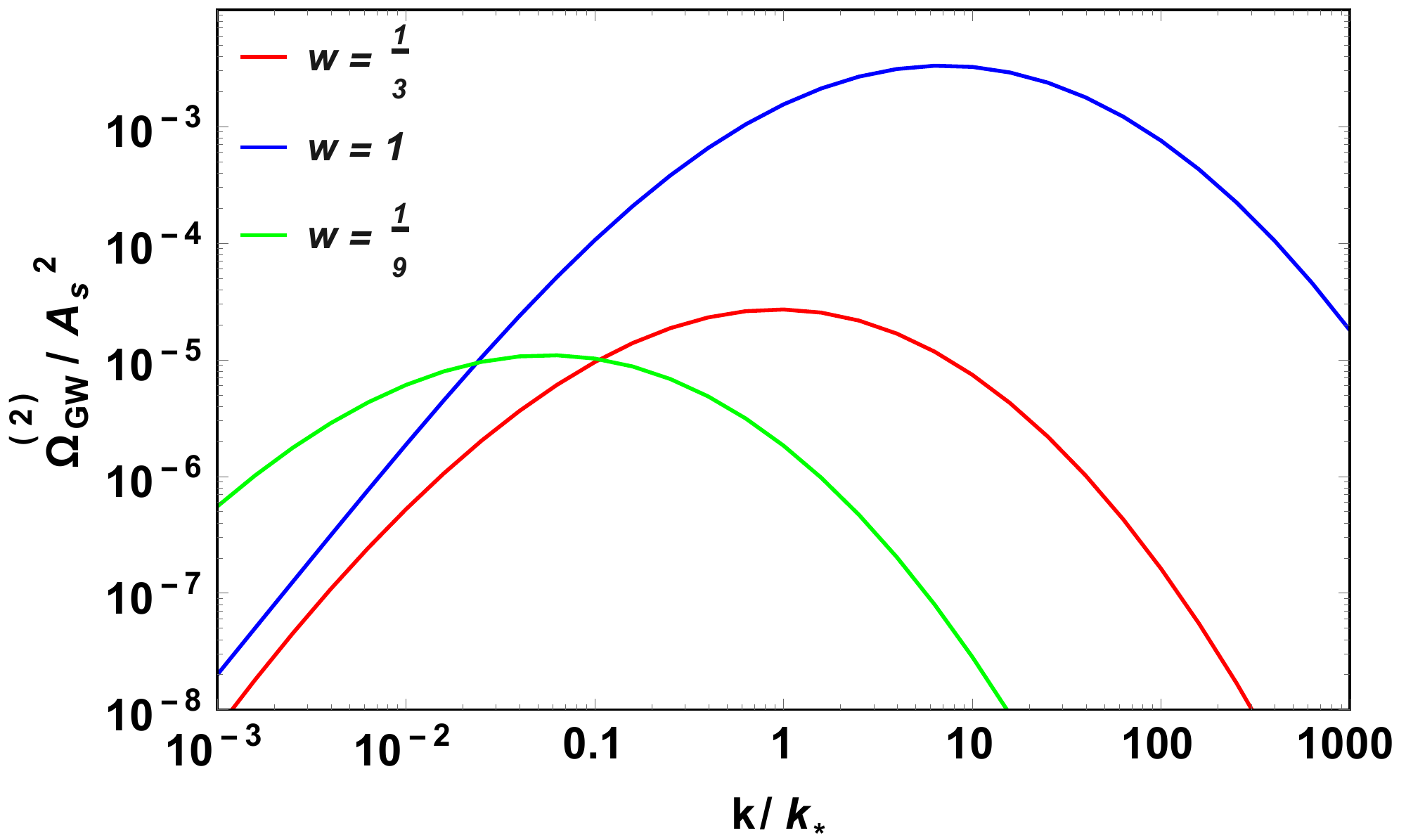}
\caption{\small{The normalised \OMG2 for the three $w$ cases are shown as a function of $k/k_*$.}}
\label{OmgA2}}
\end{figure}

A Gaussian primordial power spectrum is motivated in many single field and multi-field inflation models discussed above and is given as:
\begin{equation}
P_{\zeta}(k)=A_s\times \exp\bigg[-\frac{(\log({k/k_*}))^2}{2\sigma_p^2}\bigg]\,.
\label{PKG}
\end{equation}
The GW spectrum, normalised with $A_s^2$ is given in Fig.~\ref{OmgA2}.
%\begin{figure}[h]
%\centering
%  \begin{minipage}{.5\textwidth}
%  \centering
%%\includegraphics[width=0.55 \textwidth]{sf.pdf}
%    %\includegraphics[width=6.5cm,height=5cm]{plot1.pdf}
%    \includegraphics[width=\textwidth]{omgwbyA2_final.pdf}
%    \caption{ The scale factor as a function of the dimensionless time.}
%    \label{figS}
%  \end{minipage}\qquad
%  \begin{minipage}{.46\textwidth}
%  \centering
%%\includegraphics[width=0.55 \textwidth]{Erad.pdf}
%    %\includegraphics[width=6.5cm,height=5cm]{plot2.pdf}
%    \includegraphics[width=\textwidth]{psi_PBH.pdf}
%    \caption{The energy density in radiation  as a function of the dimensionless time}
%    \label{frad}
%  \end{minipage}
%\end{figure}
%%
Evidently, for $w=1$, there is an enhancement in the normalised GW spectra and for $w=1/9$, it is diminished. Moreover, the peak of the spectrum shifts towards $k>k_*$ for $w=1$ and towards $k<k_*$ for $w=1/9$. Now, the \NG analysis fits the observed stochastic spectrum in the frequency range $f_0\in (2.5\times10^{-9},1.2\times10^{-8})$ Hz. Thus, if we assume that the \OMG2 peak for each $w$ occurs at the scale (wavenumber $k_0=2\pi f_0$) observed by the \NG, then for $w=1/3$, $k_0/k_*\simeq 1$; for $w=1$, $k_0/k_*>1$, and for $w=1/9$, $k_0/k_*<1$. However, in that case, the amplitudes required for each case to fit the \NG result will be different: $A_s(w=1/3)=6\times 10^{-3}$, $A_s(w=1)=5.5\times 10^{-4}$, and $A_s(w=1/9)=5.1\times 10^{-3}$ to have \OMG2$_{\rm ,peak}\simeq 10^{-9}$.
\begin{figure}[!htbp]
\centering{
\includegraphics[width=0.6\textwidth]{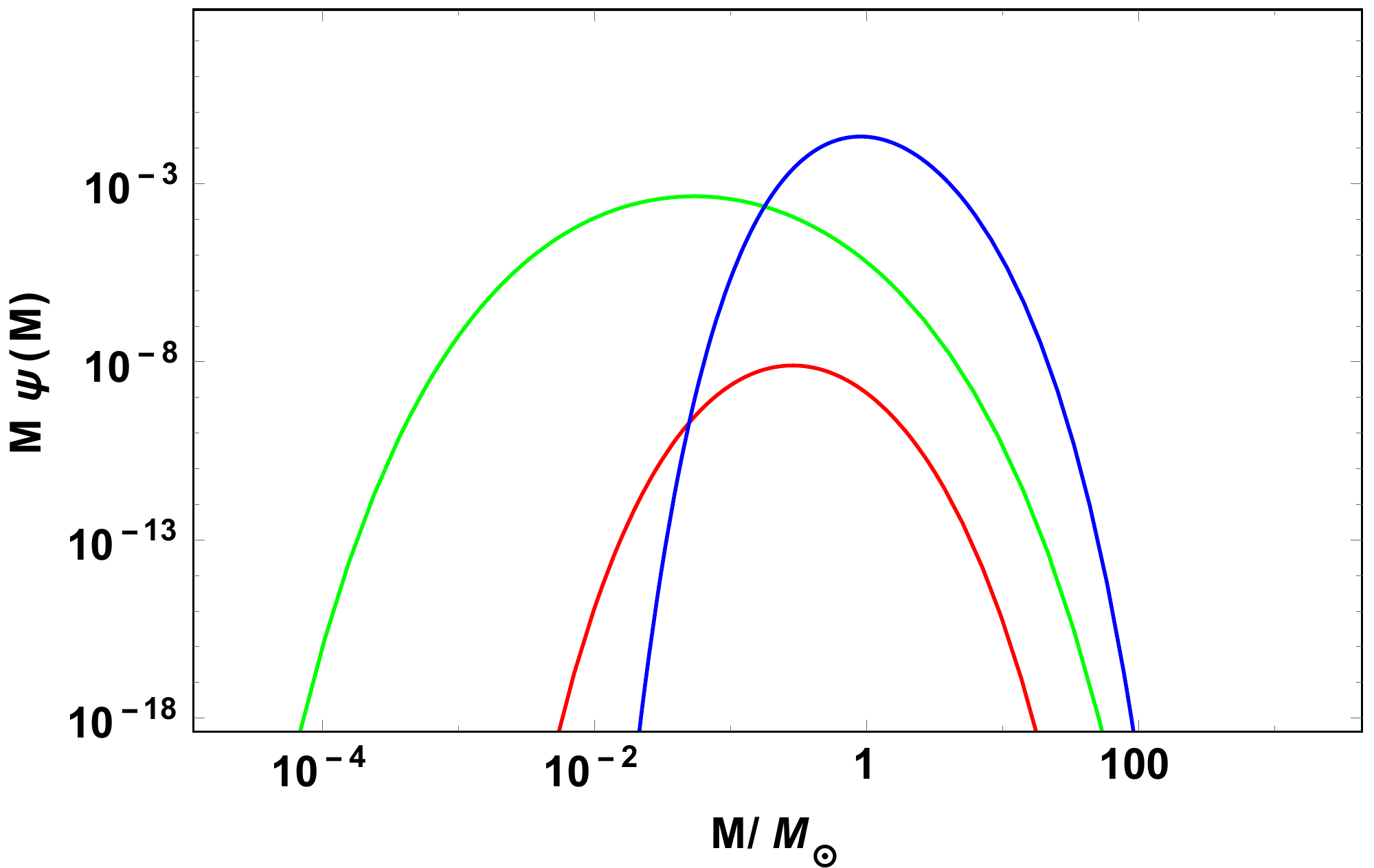}
\caption{\small{PBH mass functions for different $w$ epochs are shown for a typical $A_s=0.007$. Colour specifications are same as Fig.~\ref{OmgA2}.}}
\label{psiPBH}}
\end{figure}
%\suk{I will motivate PBH at this point and plot the mass functions for the same Gaussian spectrum. There will also be shifts in the peak for KD and $w=1/9$. We need to check if these shifts are (anti) correlated.}
%Fig.~\ref{OmgA2} also gives us an idea about the required amplitude $A_s$ for each case such that the peaks can explain the \NG observation.

The PBH mass function $\psi_w(M)$ using this $P_{\zeta}(k)$ for $w=1/3$, $1$, and $1/9$ are evaluated and given in Fig.~\ref{psiPBH} for $A_s=0.007$. This figure shows an enhancement in PBH abundance for the nonRD epochs and peak shift with respect to peak position in the RD. These peak shifts are correlated with the shifts in GW in Fig.~\ref{OmgA2}. 
\begin{figure}[!htbp]
\centering{
\includegraphics[width=0.75\textwidth]{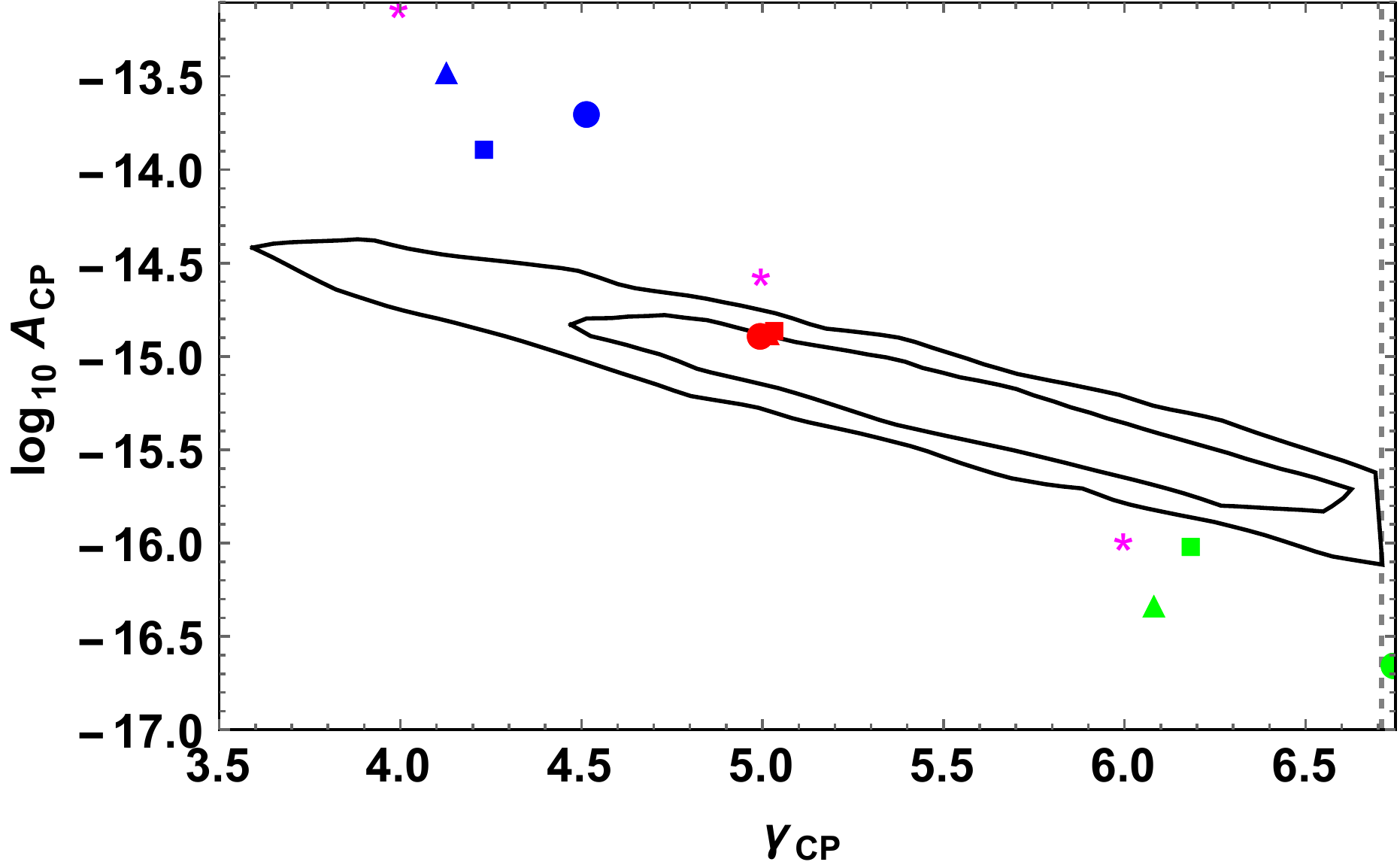}
\caption{\small{$A_{\rm CP}$ and $\gamma _{\rm CP}$ fitted for $f_{\rm PBH}=0.1$ for each case of a Gaussian primordial spectrum: $\sigma_p=1$, $2$, and $3$ are plotted with circle, square and triangles respectively;  for $w=1/3$ in red,  $w=1$ in blue and  $w=1/9$ in green. Points for $f_{\rm PBH}=0.1$ for a constant primordial power spectrum for these three $w$ are plotted as pink stars. This constant power spectrum generates PBH masses in the range: $0.1<M/M_{\odot}<5$. The dashed gray line signifies the open edge of the $2\sigma $ contour in the \NG analysis (orange contour in Fig.1 of~\cite{Arzoumanian:2020vkk}).}}
\label{dotplot}}
\end{figure}
For the curvature power spectrum given in Eq.~\eqref{PKG} with $\sigma_p=$1, 2, and 3, we evaluate \OMG2 with $k_*=3.6\times 10^6$ Mpc$^{-1}$ which corresponds to $f_*=5.5$ nHz, and fit the results in the frequency range of the \NG for each of the three $w$ epochs. For each case, we chose $A_s$ such that $f_{\rm PBH}=0.1$. The resulting $A_{\rm CP}$ and $\gamma _{CP}$ are plotted with the \NG $1\sigma$ and $2\sigma$ contours in Fig.~\ref{dotplot}.  For RD (red circle, square, and triangle), the resulting three points  for different $\sigma_p$ are close together inside the contour, therefore, abundant PBH production is consistent with the \NG observation. However, for the other two cases ($w=1$ in blue and $w=1/9$ in green), points for different $\sigma_p$ are far apart with the value closest to the contours given by $\sigma_p=2$ for both the cases. Therefore, in the rest of our analysis, we use $\sigma_p=2$ to check the consistency of abundant PBH formation in these nonRD epochs with the \NG result. In this figure, we also show the points corresponding to a constant power spectrum of the following form:
\begin{equation}
P_{\zeta}(k)=A_s\times \Theta (k-k_l)\times \Theta (k_s-k),
\label{constpow}
\end{equation}
where $k_l$ and $k_s$ correspond to the scales between which the power spectrum is constant. We have chosen $k_l$ and $k_s$ values for each $w$ case such that, the resulting PBHs from~\eqref{constpow} are always in the mass range $M/M_{\odot}\in (0.1,5)$. We find that for the constant power spectrum, PBH production with $f_{\rm PBH}=0.1$ (stars in Fig.~\ref{dotplot}) is not consistent with the \NG result in these nonstandard epochs under consideration.
\begin{table}[!htbp]
\caption{Range of amplitude $A_s$ chosen for the three cases with different EoS used in Fig.~\ref{regcont}.}
\begin{center}
\label{TT1}
\resizebox{0.4\textwidth}{!}{
\begin{tabular}{|c|c|c|c|}
\hline 
 & $w=1/3$ & $w=1$ & $w=1/9$ \\
\hline
\hline 
$A_s^{\rm max} $ & 0.015 & 0.007 & 0.0082\\
\hline 
$A_s^{\rm min}$ & 0.002 & 0.001 & 0.003\\
\hline
%\hline
%$6\times 10^{12} $ & $1/3$ & 0.013 & 0.016 & 0.0163\\
%\hline
%$6 \times 10^{12} $ & $1$ & 0.0048 & 0.0067 & 0.006\\
%\hline
\end{tabular}}
\end{center}
\end{table}

\begin{figure}[!htbp]
\centering{
\includegraphics[width=0.75\textwidth]{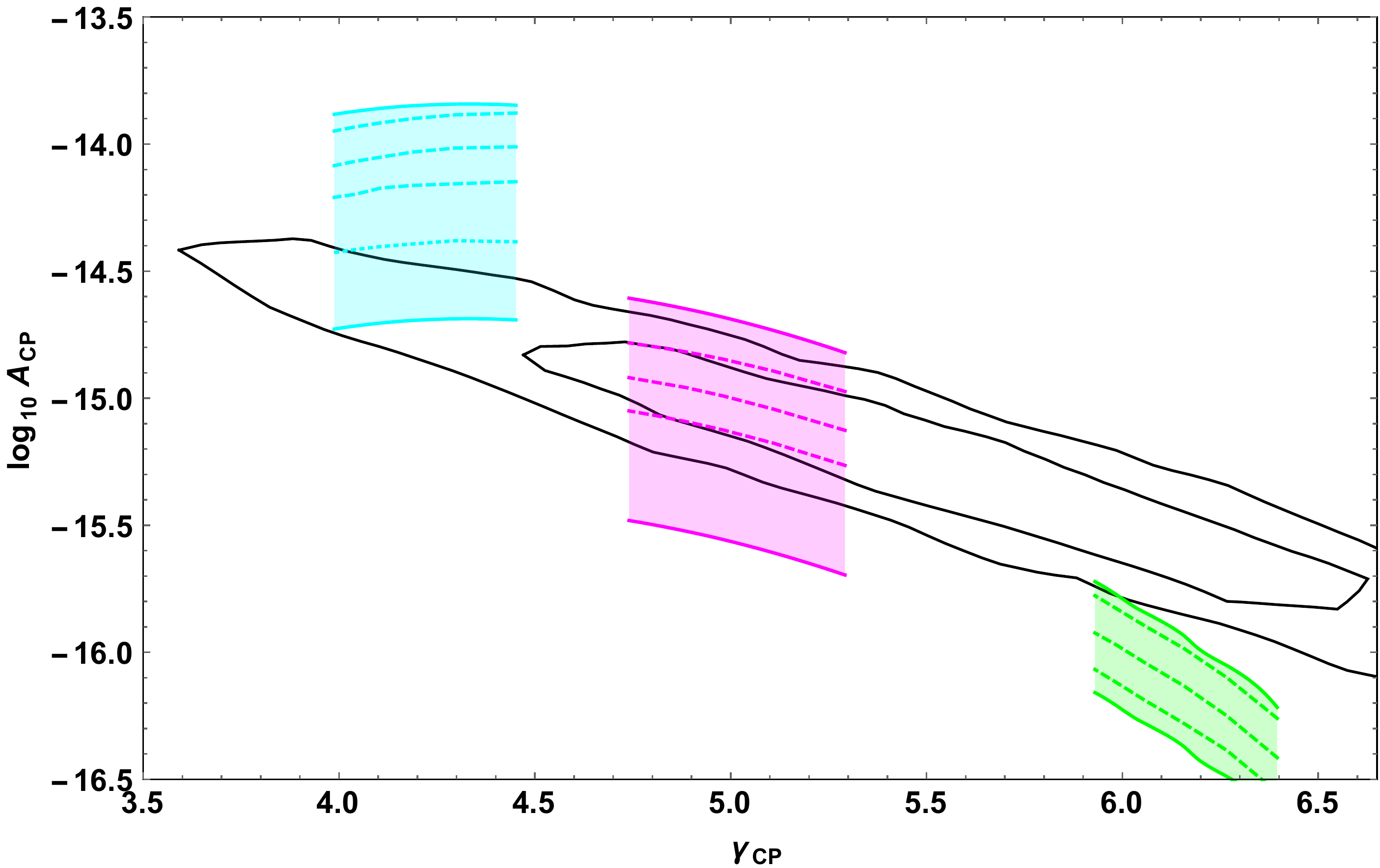}
\caption{\small{$A_{\rm CP}$ and $\gamma _{\rm CP}$ fitted for $A_s$ varied in different ranges for $w=1/3$ (magenta shaded), $w=1$ (cyan shaded), and $w=1/9$ (green shaded) with variation in the pivot scale in the range $k_*=2\times 10^6$ Mpc$^{-1}$ to $k_*=7\times 10^6$ Mpc$^{-1}$ for the Gaussian primordial spectrum with $\sigma_p=2$. The dashed lines inside each shaded region are contours for $f_{\rm PBH}=0.1$, $10^{-5}$ and $10^{-10}$ from top to bottom. The additional dotted cyan line in the $w=1$ case just enters the 2$\sigma$ contour of the \NG observation, but does not motivate abundant PBH (here $f_{\rm PBH}\simeq 10^{-22}$). For the $w=1/9$ case, the upper edge of the green shaded region motivates $f_{\rm PBH}=1$, but the observations constrain $f_{\rm PBH}$ to a much lower value.}}
\label{regcont}}
\end{figure}

\begin{figure}[!htbp]
\centering{
\includegraphics[width=0.75\textwidth]{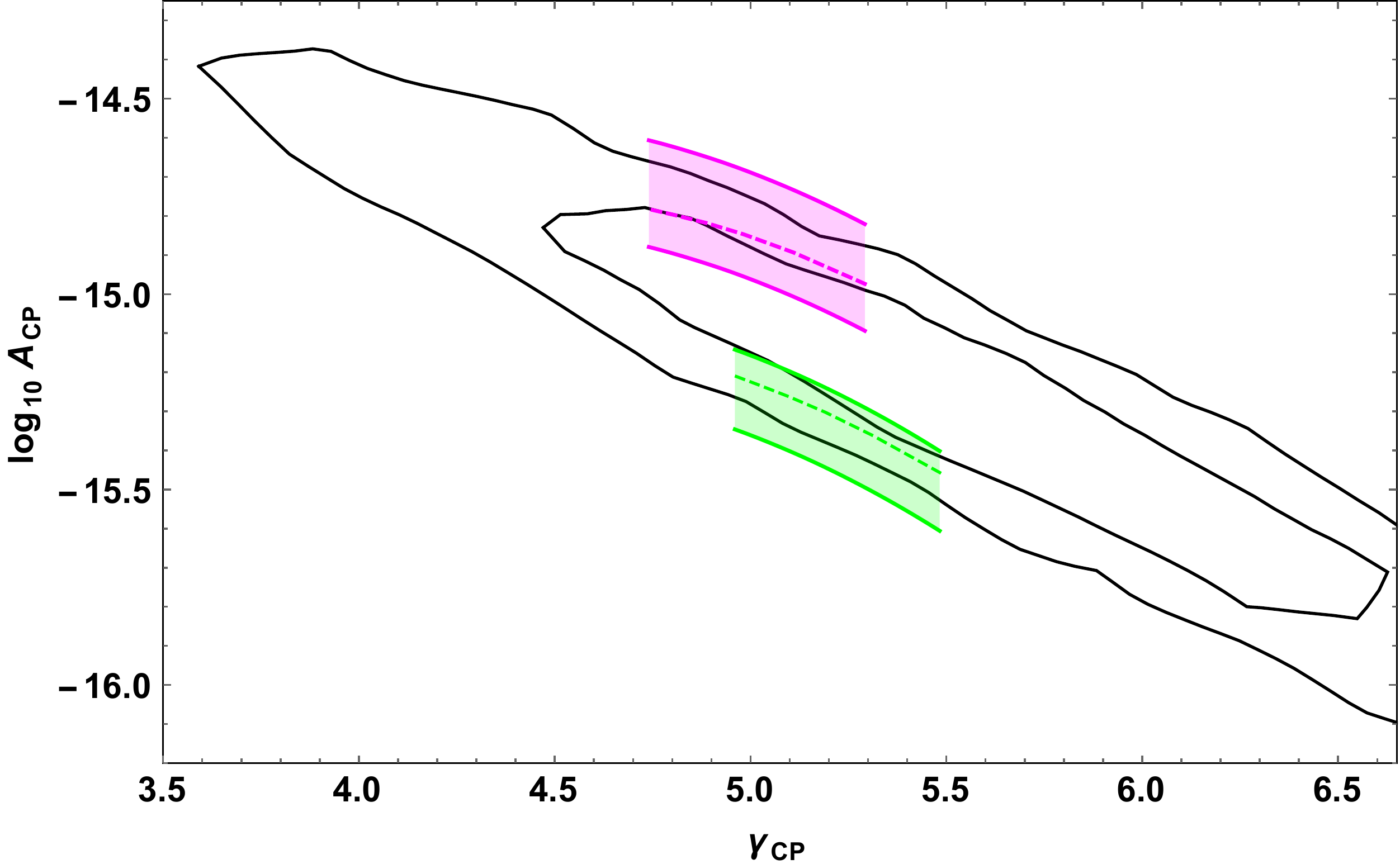}
\caption{\small{For a Gaussian primordial power spectrum, $A_{\rm CP}$ and $\gamma _{\rm CP}$ fitted for $A_s$ varied in different ranges for $w=1/3$ (magenta shaded) and $w=1/9$ (green shaded) with variation in the pivot scale in the range $k_*=2\times 10^6$ Mpc$^{-1}$ to $k_*=7\times 10^6$ Mpc$^{-1}$ for RD and $k_*=2\times 10^7$ Mpc$^{-1}$ to $k_*=7\times 10^7$ Mpc$^{-1}$ for $w=1/9$. The dashed line inside each shaded region is the $f_{\rm PBH}=0.1$ contour. The value of $\sigma_p$ is fixed to be 2.}}
\label{regcont2}}
\end{figure}

Fixing $\sigma_p=2$, we vary $A_s$ (ranges given in Table~\ref{TT1}) and vary $k_*$ in the range ($2\times 10^6$ Mpc$^{-1}$ - $7 \times 10^6$ Mpc$^{-1}$) for each $w$ scenario. The corresponding variations in $A_{\rm CP}$-$\gamma_{\rm CP}$ space are given by magenta ($w=1/3$), cyan ($w=1$), and green ($w=1/9$) shaded regions in Fig.~\ref{regcont}. We show the contours for PBH abundance in dashed lines inside these contours with $f_{\rm PBH}=0.1$, $10^{-5}$, and $10^{-10}$ from top to bottom. It is evident from Fig.~\ref{regcont} that $10\%$ PBH as dark matter is possible for RD. For $w=1$, All these dashed contours remain outside of the \NG $2\sigma$ contour; the most hopeful case in this $k_*$ range is represented by the cyan dotted contour at the bottom, which represents $f_{\rm PBH}\simeq 10^{-22}$. Therefore, abundant PBH production in $w=1$ epoch is not consistent with the \NG observation. Increasing $k_*$ will lead to the \OMG2 peak being closer to the \NG frequency range, and the $f_{\rm PBH}$ contours come down to lower $A_{\rm CP}$ values. However, the slope of \OMG2 is such that $f_{\rm PBH}=0.1$ contour comes within the \NG $A_{\rm CP}$ range only when $\gamma_{\rm CP}$ is very small. Moving towards the lower values of $k_*$ does not help since power at $k_0$ goes further down and will also influence the power spectrum at CMB scales for a broad primordial spectrum. 
For $w=1/9$, the $f_{\rm PBH}=0.1$ contour enters the \NG contour only for $k_*\simeq 7 \times 10^6$ Mpc$^{-1}$, and therefore it is possible to achieve abundant PBH production for the $w=1/9$ case for $k_*$ larger than the \NG probed range.
Therefore, we proceed with the case $w=1/9$ now and vary $k_*$ in the range ($2\times 10^7$ Mpc$^{-1}$ - $7 \times 10^7$ Mpc$^{-1}$) for a variation in $A_s$ in the range $0.005-0.0082$ for $w=1/9$. For the RD case, $A_s$ is varied in the same $k_*$ range as before, but for a smaller range in $A_s$ between $0.008-0.015$. The resulting variation in $A_{\rm CP}$-$\gamma_{\rm CP}$ space are shown in Fig.~\ref{regcont2} in green for $w=1/9$ and in magenta for $w=1/3$. The relevant PBH contours for $f_{\rm PBH}=0.1$ are shown in dashed lines. We conclude that the PBH and the second order GW production in a nonstandard epoch with EoS $w=1/9$ is consistent with the \NG result when $k_*$ is away from the \NG probed range. The gain in the $w=1/9$ over RD epoch is that for this nonstandard epoch, abundant PBH and observed GW can both be explained with a lower value of $A_s$: $A_s(w=1/9)\vert _{f_{\rm PBH}=0.1}\simeq 0.007$ and $A_s(w=1/3)\vert _{f_{\rm PBH}=0.1}\simeq 0.01$. 
The Gaussian power spectrum with $\sigma_p=2$ has a broad peak, therefore we quote a range of the PBH mass near the peaks for each case for the lowest and highest $k_*$ under consideration in Table~\ref{TT2}.
\begin{table}[!htbp]
\caption{Ranges of PBH mass $M$ for each EoS}
\begin{center}
\label{TT2}
\resizebox{0.85\textwidth}{!}{
\begin{tabular}{|c|c|c|c|}
\hline 
 & $w=1/3$ & $w=1$ & $w=1/9$ \\
\hline
\hline 
Range of $k_*$ in Mpc$^{-1}$ & $2\times 10^6$ - $7 \times 10^6$ & $2\times 10^6$ - $7 \times 10^6$  & $2\times 10^7$ - $7 \times 10^7$\\
\hline 
Range of $M/M_{\odot}$ at $k_{*,{\rm min}}$ & 0.2-2 & 1-10 & $10^{-4}$ - $5\times 10^{-3}$\\
\hline
Range of $M/M_{\odot}$ at $k_{*,{\rm max}}$ & 0.01 - 0.33 & 0.08-2 & $3\times 10^{-6}$ - $3\times 10^{-4}$\\
\hline
%$6\times 10^{12} $ & $1/3$ & 0.013 & 0.016 & 0.0163\\
%\hline
%$6 \times 10^{12} $ & $1$ & 0.0048 & 0.0067 & 0.006\\
%\hline
\end{tabular}}
\end{center}
\end{table} 
\subsubsection{Lognormal power spectrum with an exponential cut off}
\label{PLN}
\suk{The lognormal power spectrum with an exponential cutoff is motivated in~\cite{Braglia:2020eai,Vaskonen:2020lbd} and is given as
\begin{equation}
P_{\zeta}(k)=A_s\times \exp\bigg[\beta \bigg(1-\frac{k}{k_*}+\log \bigg(\frac{k}{k_*}\bigg) \bigg)-\alpha \log ^2 \big(\frac{k}{k_*}\big) \bigg]\,,
\label{PKLN}
\end{equation}
where $\alpha=0.17$ and $\beta=0.62$. The second order GW spectra normalized by $A_s^2$ for this power spectrum are shown in Fig.~\ref{OmgA2_LN} for the standard RD epoch and other two nonstandard post-inflationary epochs with $w=1$ and $w=1/9$. 
\begin{figure}[!htbp]
\centering{
\includegraphics[width=0.6\textwidth]{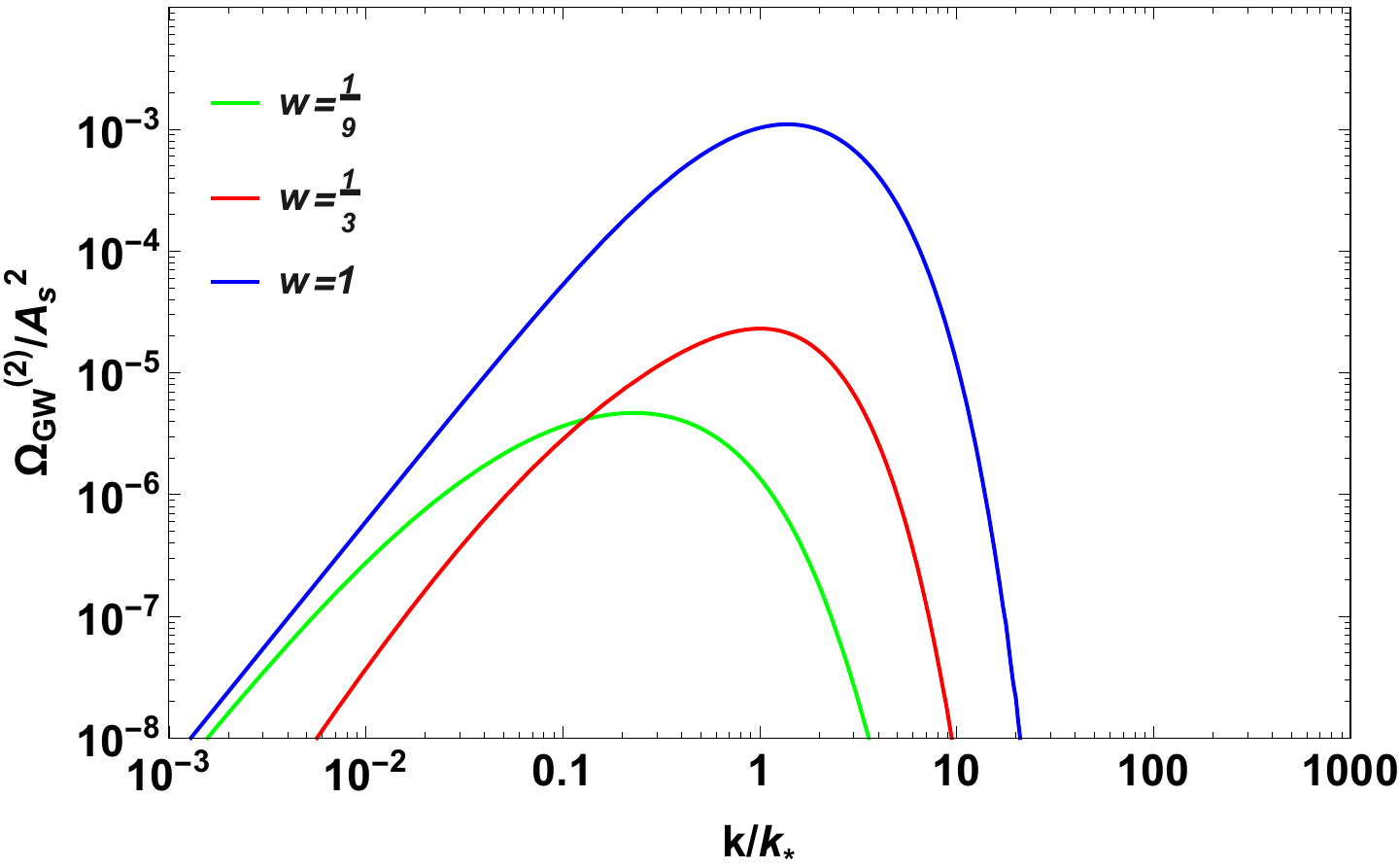}
\caption{\small{The normalised \OMG2 for the three values of $w$ are shown as a function of $k/k_*$ using the lognormal power spectrum with an exponential cutoff.}}
\label{OmgA2_LN}}
\end{figure}
The profiles of the normalised \OMG2in Fig.~\ref{OmgA2_LN} show that the peaks occur at $k=k_*$ for RD, $k<k_*$ for $w=1/9$, and $k\simeq 1.38k_*$ for $w=1$. Assuming that the location of the peaks in the \OMG2 spectra are within the \NG frequency range, the amplitude of $P_{\zeta}(k)$ required for each $w$ to explain the observed GW amplitude of $10^{-9}$ are $A_s (w=1/3)=6.6\times 10^{-3}$, $A_s (w=1)=9.5\times 10^{-4}$, and $A_s (w=1/9)=0.0146$.

The PBH mass function for this $P_{\zeta}(k)$ is plotted in Fig.~\ref{psiPBH_LN} for the three $w$ values, which show that each of them corresponds to a broad range of PBH mass. Here, with the given choices of $k_*=k_0$ and $A_s$, the PBH masses relevant to the binary merger events observed in LIGO/Virgo are produced with the most significant abundance only for $w=1$, whereas the $w=1/9$ case permits a considerable abundance of the planetary mass PBHs.
%The peak positions in Fig.s~\ref{OmgA2_LN} and~\ref{psiPBH_LN} are correlated for $w=1/9$, however, for the nonstandard case of $w=1$, the correlation is not straightforward. 
\begin{figure}[!htbp]
\centering{
\includegraphics[width=0.6\textwidth]{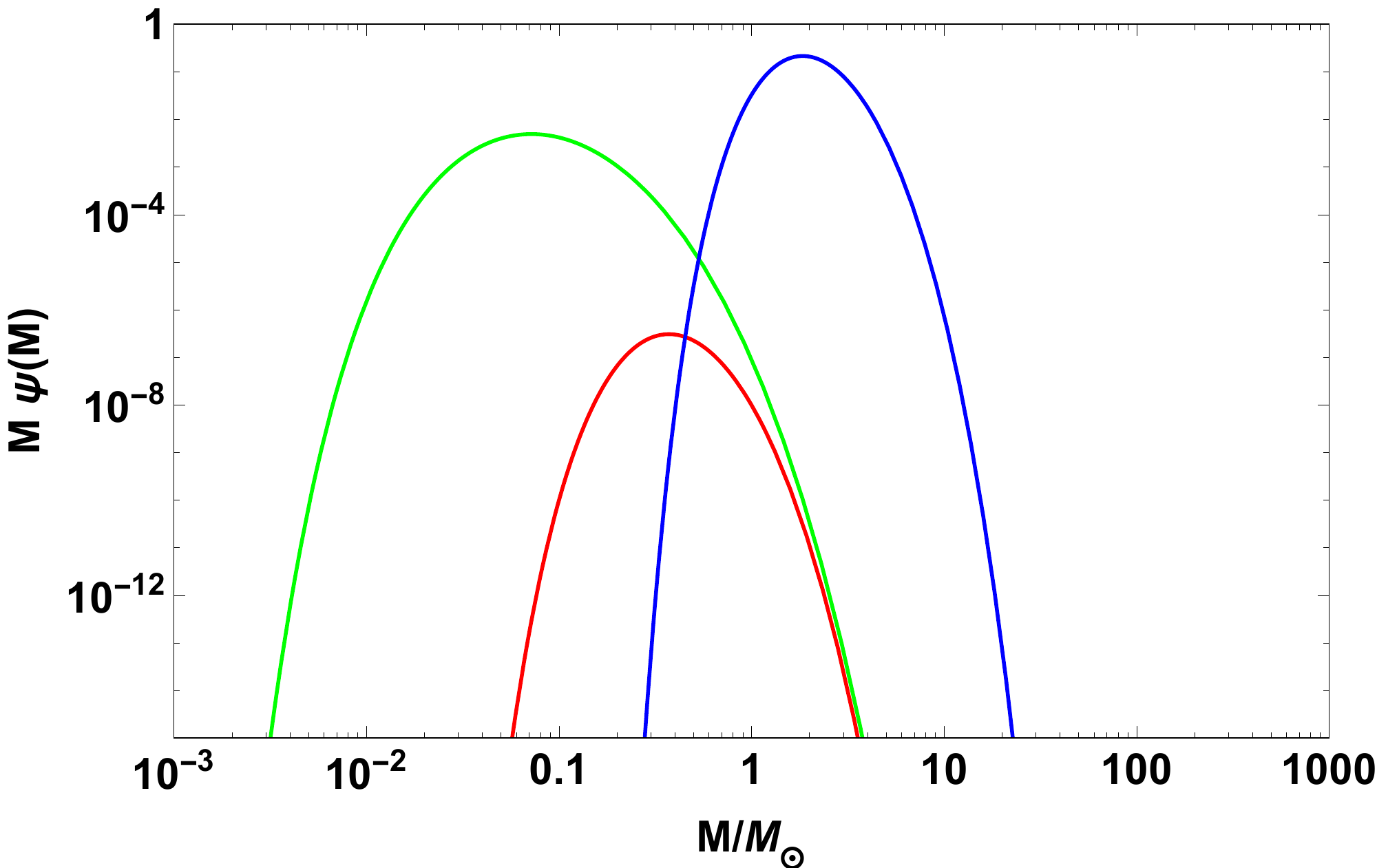}
\caption{\small{PBH mass functions for different $w$ epochs are shown for a typical primordial amplitude $A_s=0.007$ using the power spectrum given in Eq.~\ref{PKLN}. Colour specifications are same as Fig.~\ref{OmgA2_LN}.}}
\label{psiPBH_LN}}
\end{figure}

A similar analysis as in Fig.~\ref{regcont} in Sec.~\ref{GPP} is carried out here for the power spectrum in Eq.~\ref{PKLN} where $k_*$ is varied in the range $(10^6 - 10^7)$ Mpc$^{-1}$ and $A_s$ is varied in the range $(0.002 - 0.02)$. The resulting scanned parameter space in the $A_{\rm CP} - \gamma_{{\rm CP}}$ plane is shown in Fig.~\ref{LNcont} for $w=1/3$ (magenta), $w=1$ (cyan), and $w=1/9$ (green), where the dark shaded region in each case corresponds to the exact frequency range where the \NG analysis fits the observed spectrum as a common power law. The coloured dashed contours in each case signify PBH abundance of $0.1$ (upper) and $10^{-5}$ (lower). Embedding the $1\sigma$ and $2\sigma$ contours for $A_{\rm CP} - \gamma_{{\rm CP}}$ from the \NG observation in this plot (black contours), it is evident that for RD, $10\%$ PBH abundance for $P_{\zeta}(k)$ given in Eq.~\ref{PKLN} is consistent with the \NG 12.5 year data at $2\sigma$ confidence level. The nonstandard scenario with $w=1/9$ is also consistent with the \NG data where PBH is produced with $\gtrsim 10\%$ abundance, but only for $k_*\gtrsim 5\times 10^6$ Mpc$^{-1}$. On the other hand, the second order GW produced in the nonstandard $w=1$ epoch is consistent with the \NG observation only when the PBH abundance is negligible.
%The dependence of $\Omega_{\rm GW}^{(2)}/A_s^2$ on $k/k_*$ and the profiles of the mass functions for different values of $w$ is then studied in the 
%WILL WRITE ABOUT THE PLOT BELOW.
\begin{figure}[!htbp]
\centering{
\includegraphics[width=0.75\textwidth]{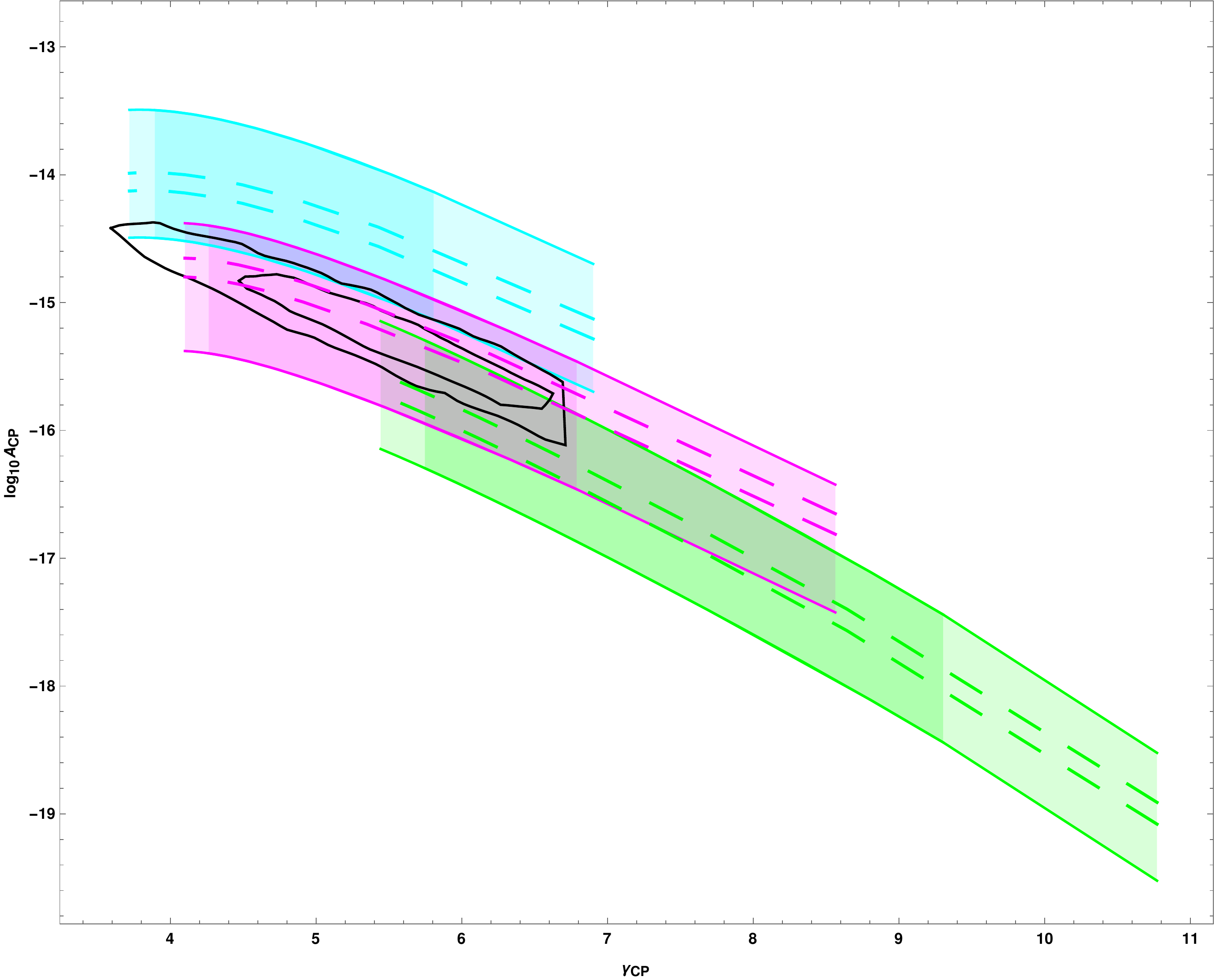}
\caption{\small{For a lognormal primordial power spectrum with an exponential cut off, $A_{\rm CP}$ and $\gamma _{\rm CP}$ fitted for $A_s$ varied in the range $0.002 - 0.02$ for $w=1/3$ (magenta shaded), $w=1$ (cyan shaded), and $w=1/9$ (green shaded). For dark shaded regions, the pivot wavenumber runs from $k_*=1.615\times 10^6$ Mpc$^{-1}$ to $k_*=7.756\times 10^6$ Mpc$^{-1}$ from right to left, where \NG fits their observation to a common power law power spectrum. The light shaded regions are shown for the pivot wavenumber $k_*=10^6$ Mpc$^{-1}$ to $k_*=10^7$ Mpc$^{-1}$ from right to left. The dashed contours are for the PBH abundance of $0.1$ (above) and $10^{-5}$ (below) for each case. Black solid contours represent the $1\sigma$ and $2\sigma$ confidence levels for the parameters $A_{\rm CP}$ and $\gamma _{\rm CP}$ observed in the \NG .}}
\label{LNcont}}
\end{figure}}

\subsubsection{Broken power law power spectrum}
\label{PBPL}
\suk{The primordial power spectrum can also have the form of a broken power law~\cite{Ballesteros:2017fsr,Byrnes:2018txb,Vaskonen:2020lbd} given by
\begin{equation}
P_{\zeta}(k)=A_s\times \frac{\alpha +\beta}{\beta \bigg(\frac{k}{k_*}\bigg)^{-\alpha}+\alpha \bigg(\frac{k}{k_*}\bigg)^{\beta}}\, ,
\label{PKBPL}
\end{equation}
where $\alpha=4$ and $\beta=0.6$. The second order GW spectra normalized by $A_s^2$ for this power spectrum are shown in Fig.~\ref{OmgA2_BPL} for the epochs $w=1/3$, $1$, and $1/9$. Unlike the power spectra discussed in Sec.~\ref{GPP} and Sec.~\ref{PLN}, this form of $P_{\zeta}(k)$ has a sharp peak, which is reflected in the wedge-like shapes of the normalized \OMG2 in Fig.~\ref{OmgA2_BPL}. Moreover, the difference in slopes of \OMG2 among the three $w$ cases clearly manifest the power-law dependence of the integration kernel $\overline{I^2}(v,u,x)$ in Eq.~\ref{eq:pgamma} on $w$.  The peak positions in this figure are $k=k_*$ for $w=1/3$, $k\simeq 3k_*$ for $w=1$, and $k\simeq 0.5k_*$ for $w=1/9$ respectively. Assuming that for each $w$, the peaks occur in the \NG observed frequency range, the peak amplitudes required to explain the \NG observation with \OMG2 $\sim 10^{-9}$ are $A_s=6.9\times 10^{-3}$, $8.7\times 10^{-4}$, and $0.0187$ for $w=1/3$, $1$, and $1/9$ respectively.

\begin{figure}[!htbp]
\centering{
\includegraphics[width=0.6\textwidth]{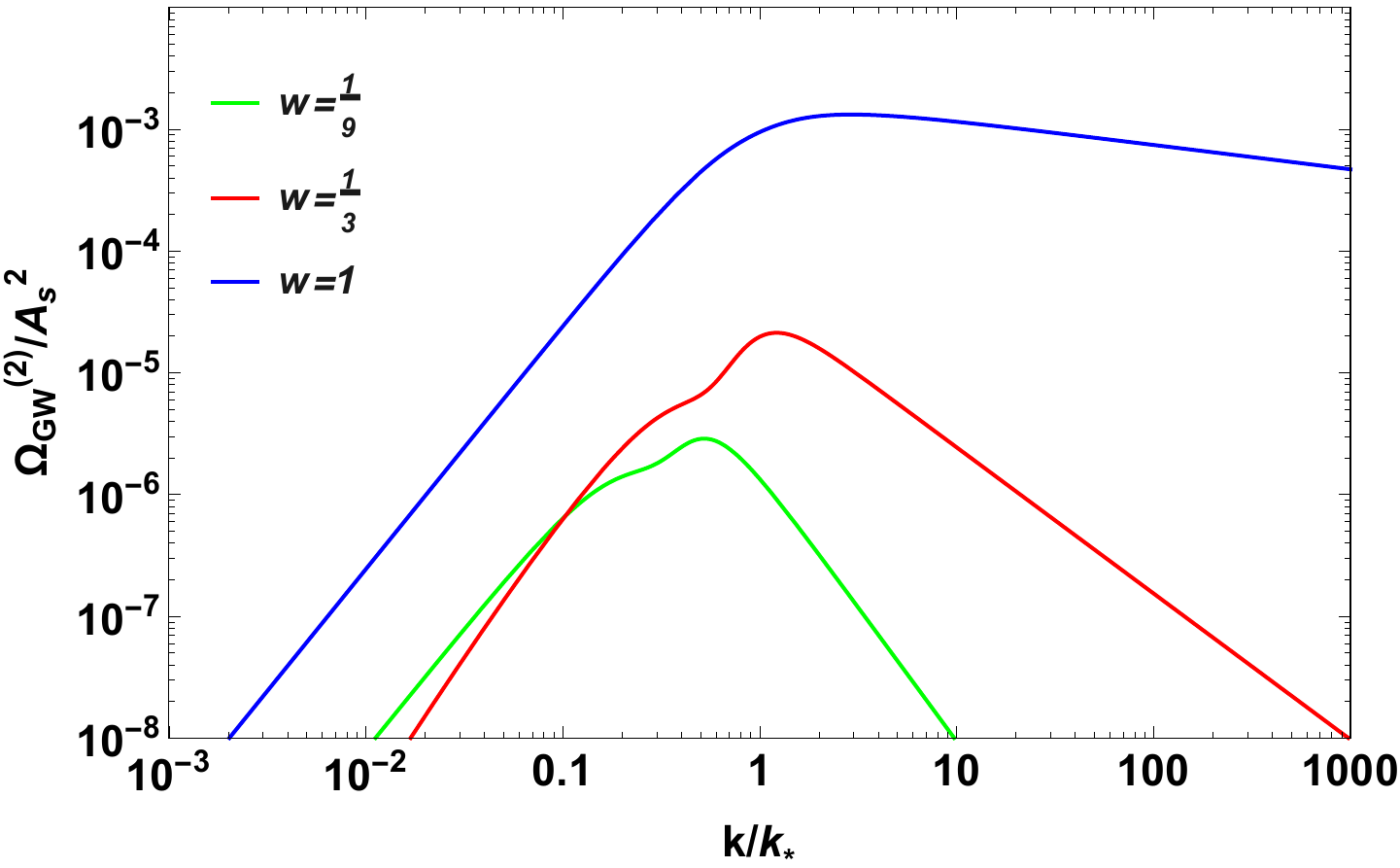}
\caption{\small{The normalised \OMG2 for the three values of $w$ are shown as a function of $k/k_*$ using the broken power law power spectrum.}}
\label{OmgA2_BPL}}
\end{figure}
% Assuming that the peaks in the \OMG2 spectra coincide with the \NG frequency range, the amplitude of $P_{\zeta}(k)$ required for each $w$ to explain the observed GW amplitude of $10{-9}$ are $A_s (w=1/3)=6.6\times 10^{-3}$, $A_s (w=1)=9.5\times 10^{-4}$ and $A_s (w=1/9)=0.0146$.
%PBH abundance is negligible.
\begin{figure}[!htbp]
	\centering{
		\includegraphics[width=0.6\textwidth]{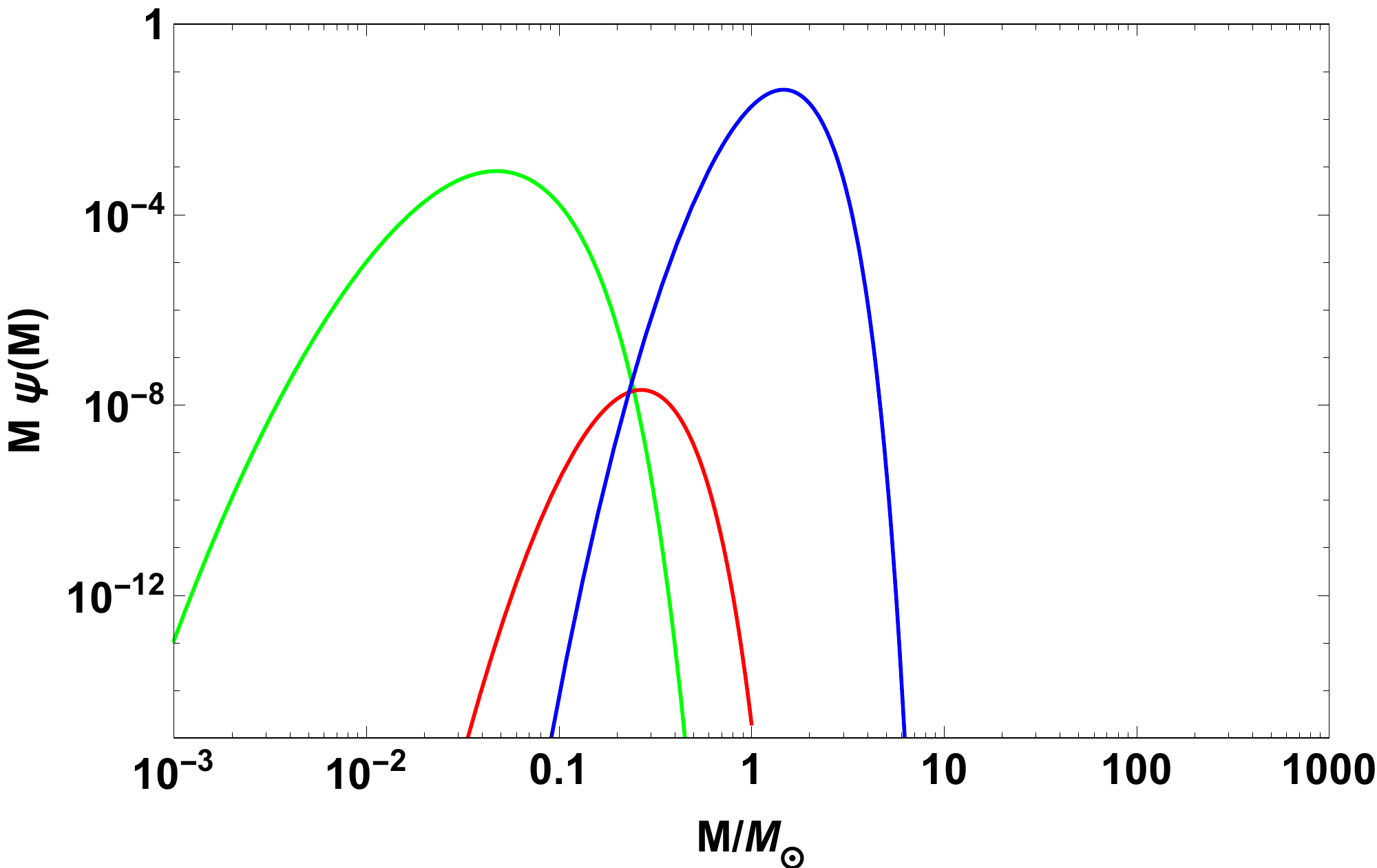}
		\caption{\small{PBH mass functions for different $w$ epochs are shown for a typical primordial amplitude $A_s=0.007$ using the power spectrum given in Eq.~\ref{PKBPL}. Colour specifications are same as Fig.~\ref{OmgA2_BPL}.}}
		\label{psiPBH_BPL}}
\end{figure}
The corresponding PBH mass function is shown in Fig.~\ref{psiPBH_BPL} for all the epochs under consideration. It is evident from Fig.~\ref{psiPBH_BPL} that PBHs are produced in different mass ranges for each case. Similar to the results in Sec.~\ref{GPP} and Sec.~\ref{PLN}, PBHs in the LIGO/Virgo observed mass range are produced with the large abundance only for $w=1$, whereas the planetary mass PBHs are produced in a significant number for $w=1/9$.
%The peak positions in Fig.s~\ref{OmgA2_LN} and~\ref{psiPBH_LN} are correlated for $w=1/9$, however, for the nonstandard case of $w=1$, the correlation is not straightforward. 

The implications of the broken power law power spectrum in Eq.~\ref{PKBPL} for \OMG2 and PBH abundance are shown in the $A_{\rm CP} - \gamma_{{\rm CP}}$ plane in Fig.~\ref{BPLcont} with the color and line specifications similar to Fig.s~\ref{regcont2} and~\ref{LNcont}. Varying $k_*$ in the range $(10^6 - 10^7)$ Mpc$^{-1}$ and $A_s$  in the range $(0.002 - 0.02)$ show that if the \NG observation is assumed to be due to the second order GW production in the nonstandard $w=1$ epoch, then PBH is produced negligibly. For the standard RD epoch, PBH with $10\%$ abundance can be produced here while the corresponding \OMG2 is consistent with \NG result within $1\sigma$ confidence limit. However, for the nonstandard dustlike case with $w=1/9$, even $100\%$ PBH abundance (although observationally not allowed) can be reached with an amplitude $A_s$ lower than that required to match predicted \OMG2 to be consistent with the \NG result at $2\sigma$ limit.
%A similar analysis as in Sec.~\ref{GPP} is carried out here for the power spectrum in Eq.~\ref{PKLN} where $k_*$ is varied in the range $(10^6 - 10^7)$ Mpc$^{-1}$ and $A_s$ is varied in the range $(0.002 - 0.02)$. The resulting scanned parameter space in the $A_{\rm CP} - \gamma_{{\rm CP}}$ plane is shown in Fig.~\ref{LNcont} for $w=1/3$ (magenta), $w=1$ (cyan) and $w=1/9$ (green), where the dark shaded region in each case corresponds to the exact frequency range where \NG analysis fit the observed spectrum as a common power law. The coloured dashed contours in each case signify PBH abundance of $0.1$ (upper) and $10^{-5}$ (lower). Embedding the $1\sigma$ and $2\sigma$ contours for $A_{\rm CP} - \gamma_{{\rm CP}}$ from observation in this plot (black contours), it is evident that for RD, $10\%$ PBH abundance for $P_{\zeta}(k)$ given in Eq.~\ref{PKLN} is consistent with the \NG 12.5 year data at $2\sigma$ confidence level. The nonstandard scenario with $w=1/9$ is also consistent with \NG data where PBH is produced with $\gtrsim 10\%$ abundance, but only for $k_*\gtrsim 5\times 10^6$ Mpc$^{-1}$. On the other hand, the second order GW produced in the nonstandard $w=1$ epoch is consistent with \NG observation only when the 
\begin{figure}[!htbp]
\centering{
\includegraphics[width=0.75\textwidth]{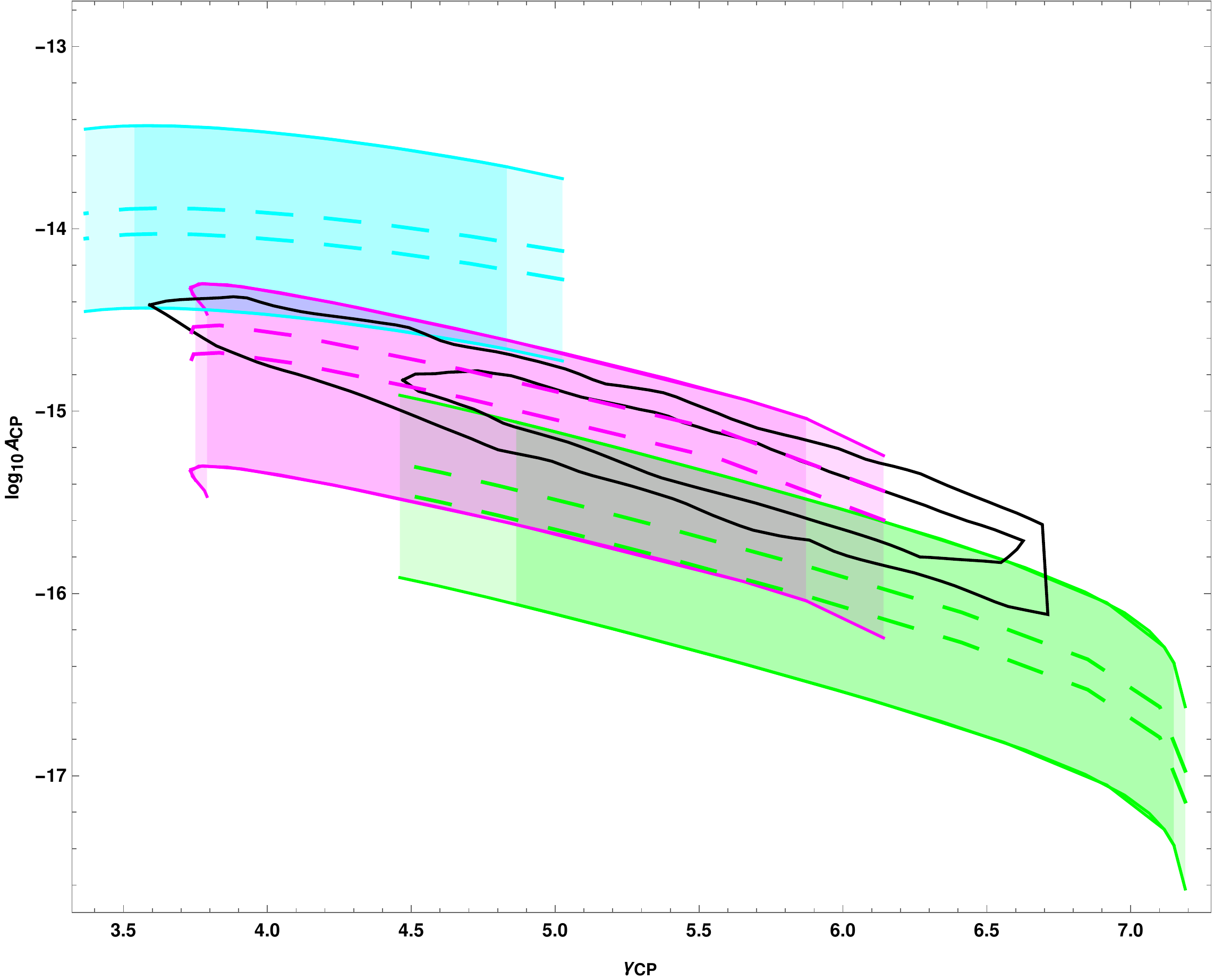}
\caption{\small{For a broken power law primordial power spectrum, $A_{\rm CP}$ and $\gamma _{\rm CP}$ fitted for $A_s$ varied in the range $0.002 - 0.02$ for $w=1/3$ (magenta shaded) $w=1$ (cyan shaded), and $w=1/9$ (green shaded). For dark shaded regions, the pivot wavenumber runs from $k_*=1.615\times 10^6$ Mpc$^{-1}$ to $k_*=7.756\times 10^6$ Mpc$^{-1}$ from right to left, where the \NG fits their observation to a common power law power spectrum. The light shaded regions are shown for the pivot wavenumber $k_*=10^6$ Mpc$^{-1}$ to $k_*=10^7$ Mpc$^{-1}$ from right to left. The dashed contours are for the PBH abundance of $0.1$ (above) and $10^{-5}$ (below) for each case. Black solid contours represent the $1\sigma$ and $2\sigma$ confidence levels for the parameters $A_{\rm CP}$ and $\gamma _{\rm CP}$ observed in the \NG .}}
\label{BPLcont}}
\end{figure}
}
\section{Discussions and conclusions}
\label{ressec}
The recent \NG 12.5 yr result can be the first putative signal for the stochastic GW originating in the early universe; further re-analysis of the data will lead to a more precise statement about whether GW has been detected or not. Therefore, it is necessary to check the early universe theories that are consistent with the \NG result interpreted as GW. 

One of the possible explanation for the \NG result can come from the stochastic GW produced
during inflation. Recently, it was shown in ref.~\cite{Vagnozzi:2020gtf} that if the NANOGrav result is explained by the GW produced during inflation with the standard thermal history, a blue tilted power spectrum will be required. However, the required parameter space to explain the NANOGrav result is entirely excluded by the GW's contribution to effective number of relativistic degrees of freedom during BBN. In this work, we have explored the idea if the \NG signal can be explained by the GWs generated during inflation with a nonstandard evolution in the post-inflationary cosmological history. While it is well established that BBN must have taken place during the RD era, there is no direct observational probe of EoS for the epoch between the inflation and BBN. In this work, we have assumed a pre-BBN epoch with an equation of state $0\leq w \leq 1$. More specifically, in the first order GW analysis, we have considered three different EoS, $w=0,1/3$, and $1$, which correspond to an early MD era, RD era (standard), and KD era respectively. 

\pri{Our main results for the first order GW are summarized in Fig.~\ref{fig:Hinf_nt_allowed_rd_md} where we have shown the \NG allowed $H_{{\rm inf}}-n_t$ parameter space along with the constraints from CMB, BBN, and LIGO for the three different EoS. It is evident from Fig.~\ref{fig:Hinf_nt_allowed_rd_mdw1} that the parameter space required to explain the \NG result in a nonstandard thermal history with a KD epoch is ruled out by the bound on $H_{{\rm inf}}$ from the BICEP2/Keck Array and Planck CMB data. It is quite clear from Fig.~\ref{fig:Hinf_nt_allowed_rd_mdw13}  that our results for standard thermal history are consistent with ref~\cite{Vagnozzi:2020gtf} as the entire parameter space required to explain the \NG result is ruled out by the constraints from BICEP2/Keck Array, Planck CMB, BBN, and LIGO observations. Interestingly, GWs in a nonstandard thermal history with an early MD epoch can explain the \NG result and can also evade the BICEP2/Keck Array, Planck CMB, BBN and LIGO bounds successfully for a finite region in the parameter space (see Fig.~\ref{fig:Hinf_nt_allowed_rd_mdw0}). This is one of the main results of our analysis for the first order GWs.}

\pri{While calculating the bounds from BBN, the upper limit of the integration given by Eq.~\eqref{eq:bbncons} was set to $10^7$ Hz, which corresponds to a reheat temperature of $10^{15}$ GeV. The upper limit is related to the scale at the end of inflation and lowering its value decreases the value of the integral.} Therefore, for a blue tilted power spectrum, this bound will be further relaxed in case of a lower value of the energy scale at the end of inflation. We have also assumed that the early MD era spans from the end of the inflation until the onset of the RD era at temperature $T_1=10$ MeV. However, there can be a scenario where this MD era lasts for a shorter period, still enveloping the \NG frequency range. In that case, this bound on GWs from BBN will only be valid if the total contribution of $\Omega_{{\rm GW}}$ from the RD era (lasting in between the end of inflation and the onset of early MD) is smaller than the total contribution from the early MD era.

The constraints from the \NG on the tilt of the primordial tensor spectrum $n_t$ for different $w$ can also be explained using the slope of the GW spectra. In the first order, the slope of the GW spectrum for a primordial tensor power spectrum with tilt $n_t$ evolving in a $w$ dominated epoch can be calculated from Eq.~\eqref{eq:gammacp} as $5-\gamma _{\rm CP}=\frac{6w-2}{3w+1}+n_t$. Thus, with the $1\sigma$ allowed range $4.5\leq\gamma _{\rm CP}\leq6.5$, the bounds on $n_t$ are: $1-21w\leq 2n_t \leq 5-9w$. Therefore for MD, RD, and KD epochs respectively, the bounds on $n_t$ arising from the \NG observed slope are: $0.5 \leq n_t \leq 2.5$, $-3 \leq n_t \leq 1$, and $-10 \leq n_t \leq -2$. However, the bounds on the observed amplitude $A_{{\rm CP}}$ at the \NG along with the bounds from BBN, CMB, and LIGO further reduce these ranges considerably to completely disallow the theoretical signal arising from the first order perturbation theory in the standard RD epoch or an early KD epoch. For an early MD epoch, however, a reduced range of $n_t$ is still consistent with all these bounds as explained in Fig.~\ref{fig:Hinf_nt_allowed_rd_mdw0}.

Our main underlying assumption is that the primordial tensor power spectrum is a pure power law for all the frequencies ranging from CMB scale to the end of inflation. Therefore, in the future, if Laser Interferometer Space Antenna (LISA), which will be observing in the frequency range $f\sim 10^{-2}$ Hz, makes some positive or null detection, that can be used to further constrain the nonstandard pre-BBN epoch. In conclusion, we have explored the possibility of \NG signal coming from the GWs generated during inflation and evolving in a nonstandard thermal history.  \pri{In the Fig.~\ref{fig:Hinf_nt_allowed_rd_mdw0acp}, we have shown the $\gamma_{{\rm CP}}-A_{{\rm CP}}$ parameter space (blue shaded region) corresponding to the allowed $H_{{\rm inf}}-n_t$ parameter space required to explain the \NG result in a nonstandard thermal history with an early MD epoch.} If the parameter space  for $\gamma_{{\rm CP}}-A_{{\rm CP}}$ shrinks further with the help of upcoming data from the \NG or other PTA observatories in the future, it will help in further narrowing down the EoS value of a pre-BBN epoch.

For the second order GW phenomenology in Sec.~\ref{GW2sec}, we consider three separate scenarios for the EoS, standard RD ($w=1/3$), early kinetic energy domination ($w=1$), and a dustlike scenario ($w=1/9$).
\pri{The PBH analysis relevant for such nonstandard scenarios with a general $w$ is discussed in Sec.~\ref{secPBH}. For the primordial power spectrum, we consider three functional forms of $P_{\zeta}(k)$ with (i) a simple Gaussian profile in Sec.~\ref{GPP}, (ii) a lognormal profile with an exponential cut-off in Sec.~\ref{PLN}, and (iii) a broken power law power spectrum in Sec.~\ref{PBPL}. For each of these cases, we quote the typical amplitudes required for each case and then proceed to the relevant implications for PBH.}

For the power spectrum in (i), we find that the analysis to attain a viable scenario that satisfies the \NG constraints on \OMG2 while also producing abundant PBH crucially depends on not only the amplitude $A_s$ of the primordial curvature power spectrum in Eq.~\eqref{PKG} but also on $\sigma_p$ and $k_*$ for a given $w$. These latter dependences mainly result from the relative shift of both \OMG2 and $\psi_w(M)$ peaks with respect to the peak of $P_{\zeta}(k)$ at $k_*$. For the RD case, both the \OMG2 peak and  $\psi_w(M)$ peak occur very close to $k_*$, therefore the dependence of the PBH abundance and \OMG2 on $\sigma_p$ is mild, as evident from the clustering of the three red points in Fig.~\ref{dotplot}. \pri{For RD, the pivot of the SGWB spectra coincides with the peak scale of the power spectrum i.e., $k_{\rm peak}=k_*$. Since the parameter $A_{\rm CP}$ corresponds to the amplitude of SGWB at the pivot frequency of the NANOGrav, for radiation domination, it only depends on the peak amplitude $A_s$ of the primordial scalar power spectrum. Therefore, the results in the $A_{\rm CP} - \gamma _{\rm CP}$ plane while varying the width $\sigma _p$ (red circle, square, and triangle) are negligibly affected.}

For $w\neq 1/3$, the peak of \OMG2 is far away from $k_*$ and therefore, if the scale of \NG observation is at $k_0=k_*$ then the peak of \OMG2 is not being observed. Hence, such cases have considerable variations in the $A_{\rm CP}$ - $\gamma _{\rm CP}$ space with change in $\sigma_p$. Now, when we put an additional constraint for obtaining $f_{\rm PBH}=0.1$, we notice the variation for points with different $\sigma_p$ for the two different nonstandard epochs under consideration as the green and blue circles, squares and triangles in Fig.~\ref{dotplot}. This variation shows that given a particular value of $k_*$ in the \NG observed range, there is a value of $\sigma_p\simeq 2$ where the nonstandard cases come closest to the \NG $2\sigma$ contour. This feature in the  $A_{\rm CP}$ - $\gamma _{\rm CP}$ space with varying $\sigma_p$ is a result of the competition between the two quantities $\frac{d A_{\rm CP}}{d\sigma_p}$ and $\frac{d f_{\rm PBH}}{d\sigma_p}$. It is analytically challenging to check the variation of these two terms since both of them involve rigorous numerical integrations.

Now, fixing $\sigma_p=2$ and varying $A_s$ as per Table~\ref{TT1} and $k_*$ in the \NG range, we show the contours (dashed lines in Fig.~\ref{regcont}) for abundant PBH production or less for the power spectrum (i). We conclude from Fig.~\ref{regcont} that the $w=1$ case cannot produce a copious amount of PBH and satisfy the \NG constraints simultaneously. We notice that for the $w=1/9$ case, the contour for $f_{\rm PBH}=0.1$ just enters the \NG $2\sigma$ contour in this range of $k_*$. Guided by our previous understanding of the relative shift in peak positions, we find that for $w=1/9$, choosing a different $k_*$ range such that \NG observes closer to the peak of the \OMG2 spectrum makes the nonstandard $w=1/9$ case consistent with the \NG $2\sigma$ contour. The gain in a $w=1/9$ case over a pure RD epoch is that both abundant PBH and \NG consistency are reached with a smaller $A_s\sim 0.007$ in the $w=1/9$ case as compared to $A_s\sim 0.01$ required in the RD case. 

The reason for the consistency of the $w=1/9$ case and the inconsistency of the $w=1$ case for the Gaussian power spectrum can also be explained in terms of the slope $\gamma _{\rm CP}$. In a general $w$-dominated epoch, the slope of the \OMG2 spectrum is $5-\gamma _{\rm CP}=\frac{6w-2}{3w+1}+2n_2$, where $n_2$ is the slope of $P_{\zeta}(k)$. For a Gaussian primordial power spectrum given in Eq.~\eqref{PKG}, $n_2=-\frac{1}{\sigma_p^2}\log_{10}(k/k_*)$. Taking the $1\sigma$ limit on $\gamma _{\rm CP}$ from the \NG $4.5\lesssim \gamma _{\rm CP}\lesssim 6.5$, we find no real limit for $\sigma_p $ for the $w=1$ case. However, for the $w=1/9$ case, the lower limit on $\gamma _{\rm CP}$ gives $\sigma_p \gtrsim 1.155$ with $k=k_0=3.6\times 10^6$ Mpc$^{-1}$ and $k_*\simeq 3.6\times 10^7$ Mpc$^{-1}$. Hence, the scenario where a nonstandard $w=1/9$ epoch is consistent with \NG result and abundant PBH production (green dashed contour in Fig.~\ref{regcont2}), $n_2 \vert_{k_0}\lesssim 0.5$. 
%\blue{Thus, the $w=1/9$ is consistent when its \OMG2 peak at $k_0$ is close to $k_*$, and rises with a small slope.}

\pri{The scalar power spectrum for the case (ii) is necessarily a tilted Gaussian in the $k$-space, and therefore, we expect our results for this case to be similar to the symmetric Gaussian power spectrum in case (i). Form Fig.~\ref{LNcont}, it is clear that $w=1$ case results in insignificant PBH abundance for the predicted \OMG2 to be consistent with the \NG data; the $w=1/3$ scenario is consistent with the \NG data within $2\sigma$ confidence limit for $f_{\rm PBH}\simeq 0.1$; the $w=1/9$ case is consistent with the \NG data for $f_{\rm PBH}> 0.1$ only for $k_* > 5\times 10^6$ Mpc$^{-1}$. Indeed, all of these results are similar to the results for the symmetric Gaussian case in Sec.~\ref{GPP}. The difference between the exact results for \OMG2 and PBH abundance between Sec.~\ref{GPP} and Sec.~\ref{PLN} is mainly due to the difference in their width. 
%The results for these two power spectra are expected to differ more for narrower profiles of $P_{\zeta}(k)$. 
For the $P_{\zeta}(k)$ considered in Sec.~\ref{PLN}, the advantage for the $w=1/9$ epoch over typical RD epoch is that the \NG consistency and $f_{\rm PBH}\gtrsim 0.1$ can be simultaneously satisfied for $A_s(w=1/9)=0.0078$ as compared to that required for the RD, i.e., $A_s(w=1/3)=0.011$.}
%\begin{figure}[!htbp]
%\centering{
%\includegraphics[width=0.75\textwidth]{slopes_LN_BPL.pdf}
%\caption{\small{Write caption and one paragraph in the text.}}
%\label{slopes}}
%\end{figure}

\pri{The third power spectrum considered in Sec.~\ref{PBPL} is a broken power law, for which the implications for PBH and second order GW are given in Fig.~\ref{BPLcont} in a consolidated manner. In this case, out of the three post-inflationary evolutions considered, abundant PBH formation with $f_{\rm PBH}=0.1$ is consistent with the \NG result only for RD. For the nonstandard cases, \OMG2 in the \NG observed range leads to a tiny value of $f_{\rm PBH}$ (for $w=1$) or it overshoots the theoretical limit $f_{\rm PBH}=1$ (for $w=1/9$). The difference in conclusion for the  $w=1/9$ case for this power spectrum as compared to the other two $P_{\zeta}(k)$ considered in Sec.~\ref{GPP} and Sec.~\ref{PLN} can be attributed to the sharp drop in power for the green curve in Fig.~\ref{OmgA2_BPL} from the peak to $k=k_*$.}

%\blue{The results for the second and third power spectra can also be discussed with the help of slopes of  $P_{\zeta}(k)$. }
%However, this $P_{\zeta}(k)$ featuring power laws provides more straightforward understanding of the slopes of \OMG2 for different $w$ values. 

\begin{figure}[!htbp]
\centering{
\includegraphics[width=0.75\textwidth]{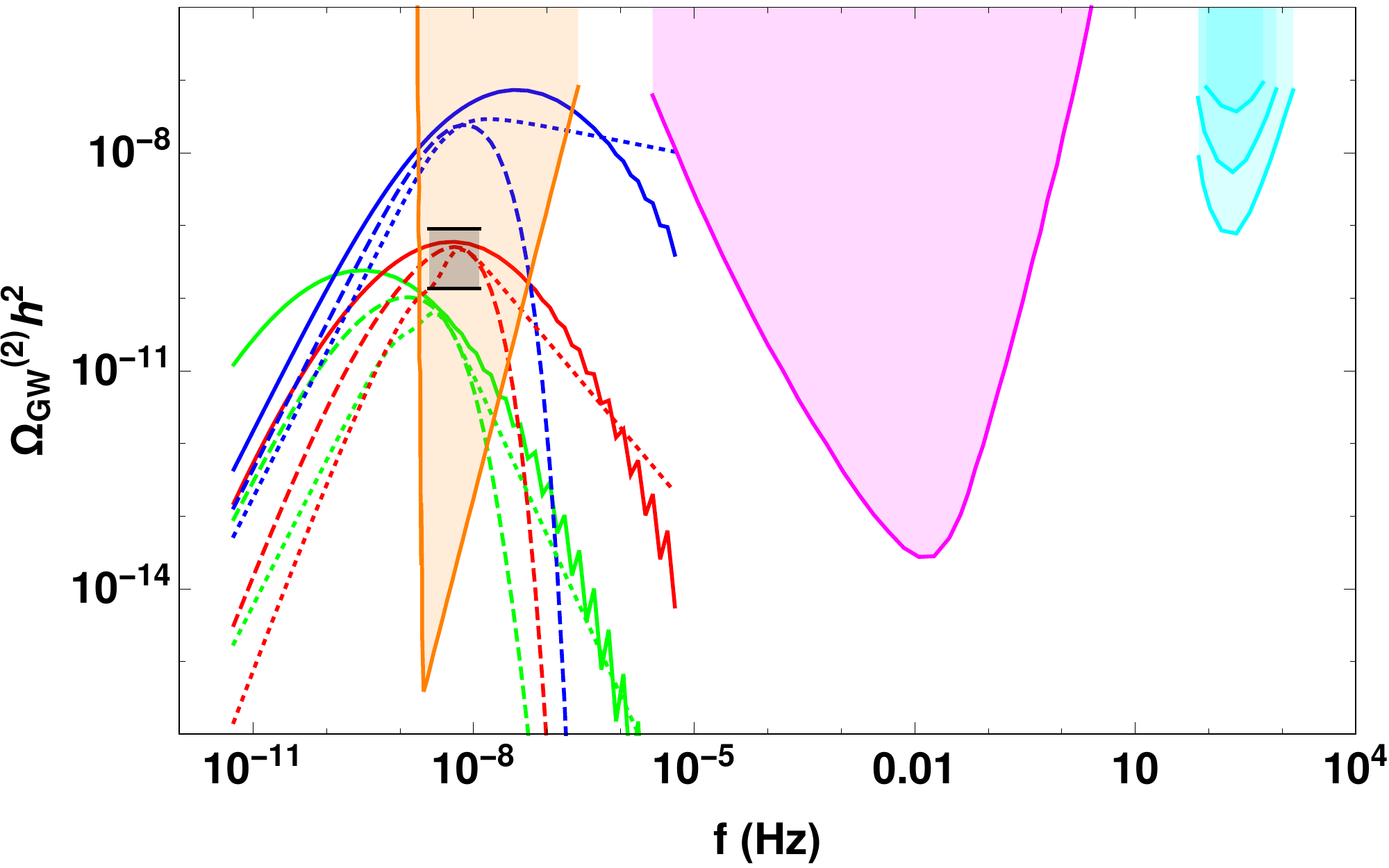}
\caption{\small{The second order GW spectra for the three values of $w$ for the Gaussian power spectrum in Eq.~\ref{PKG} (solid lines), lognormal power spectrum with an exponential cutoff in Eq.~\ref{PKLN} (dashed lines), and broken power law power spectrum in Eq.~\ref{PKBPL} (dotted lines) are shown. The red, blue and green curves represent $w=1/3$, $1$, and $1/9$ respectively. LIGO runs for O1, O2, and O5 (future) are given as cyan curves (top to bottom); magenta shade represnents the proposed sensitivity for the LISA survey; black shade region represent $1\sigma$ bound from the \NG 12.5 year observation; orange shade region represents the proposed sensitivity from the future SKA survey.}}
\label{fullomg2obs}}
\end{figure}
\pri{Similar to Table~\ref{TT2}, PBH mass ranges with different choices of $k_*$ for the second and third form of $P_{\zeta}(k)$ and for all three $w$ under consideration are presented in Table~\ref{TT4}.}
\begin{table}[!htbp]
\caption{Ranges of PBH mass $M$ produced in abundance for each EoS}
\begin{center}
\label{TT4}
\resizebox{0.85\textwidth}{!}{
\begin{tabular}{|c|c|c|c|c|}
\hline 
$P_{\zeta}(k)$ & $k_*$ in Mpc$^{-1}$ & $M/M_{\odot} (w=1/3)$ & $M/M_{\odot} (w=1)$ & $M/M_{\odot} (w=1/9)$ \\
\hline
\hline 
%Range of $k_*$ in Mpc$^{-1}$ & $2\times 10^6$ - $7 \times 10^6$ & $2\times 10^6$ - $7 \times 10^6$  & $2\times 10^7$ - $7 \times 10^7$\\
%\hline 
%Gaussian & $10^6$ & 0.2-2 & 1-10 & $10^{-4}$ - $5\times 10^{-3}$\\
%\hline
%& $10^7$  & 0.01 - 0.33 & 0.08-2 & $3\times 10^{-6}$ - $3\times 10^{-4}$\\
%\hline
Lognormal with exponential cutoff & $10^6$  & $0.8-40$ & $2-200$ &  $0.08 - 100$\\
\hline
& $10^7$  & $7\times 10^{-3} - 0.5$ & $0.07 - 8$ & $5\times 10^{-4} - 0.5$\\
\hline
Broken power law & $10^6$  & $0.2 - 10$ & $1 - 30$ & $0.02 - 8$\\
\hline
& $10^7$  & $2\times 10^{-3} - 0.2$ & $0.025 - 1$  &  $10^{-4} - 0.03$\\
\hline
%$6\times 10^{12} $ & $1/3$ & 0.013 & 0.016 & 0.0163\\
%\hline
%$6 \times 10^{12} $ & $1$ & 0.0048 & 0.0067 & 0.006\\
%\hline
\end{tabular}}
\end{center}
\end{table} 

The discussions in terms of the slopes of the second and first order GW spectrum in this section emphasise its significance in constraining pre-BBN cosmologies and primordial power spectra when a (probable) GW detection is available. We have checked that our results satisfy BBN bound on \OMG2 given in Eq.~\eqref{eq:bbncons}. The range of possible PBH masses is different for different $w$ and $k_*$ values and different $P_{\zeta}(k)$ (see Tables~\ref{TT2} and~\ref{TT4}). In case of the Gaussian power spectrum, for the nonstandard case $w=1/9$ that is finally consistent in our analysis can generate PBH in the mass range $5\times 10^{-6}M_{\odot}$ to $3\times 10^{-3}M_{\odot}$ depending on $k_*$. The observational constraints from lensing experiments such as Subaru HSC~\cite{Niikura:2017zjd}, EROS/MACHO~\cite{Tisserand:2006zx} and OGLE~\cite{Niikura:2019kqi} put upper bound on $f_{\rm PBH}$ to be 0.1 in a few points across this mass range. However, the $f_{\rm PBH}\simeq 10^{-3}$ contour will still be inside \NG $2\sigma$ in this case, which is observationally valid throughout this mass range. For the RD epoch, near solar mass PBHs can be generated with $f_{\rm PBH}=0.1$, which is the maximum observationally allowed abundance in this range, constrained mainly with EROS/MACHO~\cite{Tisserand:2006zx}.

\pri{For the lognormal power spectrum in Eq.~\ref{PKLN}, the minimum and maximum masses of PBH that can be produced for the standard RD case are $7\times 10^{-3}M_{\odot}$ and $40M_{\odot}$ respectively, which can be compared with the observational bounds discussed above in a similar manner. For this $P_{\zeta}(k)$ with $w=1/9$, both abundant PBH formation and \NG consistency are satisfied only for $k_*=10^7$ Mpc$^-1$ in Table~\ref{TT4}, where PBH masses are produced in the range $5\times 10^{-4} - 0.5 M_{\odot}$. Observational constraints on PBH abundance from EROS, OGLE, Icarus and Microlensing constaints from supernova are effective in this mass range and maximum possible abundance is $f_{\rm PBH}< 0.1$. For the broken power law power spectrum in Eq.~\ref{PKBPL}, the only viable case is RD, which can lead to abundant PBH production in the mass range $2\times 10^{-3}- 10 M_{\odot}$ and is constrained by the experiments discussed above constraining $f_{\rm PBH}\simeq 0.1$. Therefore, with the choice of parameters and $k_*$, for all of the $P_{\zeta}(k)$ under consideration, $f_{\rm PBH}= 0.1$ is a good benchmark to explore, which is shown as dashed contours in Fig.s~\ref{regcont}, ~\ref{regcont2}, ~\ref{LNcont} and~\ref{BPLcont}. As pointed out in earlier sections, the mass ranges for all theree $P_{\zeta}(k)$ discussed here only envelop a fraction of the mass range where LIGO/Virgo survey has found binary merger events. For the cases where \NG observation is consistent with abundant PBH formation (see Table~\ref{TT3}) can only lead to very small PBH abundance in the LIGO/Virgo mass range. Therefore, in this set up, it is unlikely that these the BH mergers observed in LIGO/Virgo surveys are of primordial nature~\cite{Bird:2016dcv,Clesse:2016vqa,Sasaki:2016jop}.}
%The exact number can be calculated as $f_{\rm LIGO} = \int _{0.1M_{\odot}}^{300M_{\odot}} \psi (M) dM$, which can lead to a maximum value of }

\pri{The full \OMG2$h^2$ for the three forms of $P_{\zeta}(k)$ and three values of $w$ are plotted in Fig.~\ref{fullomg2obs} with a few present and future observational bounds. In this plot, $A_s =0.007$ for all the cases, $\sigma _p=2$ for the Gaussian power spectrum (solid lines); $\alpha=0.17$ and $\beta=0.62$ for the lognormal power spectrum with an exponential cutoff (dashed lines); and $\alpha=4$ and $\beta=0.6$ for the broken power law power spectrum in (dotted lines). The red, blue and green curves represent $w=1/3$, $1$, and $1/9$ respectively. Evidently, future Square Kilometre Array (SKA)~\cite{Zhao:2013bba} data will be crucial in constraining the \OMG2 spectrum in this frequency range. The conclusions for different $w$ and different $P_{\zeta}(k)$ for $k_*=3.6\times 10^6$ Mpc$^{-1}$ are summarised in Table~\ref{TT3}. Evidently, modifying the values of the parameters $\alpha$ and $\beta$ for power spectra in Sec.~\ref{PLN} and~\ref{PBPL} will modify the conclusions of this analysis. A straightforward way to check the dependence of this analysis on these parameters is to perform a similar analysis as done in Fig.~\ref{regcont2} of Sec.~\ref{GPP}, which we hope to explore in future studies.}

\begin{table}[!htbp]
\caption{Summary: consistency of PBH formation with $f_{\rm PBH}=0.1$ while satisfying \NG 12.5 year data as second order GW spectra for different $P_{\zeta}(k)$. }
%for $k_*=3.6\times 10^6$ Mpc$^{-1}$
\begin{center}
\label{TT3}
\resizebox{0.95\textwidth}{!}{
\begin{tabular}{|c|c|c|c|c|}
\hline 
$P_{\zeta}(k)$ & $k_*$ (Mpc$^{-1}$) & $w=1/3$ & $w=1$ & $w=1/9$ \\
\hline
\hline 
Gaussian  & $10^6$ & consistent at $2\sigma$ with $A_s\simeq 0.01$ & inconsistent & inconsistent ($f_{\rm PBH}>1$)\\
\hline 
 & $10^7$ & consistent at $2\sigma$ with $A_s\simeq 0.01$ & inconsistent & consistent with $A_s\simeq 0.007$\\
\hline
Lognormal with exponential cutoff & $10^6$ & inconsistent at $2\sigma$ with $A_s\simeq 0.011$ & inconsistent & inconsistent (large $\gamma _{\rm CP}$) \\
\hline 
  & $10^7$ & consistent at $2\sigma$ with $A_s\simeq 0.011$ & inconsistent & consistent with $A_s\simeq 0.0078$ \\
\hline
Broken power law & $10^6$ & consistent at $1\sigma$ with $A_s\simeq 0.012$ & inconsistent & inconsistent ($f_{\rm PBH}>1$) \\
\hline
 & $10^7$ & consistent at $2\sigma$ with $A_s\simeq 0.0122$ & inconsistent & inconsistent ($f_{\rm PBH}>1$)\\
\hline
%$6\times 10^{12} $ & $1/3$ & 0.013 & 0.016 & 0.0163\\
%\hline
%$6 \times 10^{12} $ & $1$ & 0.0048 & 0.0067 & 0.006\\
%\hline
\end{tabular}}
\end{center}
\end{table}

After the announcement of the \NG 12.5 yr data release, many works have studied the consistency of abundant PBH formation with the \NG observation in terms of \OMG2 . References~~\cite{Vaskonen:2020lbd,DeLuca:2020agl,Kohri:2020qqd} have studied this in RD epochs. Ref.~\cite{DeLuca:2020agl} considered a wide constant primordial power spectrum and found high abundance in the PBH mass range $10^{18}-10^{21}$ gm, while still satisfying the \NG constraints. Ref.~\cite{Vaskonen:2020lbd} studied different primordial curvature power spectra. The recent work~\cite{Domenech:2020ers} discusses implications of the \NG result on the formation of planetary mass PBHs in nonstandard dust-like epochs with $-0.091<w<0.048$. They use narrow peak for their analysis, specifically a Dirac delta function, and translate the constraints on \OMG2 to that of the amplitude which generates a monochromatic mass function for PBH, whereas our $P_{\zeta}(k)$ with $\sigma_p=2$ can be noted as a broad spectrum.

\pri{Ref.~\cite{Kohri:2020qqd} motivates a narrow Gaussian curvature power spectrum using the Press-Schechter formalism to generate abundant near solar mass PBHs to be consistent with the \NG results in RD. In our work with the Gaussian power spectrum in Sec.~\ref{GPP}, we also consider Press-Schechter formalism for PBH analysis and found out $\sigma_p=1$ case to be consistent with~\cite{Kohri:2020qqd} for RD.
The lognormal power spectrum with an exponential cut off in Sec.~\ref{PLN} and the broken power law power spectrum in Sec.~\ref{PBPL} has also been studied in~\cite{Vaskonen:2020lbd} for RD epoch, however, they had discarded these scenarios because its \NG consistent amplitude does not work to generate abundant PBH. However, in our case, abundant PBH production in RD for both of these power spectra are consistent with the \NG data and therefore disagrees with the results presented in~\cite{Vaskonen:2020lbd}. The reasons for this disagreement are similar to those pointed out in the discussion section of~~\cite{Kohri:2020qqd}. Whereas ref.~\cite{Vaskonen:2020lbd} consider critical collapse formalism to evaluate PBH abundance, we note that it is only valid for exact spherical collapse, and we do not consider this formalism for PBH analysis in this paper. Other reasons for the disagreement include a choice of wider window functions for PBH analysis in our case; inclusion of the transfer function in the definition of $\sigma ^2 _\delta$, and inclusion of the nonlinear relation between density perturbation and curvature perturbation in ref.~\cite{Vaskonen:2020lbd}.}
%, including a tilted Gaussian one, which they discarded because its \NG consistent amplitude doesn not work to generate abundant PBH. However, this analysis was done using critical collapse for PBH. 

For a broad primordial power spectrum in general, the resulting \OMG2 spectrum will also be broad. Therefore, the future GW detectors close to the \NG frequency range will further constrain the width through (non-)observation of the stochastic GW background. These future constraints combined with the \NG limits will motivate a thorough understanding of the primordial (inflationary) and pre-BBN cosmologies and constrain the PBH abundances more stringently. 

For our case the mass function is extended, including both solar mass and planetary mass PBHs. The PBH analysis also depends on the formalism under consideration so that abundances from peak theory analysis or critical collapse analysis may differ considerably from the Press-Schechter analysis used here. We have used a broad window function for the Gaussian power spectrum, which substantially helps in enhancing PBH. We note here that for a broad power spectrum, the integration kernels have to be evaluated on a case by case basis for different EoS and therefore, it is challenging to show constraints as a functions of $w$ directly. In a more explicit scenario, it is possible to start from a well-motivated theoretical framework where dust-like nonstandard epochs can arise and deriving a thorough evolution for $\Omega_{\rm GW}^{(1)}$ and \OMG2 in this nonstandard epoch and RD, it is possible to constrain parameters of the theoretical interest using the \NG result and upcoming observations of stochastic GW background. Also, the \OMG2 for pure matter dominated case $w=0$ should be calculated separately for a broad primordial power spectrum since one needs to take care of the scales where non-linear growths become important. For $w=0$, the PBH formation mechanism is also quite different from $w>0$ cases due to the absence of pressure and often results in PBHs with nonzero-spins and aspherical shapes. We hope to come back to these issues in the near future. 

\section*{Acknowledgement}
The authors sincerely thank the referees for their insightful comments and suggestions that helped in improving the structure and quality of this paper. S.B. is supported by the institute postdoctoral fellowship from Physical Research Laboratory, India.
%\section{Induced GW in nonstandard thermal history}
%\section{Implications on PBH abundance}

%\section{Conclusions and discussions}

\bibliographystyle{JHEP}
\bibliography{nanograv_gw1.bib}

\providecommand{\href}[2]{#2}\begingroup\raggedright\begin{thebibliography}{100}

\bibitem{Arzoumanian:2018saf}
{\scshape NANOGRAV} collaboration, Z.~Arzoumanian et~al., \emph{{The NANOGrav
  11-year Data Set: Pulsar-timing Constraints On The Stochastic
  Gravitational-wave Background}},
  \href{https://doi.org/10.3847/1538-4357/aabd3b}{\emph{Astrophys. J.}
  {\bfseries 859} (2018) 47},
  [\href{https://arxiv.org/abs/1801.02617}{{\ttfamily 1801.02617}}].

\bibitem{Arzoumanian:2020vkk}
{\scshape NANOGrav} collaboration, Z.~Arzoumanian et~al., \emph{{The NANOGrav
  12.5-year Data Set: Search For An Isotropic Stochastic Gravitational-Wave
  Background}},  \href{https://arxiv.org/abs/2009.04496}{{\ttfamily
  2009.04496}}.

\bibitem{Sesana:2004sp}
A.~Sesana, F.~Haardt, P.~Madau and M.~Volonteri, \emph{{Low - frequency
  gravitational radiation from coalescing massive black hole binaries in
  hierarchical cosmologies}},
  \href{https://doi.org/10.1086/422185}{\emph{Astrophys. J.} {\bfseries 611}
  (2004) 623--632}, [\href{https://arxiv.org/abs/astro-ph/0401543}{{\ttfamily
  astro-ph/0401543}}].

\bibitem{Ellis:2020ena}
J.~Ellis and M.~Lewicki, \emph{{Cosmic String Interpretation of NANOGrav Pulsar
  Timing Data}},  \href{https://arxiv.org/abs/2009.06555}{{\ttfamily
  2009.06555}}.

\bibitem{Blasi:2020mfx}
S.~Blasi, V.~Brdar and K.~Schmitz, \emph{{Has NANOGrav found first evidence for
  cosmic strings?}},  \href{https://arxiv.org/abs/2009.06607}{{\ttfamily
  2009.06607}}.

\bibitem{Buchmuller:2020lbh}
W.~Buchmuller, V.~Domcke and K.~Schmitz, \emph{{From NANOGrav to LIGO with
  metastable cosmic strings}},
  \href{https://arxiv.org/abs/2009.10649}{{\ttfamily 2009.10649}}.

\bibitem{Samanta:2020cdk}
R.~Samanta and S.~Datta, \emph{{Gravitational wave complementarity and impact
  of NANOGrav data on gravitational leptogenesis: cosmic strings}},
  \href{https://arxiv.org/abs/2009.13452}{{\ttfamily 2009.13452}}.

\bibitem{Nakai:2020oit}
Y.~Nakai, M.~Suzuki, F.~Takahashi and M.~Yamada, \emph{{Gravitational Waves and
  Dark Radiation from Dark Phase Transition: Connecting NANOGrav Pulsar Timing
  Data and Hubble Tension}},
  \href{https://arxiv.org/abs/2009.09754}{{\ttfamily 2009.09754}}.

\bibitem{Addazi:2020zcj}
A.~Addazi, Y.-F. Cai, Q.~Gan, A.~Marciano and K.~Zeng, \emph{{NANOGrav results
  and Dark First Order Phase Transitions}},
  \href{https://arxiv.org/abs/2009.10327}{{\ttfamily 2009.10327}}.

\bibitem{Ratzinger:2020koh}
W.~Ratzinger and P.~Schwaller, \emph{{Whispers from the dark side: Confronting
  light new physics with NANOGrav data}},
  \href{https://arxiv.org/abs/2009.11875}{{\ttfamily 2009.11875}}.

\bibitem{Li:2020cjj}
H.-H. Li, G.~Ye and Y.-S. Piao, \emph{{Is the NANOGrav signal a hint of dS
  decay during inflation?}},
  \href{https://arxiv.org/abs/2009.14663}{{\ttfamily 2009.14663}}.

\bibitem{Vagnozzi:2020gtf}
S.~Vagnozzi, \emph{{Implications of the NANOGrav pulsar timing results for
  inflation}},  \href{https://arxiv.org/abs/2009.13432}{{\ttfamily
  2009.13432}}.

\bibitem{Vaskonen:2020lbd}
V.~Vaskonen and H.~Veermäe, \emph{{Did NANOGrav see a signal from primordial
  black hole formation?}},  \href{https://arxiv.org/abs/2009.07832}{{\ttfamily
  2009.07832}}.

\bibitem{DeLuca:2020agl}
V.~De~Luca, G.~Franciolini and A.~Riotto, \emph{{NANOGrav Hints to Primordial
  Black Holes as Dark Matter}},
  \href{https://arxiv.org/abs/2009.08268}{{\ttfamily 2009.08268}}.

\bibitem{Kohri:2020qqd}
K.~Kohri and T.~Terada, \emph{{Possible Solar-Mass Primordial Black Holes for
  NANOGrav Hint of Gravitational Waves}},
  \href{https://arxiv.org/abs/2009.11853}{{\ttfamily 2009.11853}}.

\bibitem{Domenech:2020ers}
G.~Dom\`enech and S.~Pi, \emph{{NANOGrav Hints on Planet-Mass Primordial Black
  Holes}},  \href{https://arxiv.org/abs/2010.03976}{{\ttfamily 2010.03976}}.

\bibitem{Pandey:2020gjy}
A.~K. Pandey, \emph{{Gravitational waves in neutrino plasma and NANOGrav
  signal}},  \href{https://arxiv.org/abs/2011.05821}{{\ttfamily 2011.05821}}.

\bibitem{Abbott:2016blz}
{\scshape LIGO Scientific, Virgo} collaboration, B.~Abbott et~al.,
  \emph{{Observation of Gravitational Waves from a Binary Black Hole Merger}},
  \href{https://doi.org/10.1103/PhysRevLett.116.061102}{\emph{Phys. Rev. Lett.}
  {\bfseries 116} (2016) 061102},
  [\href{https://arxiv.org/abs/1602.03837}{{\ttfamily 1602.03837}}].

\bibitem{Abbott:2016nmj}
{\scshape LIGO Scientific, Virgo} collaboration, B.~P. Abbott et~al.,
  \emph{{GW151226: Observation of Gravitational Waves from a 22-Solar-Mass
  Binary Black Hole Coalescence}},
  \href{https://doi.org/10.1103/PhysRevLett.116.241103}{\emph{Phys. Rev. Lett.}
  {\bfseries 116} (2016) 241103},
  [\href{https://arxiv.org/abs/1606.04855}{{\ttfamily 1606.04855}}].

\bibitem{TheLIGOScientific:2016pea}
{\scshape LIGO Scientific, Virgo} collaboration, B.~Abbott et~al.,
  \emph{{Binary Black Hole Mergers in the first Advanced LIGO Observing Run}},
  \href{https://doi.org/10.1103/PhysRevX.6.041015}{\emph{Phys. Rev. X}
  {\bfseries 6} (2016) 041015},
  [\href{https://arxiv.org/abs/1606.04856}{{\ttfamily 1606.04856}}].

\bibitem{Abbott:2016bqf}
{\scshape LIGO Scientific, Virgo} collaboration, B.~P. Abbott et~al.,
  \emph{{The basic physics of the binary black hole merger GW150914}},
  \href{https://doi.org/10.1002/andp.201600209}{\emph{Annalen Phys.} {\bfseries
  529} (2017) 1600209}, [\href{https://arxiv.org/abs/1608.01940}{{\ttfamily
  1608.01940}}].

\bibitem{Abbott:2017vtc}
{\scshape LIGO Scientific, VIRGO} collaboration, B.~P. Abbott et~al.,
  \emph{{GW170104: Observation of a 50-Solar-Mass Binary Black Hole Coalescence
  at Redshift 0.2}},
  \href{https://doi.org/10.1103/PhysRevLett.118.221101}{\emph{Phys. Rev. Lett.}
  {\bfseries 118} (2017) 221101},
  [\href{https://arxiv.org/abs/1706.01812}{{\ttfamily 1706.01812}}].

\bibitem{Abbott:2017oio}
{\scshape LIGO Scientific, Virgo} collaboration, B.~Abbott et~al.,
  \emph{{GW170814: A Three-Detector Observation of Gravitational Waves from a
  Binary Black Hole Coalescence}},
  \href{https://doi.org/10.1103/PhysRevLett.119.141101}{\emph{Phys. Rev. Lett.}
  {\bfseries 119} (2017) 141101},
  [\href{https://arxiv.org/abs/1709.09660}{{\ttfamily 1709.09660}}].

\bibitem{Abbott:2017gyy}
{\scshape LIGO Scientific, Virgo} collaboration, B.~P. Abbott et~al.,
  \emph{{GW170608: Observation of a 19-solar-mass Binary Black Hole
  Coalescence}},
  \href{https://doi.org/10.3847/2041-8213/aa9f0c}{\emph{Astrophys. J.}
  {\bfseries 851} (2017) L35},
  [\href{https://arxiv.org/abs/1711.05578}{{\ttfamily 1711.05578}}].

\bibitem{Fernandez:2019kyb}
N.~Fernandez and S.~Profumo, \emph{{Unraveling the origin of black holes from
  effective spin measurements with LIGO-Virgo}},
  \href{https://doi.org/10.1088/1475-7516/2019/08/022}{\emph{JCAP} {\bfseries
  08} (2019) 022}, [\href{https://arxiv.org/abs/1905.13019}{{\ttfamily
  1905.13019}}].

\bibitem{Caprini:2018mtu}
C.~Caprini and D.~G. Figueroa, \emph{{Cosmological Backgrounds of Gravitational
  Waves}}, \href{https://doi.org/10.1088/1361-6382/aac608}{\emph{Class. Quant.
  Grav.} {\bfseries 35} (2018) 163001},
  [\href{https://arxiv.org/abs/1801.04268}{{\ttfamily 1801.04268}}].

\bibitem{Christensen:2018iqi}
N.~Christensen, \emph{{Stochastic Gravitational Wave Backgrounds}},
  \href{https://doi.org/10.1088/1361-6633/aae6b5}{\emph{Rept. Prog. Phys.}
  {\bfseries 82} (2019) 016903},
  [\href{https://arxiv.org/abs/1811.08797}{{\ttfamily 1811.08797}}].

\bibitem{Ade:2018gkx}
{\scshape BICEP2, Keck Array} collaboration, P.~A.~R. Ade et~al., \emph{{BICEP2
  / Keck Array x: Constraints on Primordial Gravitational Waves using Planck,
  WMAP, and New BICEP2/Keck Observations through the 2015 Season}},
  \href{https://doi.org/10.1103/PhysRevLett.121.221301}{\emph{Phys. Rev. Lett.}
  {\bfseries 121} (2018) 221301},
  [\href{https://arxiv.org/abs/1810.05216}{{\ttfamily 1810.05216}}].

\bibitem{Aghanim:2019ame}
{\scshape Planck} collaboration, N.~Aghanim et~al., \emph{{Planck 2018 results.
  V. CMB power spectra and likelihoods}},
  \href{https://doi.org/10.1051/0004-6361/201936386}{\emph{Astron. Astrophys.}
  {\bfseries 641} (2020) A5},
  [\href{https://arxiv.org/abs/1907.12875}{{\ttfamily 1907.12875}}].

\bibitem{Cook:2011hg}
J.~L. Cook and L.~Sorbo, \emph{{Particle production during inflation and
  gravitational waves detectable by ground-based interferometers}},
  \href{https://doi.org/10.1103/PhysRevD.85.023534}{\emph{Phys. Rev. D}
  {\bfseries 85} (2012) 023534},
  [\href{https://arxiv.org/abs/1109.0022}{{\ttfamily 1109.0022}}].

\bibitem{Barnaby:2011qe}
N.~Barnaby, E.~Pajer and M.~Peloso, \emph{{Gauge Field Production in Axion
  Inflation: Consequences for Monodromy, non-Gaussianity in the CMB, and
  Gravitational Waves at Interferometers}},
  \href{https://doi.org/10.1103/PhysRevD.85.023525}{\emph{Phys. Rev. D}
  {\bfseries 85} (2012) 023525},
  [\href{https://arxiv.org/abs/1110.3327}{{\ttfamily 1110.3327}}].

\bibitem{Wang:2014kqa}
Y.~Wang and W.~Xue, \emph{{Inflation and Alternatives with Blue Tensor
  Spectra}}, \href{https://doi.org/10.1088/1475-7516/2014/10/075}{\emph{JCAP}
  {\bfseries 10} (2014) 075},
  [\href{https://arxiv.org/abs/1403.5817}{{\ttfamily 1403.5817}}].

\bibitem{Kuroyanagi:2014nba}
S.~Kuroyanagi, T.~Takahashi and S.~Yokoyama, \emph{{Blue-tilted Tensor Spectrum
  and Thermal History of the Universe}},
  \href{https://doi.org/10.1088/1475-7516/2015/02/003}{\emph{JCAP} {\bfseries
  1502} (2015) 003}, [\href{https://arxiv.org/abs/1407.4785}{{\ttfamily
  1407.4785}}].

\bibitem{Ananda:2006af}
K.~N. Ananda, C.~Clarkson and D.~Wands, \emph{{The Cosmological gravitational
  wave background from primordial density perturbations}},
  \href{https://doi.org/10.1103/PhysRevD.75.123518}{\emph{Phys. Rev. D}
  {\bfseries 75} (2007) 123518},
  [\href{https://arxiv.org/abs/gr-qc/0612013}{{\ttfamily gr-qc/0612013}}].

\bibitem{Baumann:2007zm}
D.~Baumann, P.~J. Steinhardt, K.~Takahashi and K.~Ichiki, \emph{{Gravitational
  Wave Spectrum Induced by Primordial Scalar Perturbations}},
  \href{https://doi.org/10.1103/PhysRevD.76.084019}{\emph{Phys. Rev. D}
  {\bfseries 76} (2007) 084019},
  [\href{https://arxiv.org/abs/hep-th/0703290}{{\ttfamily hep-th/0703290}}].

\bibitem{Espinosa:2018eve}
J.~R. Espinosa, D.~Racco and A.~Riotto, \emph{{A Cosmological Signature of the
  SM Higgs Instability: Gravitational Waves}},
  \href{https://doi.org/10.1088/1475-7516/2018/09/012}{\emph{JCAP} {\bfseries
  09} (2018) 012}, [\href{https://arxiv.org/abs/1804.07732}{{\ttfamily
  1804.07732}}].

\bibitem{Kohri:2018awv}
K.~Kohri and T.~Terada, \emph{{Semianalytic calculation of gravitational wave
  spectrum nonlinearly induced from primordial curvature perturbations}},
  \href{https://doi.org/10.1103/PhysRevD.97.123532}{\emph{Phys. Rev.}
  {\bfseries D97} (2018) 123532},
  [\href{https://arxiv.org/abs/1804.08577}{{\ttfamily 1804.08577}}].

\bibitem{Clesse:2015wea}
S.~Clesse and J.~Garc\'\i{}a-Bellido, \emph{{Massive Primordial Black Holes
  from Hybrid Inflation as Dark Matter and the seeds of Galaxies}},
  \href{https://doi.org/10.1103/PhysRevD.92.023524}{\emph{Phys. Rev. D}
  {\bfseries 92} (2015) 023524},
  [\href{https://arxiv.org/abs/1501.07565}{{\ttfamily 1501.07565}}].

\bibitem{Garcia-Bellido:2017mdw}
J.~Garcia-Bellido and E.~Ruiz~Morales, \emph{{Primordial black holes from
  single field models of inflation}},
  \href{https://doi.org/10.1016/j.dark.2017.09.007}{\emph{Phys. Dark Univ.}
  {\bfseries 18} (2017) 47--54},
  [\href{https://arxiv.org/abs/1702.03901}{{\ttfamily 1702.03901}}].

\bibitem{Germani:2017bcs}
C.~Germani and T.~Prokopec, \emph{{On primordial black holes from an inflection
  point}}, \href{https://doi.org/10.1016/j.dark.2017.09.001}{\emph{Phys. Dark
  Univ.} {\bfseries 18} (2017) 6--10},
  [\href{https://arxiv.org/abs/1706.04226}{{\ttfamily 1706.04226}}].

\bibitem{Ballesteros:2017fsr}
G.~Ballesteros and M.~Taoso, \emph{{Primordial black hole dark matter from
  single field inflation}},
  \href{https://doi.org/10.1103/PhysRevD.97.023501}{\emph{Phys. Rev. D}
  {\bfseries 97} (2018) 023501},
  [\href{https://arxiv.org/abs/1709.05565}{{\ttfamily 1709.05565}}].

\bibitem{Bhaumik:2019tvl}
N.~Bhaumik and R.~K. Jain, \emph{{Primordial black holes dark matter from
  inflection point models of inflation and the effects of reheating}},
  \href{https://doi.org/10.1088/1475-7516/2020/01/037}{\emph{JCAP} {\bfseries
  01} (2020) 037}, [\href{https://arxiv.org/abs/1907.04125}{{\ttfamily
  1907.04125}}].

\bibitem{Mishra:2019pzq}
S.~S. Mishra and V.~Sahni, \emph{{Primordial Black Holes from a tiny bump/dip
  in the Inflaton potential}},
  \href{https://doi.org/10.1088/1475-7516/2020/04/007}{\emph{JCAP} {\bfseries
  04} (2020) 007}, [\href{https://arxiv.org/abs/1911.00057}{{\ttfamily
  1911.00057}}].

\bibitem{Ballesteros:2020qam}
G.~Ballesteros, J.~Rey, M.~Taoso and A.~Urbano, \emph{{Primordial black holes
  as dark matter and gravitational waves from single-field polynomial
  inflation}}, \href{https://doi.org/10.1088/1475-7516/2020/07/025}{\emph{JCAP}
  {\bfseries 07} (2020) 025},
  [\href{https://arxiv.org/abs/2001.08220}{{\ttfamily 2001.08220}}].

\bibitem{Palma:2020ejf}
G.~A. Palma, S.~Sypsas and C.~Zenteno, \emph{{Seeding primordial black holes in
  multifield inflation}},
  \href{https://doi.org/10.1103/PhysRevLett.125.121301}{\emph{Phys. Rev. Lett.}
  {\bfseries 125} (2020) 121301},
  [\href{https://arxiv.org/abs/2004.06106}{{\ttfamily 2004.06106}}].

\bibitem{Fumagalli:2020adf}
J.~Fumagalli, S.~Renaux-Petel, J.~W. Ronayne and L.~T. Witkowski,
  \emph{{Turning in the landscape: a new mechanism for generating Primordial
  Black Holes}},  \href{https://arxiv.org/abs/2004.08369}{{\ttfamily
  2004.08369}}.

\bibitem{Braglia:2020eai}
M.~Braglia, D.~K. Hazra, F.~Finelli, G.~F. Smoot, L.~Sriramkumar and A.~A.
  Starobinsky, \emph{{Generating PBHs and small-scale GWs in two-field models
  of inflation}},
  \href{https://doi.org/10.1088/1475-7516/2020/08/001}{\emph{JCAP} {\bfseries
  08} (2020) 001}, [\href{https://arxiv.org/abs/2005.02895}{{\ttfamily
  2005.02895}}].

\bibitem{Ozsoy:2020kat}
O.~\"Ozsoy and Z.~Lalak, \emph{{Primordial black holes as dark matter and
  gravitational waves from bumpy axion inflation}},
  \href{https://doi.org/10.1088/1475-7516/2021/01/040}{\emph{JCAP} {\bfseries
  01} (2021) 040}, [\href{https://arxiv.org/abs/2008.07549}{{\ttfamily
  2008.07549}}].

\bibitem{Aldabergenov:2020yok}
Y.~Aldabergenov, A.~Addazi and S.~V. Ketov, \emph{{Testing Primordial Black
  Holes as Dark Matter in Supergravity from Gravitational Waves}},
  \href{https://doi.org/10.1016/j.physletb.2021.136069}{\emph{Phys. Lett. B}
  {\bfseries 814} (2021) 136069},
  [\href{https://arxiv.org/abs/2008.10476}{{\ttfamily 2008.10476}}].

\bibitem{Braglia:2020taf}
M.~Braglia, X.~Chen and D.~K. Hazra, \emph{{Probing Primordial Features with
  the Stochastic Gravitational Wave Background}},
  \href{https://arxiv.org/abs/2012.05821}{{\ttfamily 2012.05821}}.

\bibitem{Hawking:1971ei}
S.~Hawking, \emph{{Gravitationally collapsed objects of very low mass}},
  {\emph{Mon. Not. Roy. Astron. Soc.} {\bfseries 152} (1971) 75}.

\bibitem{Carr:1974nx}
B.~J. Carr and S.~Hawking, \emph{{Black holes in the early Universe}},
  {\emph{Mon. Not. Roy. Astron. Soc.} {\bfseries 168} (1974) 399--415}.

\bibitem{Carr:1975qj}
B.~J. Carr, \emph{{The Primordial black hole mass spectrum}},
  \href{https://doi.org/10.1086/153853}{\emph{Astrophys. J.} {\bfseries 201}
  (1975) 1--19}.

\bibitem{Akrami:2018odb}
{\scshape Planck} collaboration, Y.~Akrami et~al., \emph{{Planck 2018 results.
  X. Constraints on inflation}},
  \href{https://doi.org/10.1051/0004-6361/201833887}{\emph{Astron. Astrophys.}
  {\bfseries 641} (2020) A10},
  [\href{https://arxiv.org/abs/1807.06211}{{\ttfamily 1807.06211}}].

\bibitem{Akrami:2019izv}
{\scshape Planck} collaboration, Y.~Akrami et~al., \emph{{Planck 2018 results.
  IX. Constraints on primordial non-Gaussianity}},
  \href{https://doi.org/10.1051/0004-6361/201935891}{\emph{Astron. Astrophys.}
  {\bfseries 641} (2020) A9},
  [\href{https://arxiv.org/abs/1905.05697}{{\ttfamily 1905.05697}}].

\bibitem{Allahverdi:2020bys}
R.~Allahverdi et~al., \emph{{The First Three Seconds: a Review of Possible
  Expansion Histories of the Early Universe}},
  \href{https://arxiv.org/abs/2006.16182}{{\ttfamily 2006.16182}}.

\bibitem{Carr:2018nkm}
B.~Carr, K.~Dimopoulos, C.~Owen and T.~Tenkanen, \emph{{Primordial Black Hole
  Formation During Slow Reheating After Inflation}},
  \href{https://doi.org/10.1103/PhysRevD.97.123535}{\emph{Phys. Rev. D}
  {\bfseries 97} (2018) 123535},
  [\href{https://arxiv.org/abs/1804.08639}{{\ttfamily 1804.08639}}].

\bibitem{Coughlan:1983ci}
G.~Coughlan, W.~Fischler, E.~W. Kolb, S.~Raby and G.~G. Ross,
  \emph{{Cosmological Problems for the Polonyi Potential}},
  \href{https://doi.org/10.1016/0370-2693(83)91091-2}{\emph{Phys. Lett. B}
  {\bfseries 131} (1983) 59--64}.

\bibitem{Banks:1993en}
T.~Banks, D.~B. Kaplan and A.~E. Nelson, \emph{{Cosmological implications of
  dynamical supersymmetry breaking}},
  \href{https://doi.org/10.1103/PhysRevD.49.779}{\emph{Phys. Rev. D} {\bfseries
  49} (1994) 779--787}, [\href{https://arxiv.org/abs/hep-ph/9308292}{{\ttfamily
  hep-ph/9308292}}].

\bibitem{deCarlos:1993wie}
B.~de~Carlos, J.~Casas, F.~Quevedo and E.~Roulet, \emph{{Model independent
  properties and cosmological implications of the dilaton and moduli sectors of
  4-d strings}},
  \href{https://doi.org/10.1016/0370-2693(93)91538-X}{\emph{Phys. Lett. B}
  {\bfseries 318} (1993) 447--456},
  [\href{https://arxiv.org/abs/hep-ph/9308325}{{\ttfamily hep-ph/9308325}}].

\bibitem{Conlon:2005jm}
J.~P. Conlon and F.~Quevedo, \emph{{Kahler moduli inflation}},
  \href{https://doi.org/10.1088/1126-6708/2006/01/146}{\emph{JHEP} {\bfseries
  01} (2006) 146}, [\href{https://arxiv.org/abs/hep-th/0509012}{{\ttfamily
  hep-th/0509012}}].

\bibitem{Cicoli:2016olq}
M.~Cicoli, K.~Dutta, A.~Maharana and F.~Quevedo, \emph{{Moduli Vacuum
  Misalignment and Precise Predictions in String Inflation}},
  \href{https://doi.org/10.1088/1475-7516/2016/08/006}{\emph{JCAP} {\bfseries
  08} (2016) 006}, [\href{https://arxiv.org/abs/1604.08512}{{\ttfamily
  1604.08512}}].

\bibitem{Bhattacharya:2017ysa}
S.~Bhattacharya, K.~Dutta and A.~Maharana, \emph{{Constraints on K\"ahler
  moduli inflation from reheating}},
  \href{https://doi.org/10.1103/PhysRevD.96.083522}{\emph{Phys. Rev. D}
  {\bfseries 96} (2017) 083522},
  [\href{https://arxiv.org/abs/1707.07924}{{\ttfamily 1707.07924}}].

\bibitem{Maharana:2017fui}
A.~Maharana and I.~Zavala, \emph{{Postinflationary scalar tensor cosmology and
  inflationary parameters}},
  \href{https://doi.org/10.1103/PhysRevD.97.123518}{\emph{Phys. Rev. D}
  {\bfseries 97} (2018) 123518},
  [\href{https://arxiv.org/abs/1712.07071}{{\ttfamily 1712.07071}}].

\bibitem{Peebles:1998qn}
P.~Peebles and A.~Vilenkin, \emph{{Quintessential inflation}},
  \href{https://doi.org/10.1103/PhysRevD.59.063505}{\emph{Phys. Rev. D}
  {\bfseries 59} (1999) 063505},
  [\href{https://arxiv.org/abs/astro-ph/9810509}{{\ttfamily
  astro-ph/9810509}}].

\bibitem{DiMarco:2018bnw}
A.~Di~Marco, G.~Pradisi and P.~Cabella, \emph{{Inflationary scale, reheating
  scale, and pre-BBN cosmology with scalar fields}},
  \href{https://doi.org/10.1103/PhysRevD.98.123511}{\emph{Phys. Rev. D}
  {\bfseries 98} (2018) 123511},
  [\href{https://arxiv.org/abs/1807.05916}{{\ttfamily 1807.05916}}].

\bibitem{1981SvA....25..406P}
A.~G. {Polnarev} and M.~Y. {Khlopov}, \emph{{Primordial Black Holes and the ERA
  of Superheavy Particle Dominance in the Early Universe}}, {\emph{\sovast}
  {\bfseries 25} (Aug., 1981) 406}.

\bibitem{1982AZh....59..639P}
A.~G. {Polnarev} and I.~M. {Khlopov}, \emph{{Dustlike stages in the early
  universe, and constraints on the primordial black hole spectrum}},
  {\emph{\azh} {\bfseries 59} (Aug., 1982) 639--646}.

\bibitem{Hwang:2012bi}
J.-c. Hwang, H.~Noh and J.-O. Gong, \emph{{Second order solutions of
  cosmological perturbation in the matter dominated era}},
  \href{https://doi.org/10.1088/0004-637X/752/1/50}{\emph{Astrophys. J.}
  {\bfseries 752} (2012) 50},
  [\href{https://arxiv.org/abs/1204.3345}{{\ttfamily 1204.3345}}].

\bibitem{Alabidi:2013lya}
L.~Alabidi, K.~Kohri, M.~Sasaki and Y.~Sendouda, \emph{{Observable induced
  gravitational waves from an early matter phase}},
  \href{https://doi.org/10.1088/1475-7516/2013/05/033}{\emph{JCAP} {\bfseries
  05} (2013) 033}, [\href{https://arxiv.org/abs/1303.4519}{{\ttfamily
  1303.4519}}].

\bibitem{Harada:2016mhb}
T.~Harada, C.-M. Yoo, K.~Kohri, K.-i. Nakao and S.~Jhingan, \emph{{Primordial
  black hole formation in the matter-dominated phase of the Universe}},
  \href{https://doi.org/10.3847/1538-4357/833/1/61}{\emph{Astrophys. J.}
  {\bfseries 833} (2016) 61},
  [\href{https://arxiv.org/abs/1609.01588}{{\ttfamily 1609.01588}}].

\bibitem{Carr:2017edp}
B.~Carr, T.~Tenkanen and V.~Vaskonen, \emph{{Primordial black holes from
  inflaton and spectator field perturbations in a matter-dominated era}},
  \href{https://doi.org/10.1103/PhysRevD.96.063507}{\emph{Phys. Rev. D}
  {\bfseries 96} (2017) 063507},
  [\href{https://arxiv.org/abs/1706.03746}{{\ttfamily 1706.03746}}].

\bibitem{Harada:2017fjm}
T.~Harada, C.-M. Yoo, K.~Kohri and K.-I. Nakao, \emph{{Spins of primordial
  black holes formed in the matter-dominated phase of the Universe}},
  \href{https://doi.org/10.1103/PhysRevD.96.083517}{\emph{Phys. Rev. D}
  {\bfseries 96} (2017) 083517},
  [\href{https://arxiv.org/abs/1707.03595}{{\ttfamily 1707.03595}}].

\bibitem{Inomata:2019zqy}
K.~Inomata, K.~Kohri, T.~Nakama and T.~Terada, \emph{{Gravitational Waves
  Induced by Scalar Perturbations during a Gradual Transition from an Early
  Matter Era to the Radiation Era}},
  \href{https://doi.org/10.1088/1475-7516/2019/10/071}{\emph{JCAP} {\bfseries
  10} (2019) 071}, [\href{https://arxiv.org/abs/1904.12878}{{\ttfamily
  1904.12878}}].

\bibitem{Inomata:2019ivs}
K.~Inomata, K.~Kohri, T.~Nakama and T.~Terada, \emph{{Enhancement of
  Gravitational Waves Induced by Scalar Perturbations due to a Sudden
  Transition from an Early Matter Era to the Radiation Era}},
  \href{https://doi.org/10.1103/PhysRevD.100.043532}{\emph{Phys. Rev. D}
  {\bfseries 100} (2019) 043532},
  [\href{https://arxiv.org/abs/1904.12879}{{\ttfamily 1904.12879}}].

\bibitem{Figueroa:2019paj}
D.~G. Figueroa and E.~H. Tanin, \emph{{Ability of LIGO and LISA to probe the
  equation of state of the early Universe}},
  \href{https://doi.org/10.1088/1475-7516/2019/08/011}{\emph{JCAP} {\bfseries
  1908} (2019) 011}, [\href{https://arxiv.org/abs/1905.11960}{{\ttfamily
  1905.11960}}].

\bibitem{Matsubara:2019qzv}
T.~Matsubara, T.~Terada, K.~Kohri and S.~Yokoyama, \emph{{Clustering of
  primordial black holes formed in a matter-dominated epoch}},
  \href{https://doi.org/10.1103/PhysRevD.100.123544}{\emph{Phys. Rev. D}
  {\bfseries 100} (2019) 123544},
  [\href{https://arxiv.org/abs/1909.04053}{{\ttfamily 1909.04053}}].

\bibitem{Bhattacharya:2019bvk}
S.~Bhattacharya, S.~Mohanty and P.~Parashari, \emph{{Primordial black holes and
  gravitational waves in nonstandard cosmologies}},
  \href{https://doi.org/10.1103/PhysRevD.102.043522}{\emph{Phys. Rev. D}
  {\bfseries 102} (2020) 043522},
  [\href{https://arxiv.org/abs/1912.01653}{{\ttfamily 1912.01653}}].

\bibitem{Harada:2013epa}
T.~Harada, C.-M. Yoo and K.~Kohri, \emph{{Threshold of primordial black hole
  formation}}, \href{https://doi.org/10.1103/PhysRevD.88.084051}{\emph{Phys.
  Rev. D} {\bfseries 88} (2013) 084051},
  [\href{https://arxiv.org/abs/1309.4201}{{\ttfamily 1309.4201}}].

\bibitem{Zhao:2013bba}
W.~Zhao, Y.~Zhang, X.-P. You and Z.-H. Zhu, \emph{{Constraints of relic
  gravitational waves by pulsar timing arrays: Forecasts for the FAST and SKA
  projects}}, \href{https://doi.org/10.1103/PhysRevD.87.124012}{\emph{Phys.
  Rev.} {\bfseries D87} (2013) 124012},
  [\href{https://arxiv.org/abs/1303.6718}{{\ttfamily 1303.6718}}].

\bibitem{Liu:2015psa}
X.-J. Liu, W.~Zhao, Y.~Zhang and Z.-H. Zhu, \emph{{Detecting Relic
  Gravitational Waves by Pulsar Timing Arrays: Effects of Cosmic Phase
  Transitions and Relativistic Free-Streaming Gases}},
  \href{https://doi.org/10.1103/PhysRevD.93.024031}{\emph{Phys. Rev.}
  {\bfseries D93} (2016) 024031},
  [\href{https://arxiv.org/abs/1509.03524}{{\ttfamily 1509.03524}}].

\bibitem{Bernal:2019lpc}
N.~Bernal and F.~Hajkarim, \emph{{Primordial Gravitational Waves in Nonstandard
  Cosmologies}}, \href{https://doi.org/10.1103/PhysRevD.100.063502}{\emph{Phys.
  Rev.} {\bfseries D100} (2019) 063502},
  [\href{https://arxiv.org/abs/1905.10410}{{\ttfamily 1905.10410}}].

\bibitem{Bernal:2020ywq}
N.~Bernal, A.~Ghoshal, F.~Hajkarim and G.~Lambiase, \emph{{Primordial
  Gravitational Wave Signals in Modified Cosmologies}},
  \href{https://arxiv.org/abs/2008.04959}{{\ttfamily 2008.04959}}.

\bibitem{Akrami:2018vks}
{\scshape Planck} collaboration, Y.~Akrami et~al., \emph{{Planck 2018 results.
  I. Overview and the cosmological legacy of Planck}},
  \href{https://doi.org/10.1051/0004-6361/201833880}{\emph{Astron. Astrophys.}
  {\bfseries 641} (2020) A1},
  [\href{https://arxiv.org/abs/1807.06205}{{\ttfamily 1807.06205}}].

\bibitem{Aghanim:2018eyx}
{\scshape Planck} collaboration, N.~Aghanim et~al., \emph{{Planck 2018 results.
  VI. Cosmological parameters}},
  \href{https://doi.org/10.1051/0004-6361/201833910}{\emph{Astron. Astrophys.}
  {\bfseries 641} (2020) A6},
  [\href{https://arxiv.org/abs/1807.06209}{{\ttfamily 1807.06209}}].

\bibitem{Kobayashi:2010cm}
T.~Kobayashi, M.~Yamaguchi and J.~Yokoyama, \emph{{G-inflation: Inflation
  driven by the Galileon field}},
  \href{https://doi.org/10.1103/PhysRevLett.105.231302}{\emph{Phys. Rev. Lett.}
  {\bfseries 105} (2010) 231302},
  [\href{https://arxiv.org/abs/1008.0603}{{\ttfamily 1008.0603}}].

\bibitem{Baldi:2005gk}
M.~Baldi, F.~Finelli and S.~Matarrese, \emph{{Inflation with violation of the
  null energy condition}},
  \href{https://doi.org/10.1103/PhysRevD.72.083504}{\emph{Phys. Rev. D}
  {\bfseries 72} (2005) 083504},
  [\href{https://arxiv.org/abs/astro-ph/0505552}{{\ttfamily
  astro-ph/0505552}}].

\bibitem{Mukohyama:2014gba}
S.~Mukohyama, R.~Namba, M.~Peloso and G.~Shiu, \emph{{Blue Tensor Spectrum from
  Particle Production during Inflation}},
  \href{https://doi.org/10.1088/1475-7516/2014/08/036}{\emph{JCAP} {\bfseries
  08} (2014) 036}, [\href{https://arxiv.org/abs/1405.0346}{{\ttfamily
  1405.0346}}].

\bibitem{Ashoorioon:2014nta}
A.~Ashoorioon, K.~Dimopoulos, M.~M. Sheikh-Jabbari and G.~Shiu,
  \emph{{Non-Bunch\textendash{}Davis initial state reconciles chaotic models
  with BICEP and Planck}},
  \href{https://doi.org/10.1016/j.physletb.2014.08.038}{\emph{Phys. Lett. B}
  {\bfseries 737} (2014) 98--102},
  [\href{https://arxiv.org/abs/1403.6099}{{\ttfamily 1403.6099}}].

\bibitem{Gong:2014qga}
J.-O. Gong, \emph{{Blue running of the primordial tensor spectrum}},
  \href{https://doi.org/10.1088/1475-7516/2014/07/022}{\emph{JCAP} {\bfseries
  07} (2014) 022}, [\href{https://arxiv.org/abs/1403.5163}{{\ttfamily
  1403.5163}}].

\bibitem{Allen:1997ad}
B.~Allen and J.~D. Romano, \emph{{Detecting a stochastic background of
  gravitational radiation: Signal processing strategies and sensitivities}},
  \href{https://doi.org/10.1103/PhysRevD.59.102001}{\emph{Phys. Rev.}
  {\bfseries D59} (1999) 102001},
  [\href{https://arxiv.org/abs/gr-qc/9710117}{{\ttfamily gr-qc/9710117}}].

\bibitem{Smith:2006nka}
T.~L. Smith, E.~Pierpaoli and M.~Kamionkowski, \emph{{A new cosmic microwave
  background constraint to primordial gravitational waves}},
  \href{https://doi.org/10.1103/PhysRevLett.97.021301}{\emph{Phys. Rev. Lett.}
  {\bfseries 97} (2006) 021301},
  [\href{https://arxiv.org/abs/astro-ph/0603144}{{\ttfamily
  astro-ph/0603144}}].

\bibitem{Boyle:2007zx}
L.~A. Boyle and A.~Buonanno, \emph{{Relating gravitational wave constraints
  from primordial nucleosynthesis, pulsar timing, laser interferometers, and
  the CMB: Implications for the early Universe}},
  \href{https://doi.org/10.1103/PhysRevD.78.043531}{\emph{Phys. Rev.}
  {\bfseries D78} (2008) 043531},
  [\href{https://arxiv.org/abs/0708.2279}{{\ttfamily 0708.2279}}].

\bibitem{Ben-Dayan:2019gll}
I.~Ben-Dayan, B.~Keating, D.~Leon and I.~Wolfson, \emph{{Constraints on scalar
  and tensor spectra from $N_{eff}$}},
  \href{https://doi.org/10.1088/1475-7516/2019/06/007}{\emph{JCAP} {\bfseries
  1906} (2019) 007}, [\href{https://arxiv.org/abs/1903.11843}{{\ttfamily
  1903.11843}}].

\bibitem{Maggiore:1999vm}
M.~Maggiore, \emph{{Gravitational wave experiments and early universe
  cosmology}}, \href{https://doi.org/10.1016/S0370-1573(99)00102-7}{\emph{Phys.
  Rept.} {\bfseries 331} (2000) 283--367},
  [\href{https://arxiv.org/abs/gr-qc/9909001}{{\ttfamily gr-qc/9909001}}].

\bibitem{Aver:2015iza}
E.~Aver, K.~A. Olive and E.~D. Skillman, \emph{{The effects of He I $\lambda$
  10830 on helium abundance determinations}},
  \href{https://doi.org/10.1088/1475-7516/2015/07/011}{\emph{JCAP} {\bfseries
  1507} (2015) 011}, [\href{https://arxiv.org/abs/1503.08146}{{\ttfamily
  1503.08146}}].

\bibitem{Cooke:2017cwo}
R.~J. Cooke, M.~Pettini and C.~C. Steidel, \emph{{One Percent Determination of
  the Primordial Deuterium Abundance}},
  \href{https://doi.org/10.3847/1538-4357/aaab53}{\emph{Astrophys. J.}
  {\bfseries 855} (2018) 102},
  [\href{https://arxiv.org/abs/1710.11129}{{\ttfamily 1710.11129}}].

\bibitem{Hsyu:2020uqb}
T.~Hsyu, R.~J. Cooke, J.~X. Prochaska and M.~Bolte, \emph{{The PHLEK Survey: A
  New Determination of the Primordial Helium Abundance}},
  \href{https://doi.org/10.3847/1538-4357/ab91af}{\emph{Astrophys. J.}
  {\bfseries 896} (2020) 77},
  [\href{https://arxiv.org/abs/2005.12290}{{\ttfamily 2005.12290}}].

\bibitem{Aiola:2020azj}
{\scshape ACT} collaboration, S.~Aiola et~al., \emph{{The Atacama Cosmology
  Telescope: DR4 Maps and Cosmological Parameters}},
  \href{https://arxiv.org/abs/2007.07288}{{\ttfamily 2007.07288}}.

\bibitem{LIGOScientific:2019vic}
{\scshape LIGO Scientific, Virgo} collaboration, B.~Abbott et~al.,
  \emph{{Search for the isotropic stochastic background using data from
  Advanced LIGO\textquoteright{}s second observing run}},
  \href{https://doi.org/10.1103/PhysRevD.100.061101}{\emph{Phys. Rev. D}
  {\bfseries 100} (2019) 061101},
  [\href{https://arxiv.org/abs/1903.02886}{{\ttfamily 1903.02886}}].

\bibitem{Domenech:2019quo}
G.~Dom\`enech, \emph{{Induced gravitational waves in a general cosmological
  background}}, \href{https://doi.org/10.1142/S0218271820500285}{\emph{Int. J.
  Mod. Phys. D} {\bfseries 29} (2020) 2050028},
  [\href{https://arxiv.org/abs/1912.05583}{{\ttfamily 1912.05583}}].

\bibitem{Byrnes:2018txb}
C.~T. Byrnes, P.~S. Cole and S.~P. Patil, \emph{{Steepest growth of the power
  spectrum and primordial black holes}},
  \href{https://doi.org/10.1088/1475-7516/2019/06/028}{\emph{JCAP} {\bfseries
  06} (2019) 028}, [\href{https://arxiv.org/abs/1811.11158}{{\ttfamily
  1811.11158}}].

\bibitem{Niikura:2017zjd}
H.~Niikura et~al., \emph{{Microlensing constraints on primordial black holes
  with Subaru/HSC Andromeda observations}},
  \href{https://doi.org/10.1038/s41550-019-0723-1}{\emph{Nature Astron.}
  {\bfseries 3} (2019) 524--534},
  [\href{https://arxiv.org/abs/1701.02151}{{\ttfamily 1701.02151}}].

\bibitem{Tisserand:2006zx}
{\scshape EROS-2} collaboration, P.~Tisserand et~al., \emph{{Limits on the
  Macho Content of the Galactic Halo from the EROS-2 Survey of the Magellanic
  Clouds}}, \href{https://doi.org/10.1051/0004-6361:20066017}{\emph{Astron.
  Astrophys.} {\bfseries 469} (2007) 387--404},
  [\href{https://arxiv.org/abs/astro-ph/0607207}{{\ttfamily
  astro-ph/0607207}}].

\bibitem{Niikura:2019kqi}
H.~Niikura, M.~Takada, S.~Yokoyama, T.~Sumi and S.~Masaki, \emph{{Constraints
  on Earth-mass primordial black holes from OGLE 5-year microlensing events}},
  \href{https://doi.org/10.1103/PhysRevD.99.083503}{\emph{Phys. Rev. D}
  {\bfseries 99} (2019) 083503},
  [\href{https://arxiv.org/abs/1901.07120}{{\ttfamily 1901.07120}}].

\bibitem{Bird:2016dcv}
S.~Bird, I.~Cholis, J.~B. Mu\~noz, Y.~Ali-Ha\"\i{}moud, M.~Kamionkowski, E.~D.
  Kovetz et~al., \emph{{Did LIGO detect dark matter?}},
  \href{https://doi.org/10.1103/PhysRevLett.116.201301}{\emph{Phys. Rev. Lett.}
  {\bfseries 116} (2016) 201301},
  [\href{https://arxiv.org/abs/1603.00464}{{\ttfamily 1603.00464}}].

\bibitem{Clesse:2016vqa}
S.~Clesse and J.~Garc\'\i{}a-Bellido, \emph{{The clustering of massive
  Primordial Black Holes as Dark Matter: measuring their mass distribution with
  Advanced LIGO}},
  \href{https://doi.org/10.1016/j.dark.2016.10.002}{\emph{Phys. Dark Univ.}
  {\bfseries 15} (2017) 142--147},
  [\href{https://arxiv.org/abs/1603.05234}{{\ttfamily 1603.05234}}].

\bibitem{Sasaki:2016jop}
M.~Sasaki, T.~Suyama, T.~Tanaka and S.~Yokoyama, \emph{{Primordial Black Hole
  Scenario for the Gravitational-Wave Event GW150914}},
  \href{https://doi.org/10.1103/PhysRevLett.117.061101}{\emph{Phys. Rev. Lett.}
  {\bfseries 117} (2016) 061101},
  [\href{https://arxiv.org/abs/1603.08338}{{\ttfamily 1603.08338}}].

\end{thebibliography}\endgroup

\end{document}